\documentclass[hyper,a4paper]{JHEP}
   
\usepackage{epsfig}
\usepackage{latexsym}
\usepackage{graphicx}
\usepackage{amsmath}
\usepackage{amsfonts}   % if you want the fonts
\usepackage{amssymb}    % if you want extra symbols
\usepackage{float}
\usepackage{bm}
\def\tanb{\tan\beta}

\def\ee{e^+e^-\to h^0Z}

\def\lsim{\raise0.3ex\hbox{$\;<$\kern-0.75em\raise-1.1ex\hbox{$\sim\;$}}}
\def\gsim{\raise0.3ex\hbox{$\;>$\kern-0.75em\raise-1.1ex\hbox{$\sim\;$}}}
\newcommand{\captions}{\sf\caption}
\def    \be            {\begin{equation}}
\def    \ee            {\end{equation}}
\def    \bea           {\begin{eqnarray}}
\def    \eea           {\end{eqnarray}}

\def\tanb{\tan \beta}
\def\vi{\nu_i}
\def\vci{\nu^c_i}
\def\vd{v_d}
\def\vu{v_u}
\def\mHd{m_{H_d}}
\def\mHu{m_{H_u}}

\def\sw2{sin^2 \theta_w}
\def\drbar{\overline{DR}}
\def\msbar{\overline{MS}}

\def\a^tau{\alpha_{\tau}}

\def\beq{\begin{equation}}
\def\eeq{\end{equation}}
\def\beqa{\begin{eqnarray}}
\def\eeqa{\end{eqnarray}}
\def\vb#1{\vbox to #1 pt{}}
\def\smallfrac#1#2{{\textstyle{#1 \over #2}}}

\title{Analysis of the parameter space and spectrum \\of the $\mu \nu$SSM }

\author{Nicol\'as Escudero\\
        Departamento de F\'{\i}sica Te\'{o}rica C-XI
        and Instituto de F\'{\i}sica Te\'{o}rica UAM/CSIC,\\
        Universidad Aut\'{o}noma de Madrid, Cantoblanco,
        28049 Madrid, Spain\\
        E-mail: \email{nicolas.escudero@uam.es}}
\author{Daniel E. L\'{o}pez-Fogliani\\
       Department of Physics and Astronomy, University of Sheffield,\\
        Sheffield S3 7RH, England\\
        E-mail: \email{d.lopez@sheffield.ac.uk}}
\author{Carlos Mu\~noz\\
        Departamento de F\'{\i}sica Te\'{o}rica C-XI
        and Instituto de F\'{\i}sica Te\'{o}rica UAM/CSIC,\\
        Universidad Aut\'{o}noma de Madrid, Cantoblanco,
        28049 Madrid, Spain\\
        E-mail: \email{carlos.munnoz@uam.es}}
\author{Roberto Ruiz de Austri\\
        Departamento de F\'{\i}sica Te\'{o}rica C-XI
        and Instituto de F\'{\i}sica Te\'{o}rica UAM/CSIC,\\
        Universidad Aut\'{o}noma de Madrid, Cantoblanco,
       28049 Madrid, Spain\\
        E-mail: \email{rruiz@delta.ft.uam.es}}

\abstract{\small 
The $\mu\nu$SSM is a supersymmetric standard model 
that solves the $\mu$ problem of the MSSM using the $R$-parity breaking couplings
between
the right-handed 
neutrino superfields and the Higgses in the superpotential, 
$\lambda_{i} \, \hat \nu^c_i \hat H_d \hat H_u$.
The $\mu$ term is generated spontaneously through  
sneutrino vacuum expectation values,
$\mu=
\lambda_i \langle \tilde \nu^c_i \rangle$, once the electroweak symmetry is 
broken. 
In addition, the couplings 
$\kappa_{ijk} \hat \nu^c_i\hat \nu^c_j\hat \nu^c_k$
forbid a global U(1) symmetry avoiding the existence of a Goldstone boson,
and also contribute to spontaneously generate Majorana masses for neutrinos at the
electroweak scale.
Following this proposal, we have analysed in detail 
the parameter space of the $\mu\nu$SSM.
In particular, we have studied viable regions avoiding false minima and tachyons, as well as fulfilling the Landau pole constraint.
We have also computed the associated spectrum, paying special attention to the mass of the lightest Higgs. The presence of right and left-handed sneutrino vacuum expectation values leads to a peculiar structure for the mass matrices.
The most important consequence is that neutralinos are mixed with neutrinos, and 
neutral Higgses with sneutrinos.

 }

\keywords{Supersymmetric Effective Theories, Beyond Standard Model, Supersymmetry Phenomenology}
\preprint{
\rightline{FTUAM 08/16, IFT-UAM/CSIC-08-57, October 2008}
}

\begin{document}
%%%%%%%%%%%%%%%%%%%%%%%%%%%%%%%%%%%%%%%%%%%%%%%%%%
%                                                %
%    BEGINNING OF TEXT                           %
%                                                %
%%%%%%%%%%%%%%%%%%%%%%%%%%%%%%%%%%%%%%%%%%%%%%%%%%

%%%%%%%%%%%%%%%%%%%%%%%%%%%%%%%%%%%%%%%%%%%%%%%
\section{Introduction}
%%%%%%%%%%%%%%%%%%%%%%%%%%%%%%%%%%%%%%%%%%%%%%%

%The minimal supersymmetric extension of the standard model (SM), 
%the well known 
The Minimal Supersymmetric Standard Model (MSSM) \cite{mssm}
is an attractive candidate for physics beyond the Standard Model.
It not only solves many theoretical puzzles but also
one expects to find its signatures in the forthcoming large 
hadron collider (LHC). 

However, in the MSSM 
lepton and baryon number violating terms in the superpotential
like 
$\epsilon_{ab} \left(\lambda_{ijk} \hat L_i^a \hat L_j^b \hat e^c_k 
+ 
\lambda'_{ijk} \hat L_i^a \hat Q_j^b \hat d^c_k 
+ 
\mu_i  \hat L_i^a\hat H_2^b \right)$
and
$\lambda''_{ijk} \hat d^c_i \hat d^c_j \hat u^c_k$,
respectively, with
$i,j=1,2,3$ generation indices and 
$a,b=1,2$ 
$SU(2)$ indices,
%\begin{equation}
%%\begin{array}{rcl}
%\lambda d^cd^cu^c\ ,
%%\end{array}
%\label{baryon}
%\end{equation}
%%
%and
%\begin{equation}
%%\begin{array}{rcl}
%\lambda' QLd^c\ , \lambda'' LLe^c\ , 
%\epsilon LH_2\ ,
%%\end{array}
%\label{lepton}
%\end{equation}
%which break explicitly $R$-parity, 
are in principle allowed by gauge
invariance. As it is well known, to avoid too fast proton decay mediated by
the exchange of squarks of masses of the order of the electroweak scale,
the presence together of terms of the type  
$\hat L \hat Q \hat d^c$ and 
$\hat d^c \hat d^c \hat u^c$ must be forbidden, unless
we impose very stringent bounds such as e.g.
$\lambda'^*_{112}\dot \lambda''_{112} \lsim 2\times 10^{-27}$.
Clearly, these values for the couplings are not very natural, and
for constructing viable supersymmetric (SUSY) models one usually 
forbids at least one of the operators
$LQd^c$ or $u^cd^cd^c$. 
The other type of operators above are not so stringently supressed, and
therefore still a lot of freedom remains \cite{dreiner}.

One possibility to avoid the problem of proton decay in the MSSM is to impose
$R$-parity conservation (+1 for particles and -1 for superpartners).
Actually this forbids all the four operators above and thus protects the proton.
Nevertheless, the choice of $R$-parity is {\it ad hoc}. There are other discrete 
symmetries, 
like e.g. baryon triality which only forbids the baryon violating operators \cite{dreiner3}.
Obviously, for all these symmetries R-parity is violated.
Besides, in string constructions the matter superfields can be located in different sectors or have different extra $U(1)$ charges, in such a way that 
some operators violating $R$-parity can be forbidden \cite{old}, 
but others can be allowed.

The 
phenomenology of models where $R$-parity is broken differs substantially
from that of models where $R$-parity is 
conserved. Needless to mention, the LSP is no longer stable, and
therefore not all SUSY chains must yield
missing energy events at colliders.
In this context
the neutralino \cite{reviewmio} or the sneutrino \cite{sneutrino} 
are no longer candidates for the dark matter of the Universe.
Nevertheless, other SUSY particles
such as the
gravitino \cite{yamaguchi} or the axino \cite{axino} can still be used as candidates. Indeed, 
the well-known axion of the Standard Model can also be the cold dark matter. 
 
There is a large number of works in the 
literature \cite{Barbier} exploring the possibility
of $R$-parity breaking in SUSY models, and its consequences
for the detection of SUSY at the LHC \cite{dreiner2}.
%, they concentrate only on
%the usual baryon and lepton number violating terms (without righ-handed neutrinos).
%Thus they do not try to address the $\mu$ problem. 
For example, 
a popular model is the so-called Bilinear $R$-parity Violation (BRpV)
model \cite{hall},
where the bilinear terms   
$\epsilon_{ab}\ \mu_i  \hat L_i^a\hat H_2^b$
are added to the MSSM.
In this way it is in principle possible to generate neutrino masses 
through the mixing with the neutralinos without including right-handed neutrinos
in the model. One mass is generated at tree level, and the other two at one loop. 
%However, the $\mu$ problem is augmented with the three new bilinear terms.
Analyses of mass matrices \cite{mass} in the BRpV, as well as 
studies of signals at accelerators \cite{signals} 
have been extensively carried out
in the literature.
%, have been extensively analysed
%in the literature.
Other interesting models 
are those producing the spontaneous breaking of $R$-parity
through the vacuum expectation values (VEVs) of singlet fields \cite{vallee}.
In the context of the 
Next-to-Minimal Supersymmetric 
Standard Model (NMSSM) \cite{nmssm,nmssm5,nmssm2,nmssm8}, $R$-parity breaking models have also been studied 
\cite{pandita,kitano,Abada}.
For a recent review discussing the different SUSY models with and without $R$-parity proposed in the literature, see ref. \cite{variants}.

There are two strong motivations to consider extensions of the MSSM.
On the one hand, the fact that neutrino 
oscillations imply non-vanishing neutrino masses \cite{experiments}.
On the other hand, the existence of the $\mu$ problem \cite{mupb} arising
from the requirement of a SUSY mass term for the Higgs fields in the
superpotential, $\epsilon_{ab}\ \mu \hat H_d^a\hat H_u^b$, which
must be of the order of the electroweak scale in order to
successfully lead to electroweak symmetry breaking (EWSB). In the presence of a Grand Unified Theory (GUT) with a 
typical scale of the order of $10^{16}$ GeV, and/or a gravitational
theory at the Planck scale, one should explain how to obtain a mass term of the order of 
the electroweak scale.

Motivated by the above issues, two of the authors proposed \cite{NuMSSM,mulo2} to
supplement 
the superfields $\hat \nu_i$ 
%$i=1,2,3$, 
contained in the $SU(2)_L$-doublet, $\hat L_i$, with
gauge-singlet neutrino superfields $\hat \nu^c_i$ to solve the $\mu$ problem 
of the MSSM.
In addition to the usual trilinear Yukawa couplings for quarks and charged
leptons, 
and the bilinear $\mu$-term, the right-handed neutrino superfields allow the presence of new terms such as Yukawa couplings for neutrinos and possible
Majorana mass terms in the superpotential.
Besides, trilinear terms breaking $R$-parity explicitly such as 
$\epsilon_{ab} \lambda_{i} \, \hat \nu^c_i\,\hat H_d^a \hat H_u^b$
and
$\kappa_{ijk} \hat \nu^c_i\hat \nu^c_j\hat \nu^c_k$
are now also allowed by gauge invariance.
The $\mu$ term can be obtained dynamically from the former terms in the superpotential.
%$\epsilon_{ab}\ \lambda_{i} \, \hat \nu^c_i\,\hat H_d^a \hat H_u^b$.
When the electroweak symmetry is broken,
they generate the $\mu$ term spontaneously through right-handed sneutrino VEVs, 
$\mu=
\lambda_i \langle \tilde \nu^c_i \rangle$.
In addition, the terms  
$\kappa_{ijk} \hat \nu^c_i\hat \nu^c_j\hat \nu^c_k$
forbid
a global $U(1)$ symmetry in the superpotential,  
avoiding therefore the existence of a Goldstone boson.
Besides, they contribute to generate
effective Majorana masses for neutrinos at the electroweak scale.
Terms of the type
$\hat \nu^c \hat H_d \hat H_u$ and $\hat \nu^c \hat \nu^c \hat \nu^c$
have also been analysed as sources of the observed baryon asymmetry
in the Universe \cite{vallle} and of neutrino masses and bilarge 
mixing \cite{sri}, respectively.

The superpotencial including the above trilinear couplings with right-handed neutrino superfields, in addition to the
trilinear Yukawa couplings for quarks and leptons, defines the
so-called ``$\mu$ from $\nu$''
Supersymmetric Standard Model ($\mu\nu$SSM) \cite{NuMSSM}.
As discussed above, the presence of $R$-parity breaking couplings in the superpotential is not necessarily a problem, and actually the couplings 
of the $\mu\nu$SSM
are obviously harmless with respect to proton decay.
Let us also remark that, since they break explicitly lepton number, a Goldstone boson (Majoron) does not appear after spontaneous symmetry breaking.
As in the MSSM or NMSSM, the usual lepton and baryon number violating terms could also in principle be added to the superpotential. %Nevertheless, their presence
%is irrelevant for our discussion below (?????????????????????).
Actually, even if the terms  $\lambda'_{ijk} \hat L_i^a \hat Q_j^b \hat d^c_k$ 
are set to zero at the high-energy scale, one-loop corrections will generate them.
However, these contributions are very small, as we will see
in Appendix \ref{appx:rges}.

In the $\mu\nu$SSM the 
$\mu$ term is absent from the superpotential, as well as 
Majorana masses for neutrinos, and only dimensionless trilinear
couplings are present.
For this to happen we can invoke a $Z_3$ symmetry 
as it is usually done in the NMSSM. Nevertheless,
let us recall that this is actually what happens in string constructions, where 
the low-energy limit is determined by the massless string modes. Since
the massive modes are of the order of the string scale,
only trilinear couplings are present in the low-energy superpotential.
String theory seems to be relevant for the unification of
interactions, including gravity, and therefore this argument in favour of the
absence of bare mass terms in the superpotential is robust.

 Let us finally remark that since
the superpotential of the 
$\mu\nu$SSM contains only trilinear couplings, it
has a $Z_3$ symmetry, just like 
the NMSSM.
%, under which all chiral superfields transform as 
%$\Phi\to e^{2i\pi/3}\Phi$. 
Therefore, one expects to have also a 
cosmological domain wall problem \cite{wall,wall2} in this model. 
Nevertheless, 
the usual solution \cite{nowall} will also work in this 
case: non-renormalisable operators \cite{wall} in the superpotential can explicitly break the dangerous $Z_3$ symmetry, lifting the degeneracy of the 
three original vacua, and this can be done without introducing hierarchy 
problems. In addition, these operators can be chosen small enough as 
not to alter the low-energy phenomenology.

The differences between the $\mu\nu$SSM and other models proposed in
the literature to solve the $\mu$ problem are clear.
For example, in the $\mu\nu$SSM one solves
the problem without having to introduce
an extra singlet superfield as in the NMSSM, or a special form of the
Kahler potential \cite{musol}, or
superpotential couplings to the hidden sector \cite{condensados,condensados2}. 
It is also worth noticing here that previously studied $R$-parity breaking models do not try to address the $\mu$ problem. Actually, in the case of the
BRpV model the problem is augmented with the three new bilinear terms.

% On the other hand, these two type of terms break lepton number and $R$-parity 
% explicitly implying that
% the phenomenology of this model is very different from the one 
% of the MSSM/NMSSM.
% For example, the usual neutralinos are now mixed with the neutrinos.
% Since we have a generalized see-saw mechanism at the EW scale,
% for a Dirac mass of the heaviest neutrino of order the mass
% of the electron, $0.1$ MeV, an eigenvalue reproducing 
% the correct scale of the heaviest neutrino
% mass, 0.01 eV, is obtained. Playing with the hierarchies in the Dirac masses
% one can obtain the other neutrino masses.

%=========

Indeed the breaking of $R$-parity generates a peculiar structure for the mass matrices of the $\mu\nu$SSM.
The presence of right and left-handed sneutrino VEVs leads to the mixing of the
%very different from the MSSM and the NMSSM 
neutral gauginos and Higgsinos (neutralinos)
with the right and left-handed neutrinos producing a 10$\times$10 matrix. 
As discussed in ref. \cite{NuMSSM}, three
eigenvalues of this matrix are very small, reproducing the experimental 
results on neutrino masses.
Of course, other mass matrices are also modified.
This is the case for example of the Higgs boson mass matrices, where
the neutral Higgses are mixed with the sneutrinos. 
Likewise the
charged Higgses are mixed with the charged sleptons,
and the charged gauginos and Higgsinos (charginos) with the charged leptons.

The purpose of the present work 
is to extend the analysis of ref. \cite{NuMSSM}, where the
characteristics of the $\mu\nu$SSM were introduced, and
only some points concerning its phenomenology were sketched.
Several approximations were considered, and, in particular, only one generation of
sneutrinos were assumed to acquire VEVs.
Here we will work with the full three generations, analysing
the parameter space of the $\mu\nu$SSM in detail, and 
paying special attention
to the particle spectrum associated.

The outline of the paper is as follows:
%Our work is organized as follows.
In Section \ref{The model} we introduce the model, discussing in particular
its superpotential and the associated scalar potential.
%We briefly address the superpotential and the associated scalar potential of the model 
%the most relevant aspects of the $\mu \nu$SSM
%in Section \ref{The model}. 
In Section \ref{minimization} we examine
%pay special attention to 
the minimisation 
of the scalar potential.
Section \ref{sec:spectrum} is focused on the description of
the parameter space of the $\mu\nu$SSM.
% is described in Section \ref{sec:spectrum}.
In Section \ref{sec:strategy} we thoroughly discuss
the strategy followed for the analysis of the parameter space and the computation of the spectrum.
%is discussed in detail
%in Section \ref{sec:strategy}. 
Section \ref{results} is devoted to the presentation of the results.
Viable regions of the parameter space avoiding 
false minima and tachyons, as well as fulfilling the Landau pole constraint
on the couplings, are studied in detail.
The associated spectrum is then discussed,
paying special attention to the mass of the lightest neutral Higgs.
Finally, the conclusions are left for Section \ref{conclusions}.
Technical details of the model such as the mass matrices, couplings,
one-loop contributions, and relevant renormalisation group equations (RGEs),
are given in the Appendices.

\section{The model}
\label{The model}
%%%%%%%%%%%%%%%%%%%%%%%%%%%%%%%%%%%%%%%%%%%%%%%

The  superpotential of the $\mu \nu$SSM is given by \cite{NuMSSM}
\begin{align}\label{superpotential}
W = &
\ \epsilon_{ab} \left(
Y_{u_{ij}} \, \hat H_u^b\, \hat Q^a_i \, \hat u_j^c +
Y_{d_{ij}} \, \hat H_d^a\, \hat Q^b_i \, \hat d_j^c +
Y_{e_{ij}} \, \hat H_d^a\, \hat L^b_i \, \hat e_j^c +
Y_{\nu_{ij}} \, \hat H_u^b\, \hat L^a_i \, \hat \nu^c_j 
\right)
\nonumber\\
& 
-\epsilon{_{ab}} \lambda_{i} \, \hat \nu^c_i\,\hat H_d^a \hat H_u^b
+
\frac{1}{3}
\kappa{_{ijk}} 
\hat \nu^c_i\hat \nu^c_j\hat \nu^c_k \ ,
%\label{superpotential}
\end{align}
where we take $\hat H_d^T=(\hat H_d^0, \hat H_d^-)$, $\hat H_u^T=(\hat
H_u^+, \hat H_u^0)$, $\hat Q_i^T=(\hat u_i, \hat d_i)$, $\hat
L_i^T=(\hat \nu_i, \hat e_i)$, and
%$i,j,k=1,2,3$ are family indices, 
%the entries of the 3$\times$3 matrices 
$Y$, $\lambda$, $\kappa$ are dimensionless
matrices, a vector, and a totally symmetric tensor, respectively. 
%dimensionless Yukawa couplings.
%$a,b=1,2$ are
%$SU(2)_L$ indices and $\epsilon_{12}=1$. 
%It is worth noticing here that
In the following the summation convention is implied on repeated indices.
In order to discuss the phenomenology of the $\mu \nu$SSM, and
working in the framework of gravity mediated SUSY breaking, we write 
the soft terms appearing in the Lagrangian, $\mathcal{L}_{\text{soft}}$, as
\begin{eqnarray}
-\mathcal{L}_{\text{soft}} & =&
 m_{\tilde{Q}_{ij} }^2\, \tilde{Q^a_i}^* \, \tilde{Q^a_j}
+m_{\tilde{u}_{ij}^c}^{2} 
\, \tilde{u^c_i}^* \, \tilde u^c_j
+m_{\tilde{d}_{ij}^c}^2 \, \tilde{d^c_i}^* \, \tilde d^c_j
+m_{\tilde{L}_{ij} }^2 \, \tilde{L^a_i}^* \, \tilde{L^a_j}
+m_{\tilde{e}_{ij} ^c}^2 \, \tilde{e^c_i}^* \, \tilde e^c_j
\nonumber \\
&+ &
m_{H_d}^2 \,{H^a_d}^*\,H^a_d + m_{H_u}^2 \,{H^a_u}^* H^a_u +
m_{\tilde{\nu}_{ij}^c}^2 \,\tilde{{\nu}^c_i}^* \tilde\nu^c_j 
\nonumber \\
&+&
\epsilon_{ab} \left[
(A_uY_u)_{ij} \, H_u^b\, \tilde Q^a_i \, \tilde u_j^c +
(A_dY_d)_{ij} \, H_d^a\, \tilde Q^b_i \, \tilde d_j^c +
(A_eY_e)_{ij} \, H_d^a\, \tilde L^b_i \, \tilde e_j^c 
\right.
\nonumber \\
&+&
\left.
(A_{\nu}Y_{\nu})_{ij} \, H_u^b\, \tilde L^a_i \, \tilde \nu^c_j 
+ \text{c.c.}
\right] 
\nonumber \\
&+&
\left[-\epsilon_{ab} (A_{\lambda}\lambda)_{i} \, \tilde \nu^c_i\, H_d^a  H_u^b
+
\frac{1}{3}
(A_{\kappa}\kappa)_{ijk} \, 
\tilde \nu^c_i \tilde \nu^c_j \tilde \nu^c_k\
+ \text{c.c.} \right]
\nonumber \\
&-&  \frac{1}{2}\, \left(M_3\, \tilde\lambda_3\, \tilde\lambda_3+M_2\,
  \tilde\lambda_2\, \tilde
\lambda_2
+M_1\, \tilde\lambda_1 \, \tilde\lambda_1 + \text{c.c.} \right) \,.
\label{2:Vsoft}
\end{eqnarray}
%
%one obtains from (\ref{2:Vsoft}) 
%the neutral scalars develop in general
%the VEVs:
%
%
%\begin{equation}\label{2:vevs}
%\langle H_d^0 \rangle = v_d \, , \quad
%\langle H_u^0 \rangle = v_u \, , \quad
%\langle \tilde \nu_i \rangle = \nu_i \, , \quad
%\langle \tilde \nu_i^c \rangle = \nu_i^c \,,
%\end{equation}
% 

In addition to terms from $\mathcal{L}_{\text{soft}}$, the tree-level scalar potential
receives the usual $D$ and $F$ term contributions.
Thus, the tree-level neutral scalar potential is given by
\begin{equation}
V^0 = V_{\text{soft}} + V_D  +  V_F\ , 
\label{finalpotential}
\end{equation}
where
\begin{eqnarray}
V_{\text{soft}} &=& 
 m_{H_d}^{2}H^0_{d}H^{0*}_{d}+m_{H_u}^{2}H^0_{u}H_{u}^{0*}+
m_{\tilde{L}_{ij} }^2 \, \tilde{\nu}_i \, \tilde{\nu}_j^* +
m_{\tilde{\nu}^c_{ij}}^{2}\tilde{\nu}^c_{i}\tilde{\nu}^{c*}_{j}
\nonumber\\
&+&
\left(
a_{\nu_{ij}}H^0_{u}\tilde{\nu}_{i}\tilde{\nu}^{c}_{j}
%m_{\tilde{\nu}_{i}}^{2}\tilde{\nu}_{i}\tilde{\nu}_{i}^{*}
-a_{\lambda_{i}}\tilde{\nu}^c_{i}H^0_{d}H^0_{u}
+
\frac{1}{3} {a_{\kappa_{ijk}}\tilde{\nu}^c_{i}\tilde{\nu}^c_{j}\tilde{\nu}^c_{k}}
+
\text{c.c.} \right)\ ,
\label{akappa}
\end{eqnarray}
with
$a_{\nu_{ij}}\equiv (A_\nu Y_\nu)_{ij}$, 
$a_{\lambda_i}\equiv (A_\lambda\lambda)_i$,  
$a_{\kappa_{ijk}}\equiv (A_\kappa \kappa)_{ijk}$,
%Besides the potential receives the $D$ and $F$-term
%contributions
\begin{equation}
V_D  =
\frac{G^2}{8}\left(\tilde{\nu}_{i}\tilde{\nu}_{i}^* 
                   + H^0_{d}H_{d}^{0*}-H_{u}H_{u}^{0*}\right)^{2},
\end{equation}
with $G^2\equiv g_{1}^{2}+g_{2}^{2}$, and
\begin{eqnarray}
V_{F}  &=&
 \lambda_{j}\lambda_{j}^{*}H^0_{d}H_{d}^{0*}H^0_{u}H_{u}^{0*}
 +\lambda_{i}\lambda_{j}^{*}H^0_{d}H_{d}^{0*}\tilde{\nu}^c_{i}
                                                   \tilde{\nu}^{c*}_{j}
 +\lambda_{i}\lambda_{j}^*H^0_{u}H_{u}^{0*}\tilde{\nu}^c_{i}
                                                   \tilde{\nu}^{c*}_{j}
 +\kappa_{ijk}\kappa_{ljm}^{*}\tilde{\nu}^c_{i}\tilde{\nu}^{c*}_{l}
                                   \tilde{\nu}^c_{k}\tilde{\nu}^{c*}_{m}
\nonumber\\
 &-& (\kappa_{ijk}\lambda_{j}^*H_{d}^{0*}H_{u}^{0*}
                                      \tilde{\nu}^c_{i}\tilde{\nu}^c_{k}
 -Y_{\nu_{ij}}\kappa_{ljk}^{*}H^0_{u}\tilde{\nu}_{i}\tilde{\nu}^{c*}_{l}
                                                     \tilde{\nu}^{c*}_{k}
 +Y_{\nu_{ij}}\lambda_{j}^{*}H_{d}^{0*}H_{u}^{0*}H^0_{u}\tilde{\nu}_{i} 
\nonumber \\
 &+& Y_{\nu_{ij}}^{*}\lambda_{k}H^0_{d}\tilde{\nu}^c_{k} \tilde{\nu}_{i}^{*}
                                  \tilde{\nu}^{c*}_{j} 
 + \text{c.c.}) 
\nonumber \\
 &+& Y_{\nu_{ij}}Y_{\nu_{ik}}^*H^0_{u}H_{u}^{0*} \tilde{\nu}^c_{j}
                                                \tilde{\nu}^{c*}_{k}
 +Y_{\nu_{ij}}Y_{\nu_{lk}}^{*}\tilde{\nu}_{i}\tilde{\nu}_{l}^{*}
                                    \tilde{\nu}^c_{j}\tilde{\nu}^{c*}_{k}
 +Y_{\nu_{ji}}Y_{\nu_{ki}}^{*}H^0_{u}H_{u}^{0*}\tilde{\nu}_{j}\tilde{\nu}_{k}^* \, .
\end{eqnarray}

Once the electroweak symmetry is spontaneously broken, the neutral scalars develop in general the following VEVs:
\begin{equation}\label{2:vevs}
\langle H_d^0 \rangle = v_d \, , \quad
\langle H_u^0 \rangle = v_u \, , \quad
\langle \tilde \nu_i \rangle = \nu_i \, , \quad
\langle \tilde \nu_i^c \rangle = \nu_i^c \,.
%\label{esperados}
\end{equation}
In the following we will assume for simplicity that all parameters in the potential are real. 
Although in 
%\nex{'}\ne{``}multi-Higgs\nex{'}\ne{''} 
'multi-Higgs' 
models with real parameters the VEVs of the neutral scalar fields 
can be in general complex \cite{M+3}, the analysis of this possibility
is beyond the scope of this work, and we leave it for a forthcoming publication, where spontaneous CP violation will be studied in detail \cite{FLMR}. 
Nevertheless, it is worth noticing here that this assumption of real VEVs is consistent once one guarantees that the minimum with null phases is the global one.
%Neglecting the effects of the relevant SUSY terms of the potential dependent on %the phases,
%$\lambda_i \nu^c_i \lambda_j^* \nu^{*c}_j (H_u^0 H_{u}^{0*} + H_d^0 H_{d}^{0*})$,
%which might make the positive contribution of the $F$ terms smaller,
It is straightforward to see that this is guaranteed in general for the
VEVs $v_u$, $v_d$, $\nu^c_i$, imposing the conditions
 $\lambda_i>0$, $\kappa_{iii}>0 $, $A_{\lambda_i}>0$, $A_{\kappa_{iii}}<0$, and $A_{\kappa_{ijk}}=\kappa_{ijk}= 0$ if $i= j =k$ is not satisfied. 
Concerning the VEVs $\nu_i$,
% Observing that the $Y_{\nu_{ij}}$ are really 
%small,  $Y_{\nu_{ij}} \lesssim 10^{-6}$ in order 
%to reproduce neutrino masses  given by the see-saw 
%mechanism at TeV scale \cite{NuMSSM}, for the 
%sneutrino sector 
it is sufficient to impose 
$Y_{\nu_{ii}}>0$, and $Y_{\nu_{ij}}=A_{\nu_{ij}}=0$ 
for $i \neq j$, with the extra condition 
% $\lambda_i v^2_u v_d + \lambda_j \nu^c_j 
% \nu^c_i v_d -A_{\nu_{ij}} v_u \nu^c_j - 
% \kappa_{ijk} \nu^c_j \nu^c_k v_u>0$.
\begin{equation}
\lambda_i v^2_u v_d + \lambda_j \nu^c_j 
\nu^c_i v_d -A_{\nu_{ij}} v_u \nu^c_j - 
\kappa_{ijk} \nu^c_j \nu^c_k v_u>0\ .
\label{anu}
\end{equation}
%NO HABRIA QUE PONER AQUI QUE TAMBIEN HAY QUE COGER TODOS LOS VEVS POSITIVOS?
The above conditions on the signs of the parameters, together with 
(\ref{anu}), will be used for the analysis of the
parameter space and spectrum of the $\mu\nu$SSM in Section \ref{results}.

\section{Minimisation of the potential}
\label{minimization}
%%%%%%%%%%%%%%%%%%%%%%%%%%%%%%%%%%%%%%%%%%%%%%

As mentioned above, the EWSB generates the VEVs written in eq. (\ref{2:vevs}).
% Once the electroweak symmetry is spontaneously broken, the neutral scalars develop in general the following VEVs:
% \begin{equation}\label{2:vevs}
% \langle H_d^0 \rangle = v_d \, , \quad
% \langle H_u^0 \rangle = v_u \, , \quad
% \langle \tilde \nu_i \rangle = \nu_i \, , \quad
% \langle \tilde \nu_i^c \rangle = \nu_i^c \,,
%\label{esperados}
% \end{equation}
Thus one can define as usual
%and as usual one can write
%Shifting the neutral fields with non-zero VEVs as 
\begin{eqnarray}
H^0_u = h_u+\,iP_u+v_u\ , \,\,\,\,\, H_d^0=h_d+\,iP_d+v_d,
\nonumber\\ 
\, \widetilde{\nu}^c_i=
     (\widetilde{\nu}^c_i)^R+\, i(\widetilde{\nu}_i^c)^I+\nu_i^c\ ,  
\,\,\,\,\, \widetilde{\nu}_i=(\widetilde{\nu}_i)^R  +\, i (\widetilde{\nu}_i)^I
  + \nu_i\ .
\end{eqnarray}
Then, the tree-level scalar potential contains the following 
linear terms:
\begin{equation}
% V^0_{\text linear} = t^0_d h_d + t^0_u h_u+ t^0_{\vci} (\vci)^R 
%                    + t^0_{\vi} (\vi)^R\ ,
V^0_{\text linear} = t^0_d h_d + t^0_u h_u+ t^0_{\vci} (\widetilde{\nu}^c_i)^R 
                   + t^0_{\vi} (\widetilde{\vi})^R\ ,                   
\end{equation} 
where the different $t^0$ are the tadpoles at tree-level. They
are equal to zero at the minimum of the tree-level potential, and 
are given by
\begin{align}
t^0_d & = \frac{1}{4}G^2\left(\nu_{i}\nu_{i}+v_{d}^{2}-v_{u}^{2}\right)v_{d}
   + m_{H_{d}}^{2}v_{d}-a_{\lambda_i}v_{u}\nu_i^c
   + \lambda_i\lambda_{j}v_{d}\nu_i^c \nu_j^c  
\nonumber\\ 
 & + \lambda_{i}\lambda_{i}v_{d}v_{u}^{2}
   - \lambda_{j}\kappa_{ijk}v_{u}\nu_i^c \nu_k^c 
   - Y_{\nu_{ij}} \lambda_k \nu_{i}\nu_k^c \nu_j^c
   - Y_{\nu_{ij}}\lambda_{j}v_{u}^2 \nu_{i}\ , \\ 
% t^0_d & = \frac{1}{4}G^2\left(\nu_{i}\nu_{i}+v_{d}^{2}-v_{u}^{2}\right)v_{d}
%    + m_{H_{d}}^{2}v_{d}-a_{\lambda_i}v_{u}\nu_i^c
%    + \lambda_i\lambda_{j}v_{d}\nu_i^c \nu_j^c  
% \nonumber\\ 
%  & + \lambda_{i}\lambda_{i}v_{d}v_{u}^{2}
%    - \left[\lambda_{j}\kappa_{ijk}v_{u}\nu_i^c \nu_k^c 
%    + Y_{\nu_{ij}} \lambda_k \nu_{i}\nu_k^c \nu_j^c
%    + Y_{\nu_{ij}}\lambda_{j}v_{u}^2 \nu_{i}
% \right], \\ 
\nonumber
\\
t^0_u & = -\frac{1}{4}G^2\left(\nu_{i}\nu_{i}+v_{d}^{2}-v_{u}^{2}\right)v_{u}
   + m_{H_{u}}^{2}v_{u}+a_{\nu_ {ij}}\nu_i\nu_j^c  -a_{\lambda_i}\nu_i^c v_{d}
\nonumber \\ 
&  + \lambda_{i}\lambda_jv_{u}\nu_i^c\nu_j^c+\lambda_{j}\lambda_{j}v_{d}^2v_u
   - \lambda_{j}\kappa_{ijk}v_{d}\nu_i^c \nu_k^c 
   + Y_{\nu_{ij}}\kappa_{ljk}\nu_{i}\nu_l^c \nu_k^c
\nonumber \\ 
 & - 2\lambda_j Y_{\nu_{ij}}v_{d} v_u \nu_{i}
   + Y_{\nu_{ij}}Y_{\nu_{ik}}v_{u}\nu_k^c\nu_{j}^{c}
   + Y_{\nu_{ij}}Y_{\nu_{kj}}v_{u}\nu_{i}\nu_{k} \ , \\ 
\nonumber\\
t^0_{\nu^c_i} & =  m^2_{\widetilde{\nu}_{ij}^{c}}\nu_{j}^{c} + 
a_{\nu_{ji}}\nu_{j}v_{u}
   - a_{\lambda_i}v_u v_d + a_{\kappa_{ijk}}\nu_{j}^{c}\nu_{k}^{c} 
   + \lambda_i\lambda_{j}v_{u}^{2}\nu_{j}^{c}    
   + \lambda_{i}\lambda_{j}v_{d}^{2}\nu_{j}^{c}  
\nonumber \\ 
  &- 2\lambda_{j}\kappa_{ijk}v_{d}v_{u}\nu_{k}^{c}
   + 2\kappa_{lim}\kappa_{ljk}\nu_{m}^{c}\nu_{j}^{c}\nu_{k}^{c}
   - Y_{\nu_{ji}}\lambda_{k}\nu_{j}\nu_{k}^{c}v_{d}
   - Y_{\nu_{kj}}\lambda_{i}v_{d}\nu_{k}\nu_{j}^{c}
\nonumber\\
  & + 2 Y_{\nu_{jk}}\kappa_{ikl}v_{u}\nu_{j}\nu_{l}^{c}
    + Y_{\nu_{ji}}Y_{\nu_{lk}}\nu_{j}\nu_{l}\nu_{k}^{c}
    + Y_{\nu_{ki}}Y_{\nu_{kj}}v_{u}^{2}\nu^c_{j} \ , \\ 
\nonumber\\
t^0_{\nu_i} & =  \frac{1}{4}G^2(\nu_{j}\nu_{j}+v_{d}^{2}-v_{u}^{2})\nu_{i}
   + m^2_{\widetilde{L}_{ij}}\nu_{j}+a_{\nu_{ij}}v_{u}\nu_{j}^{c} 
   - Y_{\nu_{ij}}\lambda_{k}v_{d}\nu_j^c \nu_k^c
\nonumber \\
%\label{tadpo}
 &  - Y_{\nu_{ij}}\lambda_{j}v_{u}^{2}v_{d} 
   + Y_{\nu_{il}}\kappa_{ljk}v_{u}\nu^c_{j}\nu^c_{k} 
   + Y_{\nu_{ij}} Y_{\nu_{lk}}\nu_{l}\nu_j^c\nu_k^c
   + Y_{\nu_{ik}}Y_{\nu_{jk}}v_{u}^{2}\nu_j \ .
\label{eq:tadpoles}
\end{align}

As it is well known, in order to find reliable results 
for the EWSB, it is necessary to 
include the one-loop radiative corrections. 
The effective scalar potential at one-loop level is then
\begin{equation}
V = V^0 + V^1 \,,
\end{equation}
where $V^1$ includes bubble diagrams at one-loop with all kinds 
of (s)particles running in the loop \cite{Zhang:1998bm}. Minimizing 
the full potential is equivalent to the requirement that the one-loop 
corrected tadpoles, $t = t^0 + t^1$, where $t^1$ represents 
the one-loop part, vanish. 

Let us finally remark that, since minima with some or all of the VEVs in 
eq. (\ref{2:vevs}) vanishing are in principle possible, one has to check that 
the minumum breaking the electroweak symmetry, and generating
the $\mu$ term spontaneously, is the global one.
This will be studied in detail when analyzing the parameter space of the model in Section \ref{pss}.

% our candidate to be a minimun is the global one comparing with
% the others real local minimun, assuming as we said that the global one is real.]

% The next step consists in performing a check to be sure that the 
% minimum which breaks the electroweak symmetry spontaneously
% is the global one. For it, we compute the global minimun using 
% a `genetic' algorithm for global optimisation \cite{gop} which has a 
% high performance and compare it with the physical one ***

 %%%%%%%%%%%%%%%%%%%%%%%%%%%%%%%%%%%%%%%%%%%%%%%
\section{ $\mu \nu$SSM  parameter space  
\label{sec:spectrum}}
%%%%%%%%%%%%%%%%%%%%%%%%%%%%%%%%%%%%%%%%%%%%%%%

At low energy the free parameters in the neutral scalar sector
are: 
$\lambda_i$, $\kappa_{ijk}$, $\mHd$, $\mHu$, $m_{\widetilde{L}_{ij}}$, 
$m_{\widetilde{\nu}_{ij}^{c}}$,
$A_{\lambda_i}$, $A_{\kappa_{ijk}}$, and $A_{\nu_{ij}}$.
Strong upper bounds upon the intergenerational scalar mixing 
exist \cite{Gabbiani:1996hi}, so in the following we assume that 
such mixings are negligible, and therefore the sfermion soft mass 
matrices are diagonal in the flavour space.
This occurs for example in several string compactifications as a consequence of 
having diagonal
Kahler metrics, or when the dilaton is the source of SUSY breaking \cite{bim}. 
Thus using the eight minimization conditions for the neutral scalar potential 
in the previous section,
one can eliminate the soft masses 
$\mHd$, $\mHu$, $m_{\widetilde{L}_{i}}$, and $m_{\widetilde{\nu}_{i}^{c}}$
in favour of the VEVs
$\vd$, $\vu$,
$\nu_i$, and $\nu^c_i$.
On the other hand, using the Standard Model Higgs VEV,
$v\approx 174$ GeV, 
$\tanb$, and $\vi$, one can determine the SUSY Higgs 
VEVs, $\vd$ and $\vu$, through $v^2 = \vd^2 + \vu^2 + \vi^2$. 
We thus consider as independent parameters
the following set of variables:
\bea
\lambda_i, \, \kappa_{ijk},\, \tan\beta, \, \nu_i, \nu^c_i, \, A_{\lambda_i}, \, A_{\kappa_{ijk}}, \, A_{\nu_{ij}}\ .
\label{freeparameters}
\eea

It is worth remarking here that
the VEVs of the left-handed sneutrinos,
%  $\tilde \nu$, 
$\nu_i$, are in general small.
Notice that in eq. (\ref{eq:tadpoles})
$\nu\to 0$ as $Y_{\nu}\to 0$
to fulfil $t^0_{\nu_i}=0$, and since
the couplings $Y_{\nu}$ determine the Dirac masses for the
neutrinos,  $Y_{\nu}v_u\sim m_D\lsim 10^{-4}$ GeV, the $\nu$'s have to be very small.
Using this rough argument one can also get an estimate of the values,
$\nu\lsim m_D$ \cite{NuMSSM}.
Then, since $\nu_i<<v_d, v_u$ we can define the above value of $\tanb$ as usual,
$\tan\beta=\frac{v_u}{v_d}$.

Assuming for simplicity that there is no intergenerational mixing
in the parameters of the model,
%eq. (\ref{freeparameters}), 
and that they have
the same values for the three families (with the 
exception of $\nu_i$ for which 
we need at least two generations with different VEVs in order to guarantee the
correct hierarchy of neutrino masses), the low-energy free parameters in our analysis will be
\bea
\lambda, \, \kappa,\, \tan\beta, \, \nu_1, \,  \nu_3, \nu^c, \, A_{\lambda}, \, A_{\kappa}, \, A_\nu\ ,
\label{freeparameters2}
\eea
where we have chosen $\nu_1=\nu_2\neq\nu_3$, and we have defined ${\lambda} \equiv {\lambda_i}$, 
$\kappa\equiv {\kappa_{iii}}$, 
$\nu^c \equiv \nu^c_i$, $A_{\lambda} \equiv A_{\lambda_i}$, 
$A_{\kappa} \equiv A_{\kappa_{iii}}$,
$A_{\nu} \equiv A_{\nu_{ii}}$.
Nevertheless, let us remark that the formulas given in the Appendices are for 
the general case, without assuming universality of the parameters or vanishing 
intergenerational mixing.

The soft SUSY-breaking terms, namely gaugino masses,
$M_{1,2,3}$, scalar masses, $m_{\tilde Q,\tilde u^c,\tilde d^c,\tilde e^c}$,
and trilinear parameters, $A_{u,d,e}$, are also
taken as free parameters and specified at low scale.
Data on neutrino masses, and the usual 
Standard Model parameters such as fermion and gauge boson masses,
the fine structure constant $\alpha (M_Z)$, the Fermi constant from 
muon decay $G_F^\mu$, and the strong coupling constant $\alpha_s(M_Z)$,  
will be used in the computation~\cite{pdg07}. Concerning the top mass,
we will take $m_t = 172.6$ GeV~\cite{topmass:mar08}.

 %%%%%%%%%%%%%%%%%%%%%%%%%%%%%%%%%%%%%%%%%%%%%%%
\section{Strategy for the analysis 
%of the $\mu\nu$SSM  
\label{sec:strategy}}
%%%%%%%%%%%%%%%%%%%%%%%%%%%%%%%%%%%%%%%%%%%%%%%

We now show the algorithm used in the analysis of the model.
In particular in the analysis of the parameter space, and in the computation of the
spectrum. 
%computation of the spectrum.
%Standard Model parameters (fermion and gauge boson masses,
%the fine structure constant $\alpha (M_Z)$, the Fermi constant from 
%muon decay $G_F^\mu$ and $\alpha_s(M_Z)$) and data on neutrino masses 
%from oscillations are used as constraints. 
%The soft SUSY breaking parameters and the superpotential parameters 
%$\lambda^i$ and $\kappa^{ijk}$ are then free parameters. 
%One of the neutrino masses, $\tanb$  and the sneutrino 
%left and right vevs ($\vi$ and $\vci$ respectively) are used  
%as free parameters as well. 
%However, in what follows, the soft masses associated with the 
%Higgses ($\mHd$, $\mHu$) and the ones with the sneutrinos 
%($\mL$, $\mNU$) are constrained by 
%imposing that electroweak symmetry is spontaneously broken. 
%First, we describe the evolution of the low-energy Standard Model input
%parameters below $M_Z$, then detail the rest of the algorithm.
% Standard Model parameters (fermion and gauge boson masses,
% the fine structure constant $\alpha (M_Z)$, the Fermi constant from 
% muon decay $G_F^\mu$ and $\alpha_s(M_Z)$) and data on neutrino masses 
% are used as constraints. The couplings 
% $\lambda^i$, $\kappa^{ijk}$, the soft SUSY breaking parameters and 
% the sneutrino left and right VEV's ($\vi$ and $\vci$ respectively) 
% are free parameters. However, in what follows, the soft masses associated 
% with the Higgses ($\mHd$, $\mHu$) and the ones with the sneutrinos 
% ($\mL$, $\mNU$) are constrained by imposing EWSB. 
Below $M_Z$, $\alpha(M_Z)$ and $\alpha_s(M_Z)$ are first evolved to 1 GeV 
using 3 loop QCD and 1 loop 
QED Standard Model $\beta$-functions \cite{Gorishnii:1990zu}. Then 
the two  gauge couplings and all Standard Model fermion masses except
the top quark mass are run to $M_Z$. The $\beta$-functions of fermion masses are
taken to be zero at renormalisation scales below their running masses.
The parameters at $M_Z$ are used as the low energy boundary condition in the
rest of the evolution.

%\subsection{Below $M_Z$}

%$\alpha(M_Z)$, $\alpha_s(M_Z)$ are first evolved to 1 GeV using 3 loop QCD and
%1 loop QED~\cite{Gorishnii:1990zu,Tarasov:1980au,Gorishnii:1984zi} with
%step-function decoupling of fermions at their running masses. 
%The contribution from 2-loop matching~\cite{Chetyrkin:1997sg} is negligible; 
%the effect of 3-loop terms in the renormalisation group equations is an order 
%of magnitude larger.
%Then, the two gauge couplings and all Standard Model fermion masses except
%the top mass are run to $M_Z$. The $\beta$ functions of fermion masses are
%taken to be zero at renormalisation scales below their running masses.
%The parameters at $M_Z$ are used as the low energy boundary condition in the
%rest of the evolution.

%We work in the dimensional reduction ($\drbar$) scheme 
%which preserves supersymmetry \cite{} and in which the 
We work in the dimensional reduction ($\drbar$) scheme 
\cite{Bardeen:1978yd} in which the counterterms cancel only the 
divergent pieces of the self-energies required to obtain the pole masses. 
Thus, they become finite depending on an arbitrary scale $Q$ 
and the tree level masses are promoted to running masses in order to 
cancel the explicit scale dependence of the self-energies. It implies 
that all the parameters entering in the tree-level 
masses (couplings and soft masses) are $\drbar$ running quantities.

%\subsection{Initial estimate \label{subsec:initial}}
The algorithm proceeds via the iterative method, and therefore an 
approximate initial guess of the $\mu \nu$SSM parameters is required. 
As explained above,
%Starting from the Standard Model Higgs VEV $v=246.22/\sqrt{2}$ GeV, 
from $\tanb$ and $M_Z$
%$\vi$ 
one can determine the Higgs VEVs,
%VEV's 
$\vd$ and $\vu$,
and from these
% through $v^2 = \vd^2 + \vu^2 + \vi^2$, where 
%the subindex $i$ runs over the three families of sneutrinos. 
%From this, 
the third family $\drbar$ Yukawa couplings can be approximated
as
\begin{equation} \label{eq:massaway}
Y_t(Q) = \frac{m_t(Q)}{v_u}\ , \qquad
Y_{b,\tau}(Q) = \frac{m_{b,\tau}(Q)}{v_d}\ ,
\end{equation}
where $Q=m_t(m_t)$ is the renormalisation scale. 
The $\msbar$ values of fermion masses are used for this initial
estimate. 
%Notice that in the initial iteration, $Y_{\tau}$ is obtained 
%directly from the $\tau$ lepton mass. 
The fermion masses and $\alpha_s$ at the top mass scale are obtained by 
evolving the previously obtained fermion masses and gauge couplings 
from $M_Z$ to $m_t$ (with the same accuracy). 
%The $\vci$ VEV's and the rest of Yukawa couplings $\lambda_i$ and 
%$\kappa_{ijk}$ are inputs where, for simplicity, we assume universality 
%for ****
% The electroweak gauge couplings are estimated by $\alpha_1(M_Z)
% = 5 \alpha(M_Z) / (3 c_W^2)$, $\alpha_2(M_Z) = \alpha(M_Z) / s_W^2$. 
% Here, $s_W$ is taken to be the on-shell value. These two gauge couplings 
% are then evolved to $m_t$ with one-loop Standard Model $\beta$-functions.
The electroweak gauge couplings are estimated by $\alpha_1(M_Z)
= 5\alpha(M_Z) /3 \cos^2\theta_W$, $\alpha_2(M_Z) = \alpha(M_Z) /\sin^2\theta_W$. 
Here, $\sin\theta_W$ is taken to be the on-shell value. These two gauge couplings 
are then evolved to $m_t$ with one-loop Standard Model $\beta$-functions.

The gauge and Yukawa couplings and the VEVs are then evolved 
to the scale (in the first iteration we guess $M_S$)
\begin{equation}
M_S \equiv \sqrt{m_{\tilde t_1}(M_S) m_{\tilde t_2}(M_S)} \label{msusy}\ ,
\end{equation}
where  the scale dependence of the electroweak breaking 
conditions is smallest \cite{Casas:1998vh}. For it we employ the one-loop 
$\drbar$  $\beta$-functions given in Appendix E. The supplied boundary 
conditions on the soft terms are then applied.

At this point we determine the neutrino Yukawa couplings through the $10\times 10$ neutral fermion mass matrix which can be written as %\cite{fogliani:2005m} 
\cite{NuMSSM}
\begin{equation}
{\cal M}_n \ = \ \left( \begin{array}{cc} 
                           M   &    m \\
                           m^T &    0
                         \end{array}
                  \right)\ ,
\label{mneut}
\end{equation}
where $M$ is a $7\times 7$ matrix composed by the MSSM 
neutralino mass 
matrix and its mixing with the $\nu^c_i$, while $m$ is a $7\times 3$ matrix
containing the mixing of the $\nu_i$ with MSSM neutralinos and the 
$\nu^c_i$. The full matrix is written in Appendix \ref{nfms}.
%A, eq. 
%(\ref{matrixneutralinos}).

The above matrix is of the see-saw type giving rise 
to the neutrino masses which in order to account for 
the atmospheric neutrino anomaly have to be very small. 
This is the case since the entries of the matrix $M$ are much
larger than the ones in the matrix $m$.
Notice in this respect that the entries of $M$ are of the order of the 
electroweak scale while the ones in $m$ are of the order 
of the Dirac masses for the neutrinos. Therefore 
in a first approximation the effective neutrino mixing mass matrix 
can be written as
\begin{equation}
m_{\text{eff}} = -m^T \cdot M^{-1} \cdot m\ . 
\end{equation}
Because $m_{\text{eff}}$ is symmetric and $m^{\dag}_{\text{eff}} m_{\text{eff}} $ is Hermitian, 
one can diagonalise them by a unitary transformation
\begin{eqnarray}
U_{\text{MNS}}^T\ m_{\text{eff}}\ U_{\text{MNS}} = 
\text{diag}\ (m_{\nu_1}, m_{\nu_2}, m_{\nu_3})\ ,
\\
%\label{nudiag}
%\end{equation}
%\begin{equation}
U_{\text{MNS}}^\dag \ m_{\text{eff}}^{\dag} m_{\text{eff}} \  U_{\text{MNS}} = 
\text{diag}\ (m_{\nu_1}, m_{\nu_2}, m_{\nu_3})\ .
\label{nudiag}
\end{eqnarray}
The masses are connected with experimental measurements through 
\begin{equation}
m_{\nu_2} = \sqrt{m_{\nu_1}^2 + \Delta m_{sol}^2}\ , \qquad
m_{\nu_3} = \sqrt{m_{\nu_1}^2 + \Delta m_{atm}^2}\ . 
\label{neutmass}
\end{equation}
To determine the neutrino Yukawa couplings we choose the basis 
where $Y_\nu$ is diagonal. Then we employ a numerical procedure 
which consists in solving three non-linear coupled equations in 
$Y_{\nu_{ii}}$ determined by the diagonalisation of $m_{eff}$.
Another way would consist in fixing the neutrino Yukawa couplings 
as inputs giving the left-handed sneutrino VEVs as outputs. However the 
method employed is appropriated from the numerical stability point of view. 
%For the MNS matrix we follow the standard parametrisation
%The computation of the neutrino mixing angles arises  
%from Eq.~(\ref{nudiag}). For it we follow the standard 
%parametrization of the **** 
%After that, using Eq.~(\ref{nudiag}) one computes the mixing angles. 
%With the aim of obtaining neutrino mixing angles which 
%can be compared, directly, with experimental results we follow 
%the procedure outlined in the Mixing Parameter Tools (MPT) 
%mathematica package~\cite{mpt}. It follows the standard parametrization 
%\begin{eqnarray}\label{StandardParametrizationU}
% U_{MNS} & = &
%    \diag(e^{\I\delta_{e}},e^{\I\delta_{\mu}},e^{\I\delta_{\tau}}) \cdot V \cdot 
%    \diag(e^{-\I\varphi_1/2},e^{-\I\varphi_2/2},1)\ ,
%\end{eqnarray}
%where 
%\begin{equation}
% V=\left(
% \begin{array}{ccc}
% c_{12}c_{13} & s_{12}c_{13} & s_{13}e^{-\I\delta}\\
% -c_{23}s_{12}-s_{23}s_{13}c_{12}e^{\I\delta} &
% c_{23}c_{12}-s_{23}s_{13}s_{12}e^{\I\delta} & s_{23}c_{13}\\
% s_{23}s_{12}-c_{23}s_{13}c_{12}e^{\I\delta} &
% -s_{23}c_{12}-c_{23}s_{13}s_{12}e^{\I\delta} & c_{23}c_{13}
% \end{array}
% \right)\ , 
%\end{equation}
%being $s_{ij} \equiv sin \, \theta_{ij}$, $c_{ij} \equiv cos \, \theta_{ij}$.
%The conventions used for extracting the mixing parameters from eq. (4.8) 
%are outlined in Ref. \cite{Antusch:2003kp}.
%ES CONVENIENTE ENTRAR EN ESTE DETALLE, DADO QUE NOSOTROS NO TRATAMOS DE REPRODUCIR LOS MIXINGS Y SOLO LAS MASAS ???

The determination of the charged lepton Yukawa couplings should 
follow a similar procedure through the charged fermion mass matrix written
in Appendix \ref{cfms}.
%A, eq. (\ref{matrixcharginos}).
In this matrix the charginos are mixed with 
the charged leptons. However, because it turns out that 
$Y_{\nu_{ij}} \lsim 10^{-6}$ in order to achieve the smallness of the 
neutrino masses (and also $\nu_i \lsim 10^{-4}$ GeV 
as discussed in Section \ref{sec:spectrum}),
the $2\times 2$ chargino submatrix is basically decoupled from the 
$3\times 3$ charged lepton submatrix.
Thus the charged lepton 
Yukawas can be determined directly from the charged lepton masses 
\cite{NuMSSM}, as it is stated above in eq.~(\ref{eq:massaway}).

%%%%%%%%%%%%%%%%%%%%%%%%%%%%%%%%

%\subsection{Electroweak symmetry breaking}

At the $M_S$ scale the tree-level tadpoles, eqs.~(3.4-3.7), are 
set to be zero to guarantee the EWSB. As discussed above,
at this scale the scale 
dependence of the EWSB parameters is smallest. 
In this way the soft masses  $m^2_{H_d}(M_S)$, $m^2_{H_u}(M_S)$, 
$m^2_{\widetilde{L}_i}(M_S)$, and $m^2_{\widetilde{\nu}_i^c}(M_S)$, are derived.
% just solving a set of linear equations ***
% Strong upper bounds upon the intergenerational scalar mixing 
% exist \cite{Gabbiani:1996hi} so in the following we assume that 
% such mixings are negligible and therefore the sfermion softmass 
% matrices are diagonal in the flavour space.
%$\mHd$, $\mHu$, $m^2_{\widetilde{L}_{i}}$, and $m^2_{\widetilde{\nu}_{i}^{c}}$,

The next step consists of performing a check in order to ensure that the 
minimum which breaks the electroweak symmetry spontaneously
is the global one. For it, we compute the global minimun using 
a `genetic' algorithm for global optimisation \cite{gop} which has a 
high performance. Then we compare it with the physical one.

%%%%%%%%%%%%%%%%%%%%%%%%%%%%%%%%

%\subsection{Spectrum}

In the final step the $\drbar$ (tree-level) superparticle mass spectrum 
consisting of squarks, CP-even (odd) neutral scalars, charged scalars, 
neutral fermions and charged 
fermions (see Appendix A) is determined at the $M_S$ scale. 
Notice that once the tree-level mass spectrum is known, radiative 
corrections to the neutral scalar potential and the tadpoles 
as well as for computing pole masses are calculable. 

% At the weak scale, we take as free parameters in the Higgs sector:
% \bea
% \lambda_i, \, \kappa_{ijk},\, \tan(\beta), \, \nu_i, \nu^c_i, \, 
%A_{\lambda_i}, \, A_{\kappa_{ijk}}, \, A_{\nu_i}.
% \eea

% We observe that since $\nu_i<<v_d<v_u$ we can defined 
%$tan(\beta)=\frac{v_u}{v_d}$ as usual.

% We take all the families equal and without mixing, with the only 
%exception of the $\nu_i$ where we take the third family different 
%to the others , then the parameters are: 

% \bea
% \lambda, \, \kappa,\, \tan(\beta), \, \nu_1, \,  \nu_3, \nu^c, \, 
%A_{\lambda}, \, A_{\kappa}, \, A_\nu.
% \eea

In order to check the absence of a Landau singularity 
(by requiring any Yukawa coupling to be less than $\sqrt{4\, \pi}$) 
the Yukawa couplings are evolved to the GUT scale. 
Finally, Yukawa couplings, gauge couplings, and VEVs, 
are evolved back down to $M_Z$, and SUSY one-loop thresholds containing 
squark/gluino in the loop are added to the third family of quark 
Yukawa couplings and to the strong coupling constant \cite{Pierce:1996zz}. 
The whole process is iterated, as it is sketched in 
Fig.~\ref{fig:algorithm}, with the inclusion of one-loop corrections 
to the neutral scalar potential. It is equivalent to add  
the one-loop tadpoles to eqs. (3.4-3.7). Then the global minimun 
is computed following the procedure described above. For this work we 
have computed the leading one-loop contributions to the tadpoles, 
which come from (s)quarks in the loops, in the $\drbar$ scheme. 
The results are given in Appendix C. For the neutral scalar potential we 
employ the results in ref. \cite{Zhang:1998bm}.

%\vspace{0.5cm}

\FIGURE{\label{fig:algorithm}
\begin{picture}(323,250)
\put(10,0){\makebox(280,10)[c]{\fbox{Run to $M_S$. Calculate sparticle pole masses}}}
\put(150,37.5){\vector(0,-1){23}}
\put(10,40){\makebox(280,10)[c]{\fbox{Run to $M_Z$}}}
\put(150,76.5){\vector(0,-1){23}}
\put(10,80){\makebox(280,10)[c]{\fbox{Run Yukawas to $M_{GUT}$. Check for Landau Poles}}}
\put(150,116.5){\vector(0,-1){23}}
\put(10,120){\makebox(280,10)[c]{\fbox{Neutrino Yukawas, EWSB}}}
\put(150,156.5){\vector(0,-1){23}}
\put(10,160){\makebox(280,10)[c]{\fbox{Apply soft SUSY-breaking
boundary conditions}}}
\put(150,195.5){\vector(0,-1){23}}
\put(10,199){\makebox(280,10)[c]{\fbox{Run to $M_S$}}}
\put(150,235){\vector(0,-1){23}}
\put(10,240){\makebox(280,10)[c]{\fbox{Set VEVs, Yukawas, and add SUSY 
rad. corr. to $g_s(M_Z)$, $h_{t,b}(M_Z)$}}}
\put(183,45){\line(1,0){160}}
\put(343,45){\line(0,1){200}}
\put(343,245){\vector(-1,0){27}}
\end{picture}
\vspace{0.5cm}
\caption{Iterative algorithm used to calculate the SUSY spectrum. Each step
(represented by a box) is detailed in the text. The initial step is the
uppermost one. $M_S$ is the scale at which the EWSB
conditions are imposed, as discussed in the text.}}

Once the $\drbar$ sparticle masses all converge to better than the 
desired fractional accuracy, the computation of the physical masses requires 
the addition of loop corrections.
It is well known that the role of the radiative corrections to the 
lightest CP-even Higgs boson mass is extremely important
(see ref. \cite{higgs} for studies of this effect in the NMSSM). 
The leading ones come from an incomplete cancellation of the 
quark and squark loops. Following this we have added those corrections
as described below. The rest of the masses are tree-level 
$\drbar$ running masses.

The gluino mass is then given by 
\begin{equation}
m^{tree}_{\tilde g} = M_3(M_S)\ .
\label{glumass}
\end{equation}

The rest of  SUSY particles mix in the interaction basis, and a rotation 
to their mass states basis is required.
The scalar sector includes the squarks which are MSSM-like 
and therefore their masses and mixing angles are the result of 
performing a Jacobi $2\times2$ rotation of the matrices
in Appendix \ref{squarkss}. Charged and neutral
fermion masses are the result of diagonalising 
their mass matrices which are given in Appendices \ref{cfms} and \ref{nfms}
% A.2 and A.3 
respectively.

It is worth mentioning that a final check is required 
to see if the procedure used above to compute neutrino 
Yukawa couplings is consistent with the final lightest eigenvalues of the 
neutralino mass matrix eq. (\ref{mneut}). 

The CP-even scalar masses are obtained from the real 
parts of the poles of the propagator matrix
\begin{equation}
Det\left [p^2_i {\bf 1} - \mathcal{M}_{S^0}(p^2_i) \right ] = 0\ , 
\hspace{0.5in} m_i^2 \equiv \mathcal{R}e(p^2_i)\ ,
\end{equation}
where 
\begin{equation}
\mathcal{M}_{S^0}(p^2) = \mathcal{M}_{S^0}^{\drbar}(Q) 
%     + \delta \mathcal{M}_{ij},
   + \Pi_{S^{o}} (p^2, Q)\ ,
\end{equation}
with $\Pi^{S^{o}}$ being the matrix of the renormalised 
self-energies in the $\drbar$ scheme of the CP-even scalars.
The ones involving quarks and squarks in the loop are shown 
in Appendix D.
We diagonalise  the matrix $\mathcal{M}_{S^0}(p^2_i)$ at an external 
momentum scale equal to its pole mass $p^2_i = m_i^2$ through 
an iterative procedure.

%One of the straight consecuences of the model is that 
%the ${\tilde \nu}_i$ are eigenvectors of the scalar neutral 
%mass matrix. Therefore their masses are $(m^2_{\tilde{L}})^i$. To avoid 
%unwanted tachyons in the spectrum, it imposes contraints 
%on the parameters involved in the ${\tilde \nu}_i$ directions 
%minimization conditions.

Finally, the quark Yukawa couplings, gauge couplings, and VEVs 
are evolved back down to $M_Z$.

%%%%%%%%%%%%%%%%%%%%%%%%%%%%%%%%%%%%%%%%%%%%%%%
\section{Results and discussion}
\label{results}
%%%%%%%%%%%%%%%%%%%%%%%%%%%%%%%%%%%%%%%%%%%%%%%

Using the results of the previous Sections and Appendices, we will study in detail the parameter space
and spectrum of the $\mu\nu$SSM.

\subsection{Analysis of the parameter space}
\label{pss}
%*** Also say about neutrino hierarchical ****

% In this subsection the parameter space and spectrum of the 
% $\mu\nu$SSM will be studied.
% Although the free parameters in our model have already been 
% presented in Section \ref{sec:spectrum}, it is worth recalling that
% the Higgs 
%and neutralino sectors 
% sector of the theory are specified by
% \bea
% \lambda, \, \kappa,\, \tan\beta, \, \nu_1, \,  \nu_3, \nu^c, \, A_{\lambda}, \, A_{\kappa}, \, A_\nu, \, M_2\ .
% \label{freeparameters2}
% \eea
% As aforementioned, we take these parameters to be free at the electroweak (CHEQUEAR) scale.
% For simplicity the low-energy squark masses and trilinear couplings,
% which are not specially relevant for our analysis, are taken
% to be degenerate with values
% \bea
% m_{\tilde Q}=m_{\tilde u^c}=m_{\tilde d^c}=m_{\tilde e^c}=1 \, \text{TeV}\ ,
% \eea
% and
% \bea
% A_u=A_d=A_e=1 \, \text{TeV}\ .
% \eea
% We also choose 
% \bea
% M_2=1 \, \text{TeV}\ .
% \eea
% Later on we will address variations of these
% values (RECORDAR QUE VARIAR M2 INFLUYE EN EL ESPECTRO DE CHARGINOS Y NEUTRALINOS, VARIAR LA MASA DEL SELECTRON RIGHT INFLUYE EN EL ESPECTRO DE LOS ESCALARES CARGADOS).
% Throughout this section,
% we will fix the values (DECIR POR QUE NO SON RELEVANTES SUS VARIACIONES)
% \bea
% \nu_1=1.4\times 10^{-5}\ \text{GeV}, \,  \nu_3=1.4\times 
% 10^{-4}\ \text{GeV},
% \, A_\nu= -1 \, \text{TeV}\ ,
% \eea
% and
% consider several
% choices for the values
% of 
% \bea
% \lambda, \, \kappa,\, \tan\beta, \, \nu^c, \, A_{\lambda}, \, A_{\kappa}\ ,
% \eea
% and for each case, we will study the associated phenomenology (LO HAREMOS?).

In this subsection the parameter space of the 
$\mu\nu$SSM will be studied. We will see that avoiding the existence of false 
minima and tachyons, as well as imposing perturbativity (Landau pole condition) on the couplings of the model,
important constraints on the parameter space will be found.

The free parameters of our model have already been 
presented in eq. (\ref{freeparameters2}).
As aforementioned, we take them to be free at the electroweak scale.
As discussed in Section \ref{sec:strategy},
we will determine the neutrino Yukawa couplings through 
the experimental data on neutrino masses.
We will use the direct hierarchical difference of masses, taking
the typical values
$m_{\nu_1}=10^{-12}$ GeV, $m_{\nu_2}=9.1\times 10^{-12}$ GeV and $m_{\nu_3}=4.7\times 10^{-11}$ GeV.
Finally, as discussed in Section \ref{sec:spectrum}, it is sufficient
to work with only two different left-handed sneutrino VEVs. In particular, we choose 
$\nu_1=\nu_2=1.4\times 10^{-5}$ GeV and $\nu_3=1.4\times 
10^{-4}$ GeV, which are typical values in order to satisfy the minimum equations 
(\ref{eq:tadpoles}) and data on neutrino masses through the see-saw mechanism (\ref{mneut}). Possible variations of these values will not modify qualitatively our results below.

Throughout this section
we will consider several choices for the values of 
\bea
\lambda, \, \kappa,\, \tan\beta, \, \nu^c, \, A_{\lambda}, \, A_{\kappa}\ ,
A_{\nu}\ ,
\label{freeparameters25}
\eea
using the sign conditions explained in Section \ref{The model}.
Besides, we work with a negative value of $A_{\nu}$ in order to fulfill 
condition (\ref{anu}) more easily.

Concerning the rest of the soft parameters we will take for simplicity in the computation
$m_{\tilde Q,\tilde u^c,\tilde d^c,\tilde e^c}=1$ TeV,
$A_{u,d,e}=1$ TeV, and for the gaugino masses only $M_2=1$ TeV will be 
used as input, whereas the others 
will be determined by the approximate GUT relations 
$M_1 = \frac{\alpha_1^2}{\alpha_2^2} M_2$,
$M_3 =  \frac{\alpha_3^2}{\alpha_2^2} M_2$,
implying $M_1\approx 0.5 M_2$, $M_3\approx 2.7 M_2$.

Let us first discuss when the minimum we find following 
Sections \ref{The model} and \ref{minimization}
is the global one. In particular, one has to be sure that it is deeper than the local minima
with some or all of the VEVs in 
eq. (\ref{2:vevs}) vanishing.
Concerning the latter one can check that the most relevant minima are 
the solutions with only
$v_u$ or $\nu^c$ different from zero (in some special situations also the case
with all VEVs vanishing can be relevant).
For example, for a given value of $\nu^c$
the term  proportional to 
$a_{\kappa}$ in (\ref{akappa}) turns out to be important:
the more negative the value of $A_{\kappa}$, the deeper the minimum becomes.
This might in principle give rise to a value of the potential
(\ref{finalpotential}) in the direction with only $\nu^c \neq 0$, more negative than
the one produced in the realistic direction with all VEVs non vanishing.
In that case the associated points in the parameter space would be excluded by the
existence of false minima. 
Notice that $m^2_{H_u}$ is independent on the value of $A_\kappa$ as can be deduced 
from eq. (3.4) with $t^0_u=0$. Thus although $m^2_{H_u}$ will contribute to the realistic direction,
it plays no role in the above argument.

On the other hand, we can also deduce from eq. (3.4) 
that for reasonable values of the parameters
the larger $\nu^c$, the smaller $m^2_{H_u}$ becomes in order to cancel 
$t^0_u$.
%(NO ESTA CLARO COMO DE GENERAL ES ESTO, DEPENDERA DE LAMBDA, KAPPA). 
As a consequence, the realistic direction becomes  
deeper,
%(see also the negative contribution proportional to $a_{\lambda}$ in (\ref{akappa}) NO SE SI HAY QUE PONER ESTO), 
and the associated
points in the parameter space are allowed. 

\begin{figure}[t!]
 \begin{center}
\hspace*{-8mm}
    \begin{tabular}{cc}
    %\epsfig{file=finalfigures/elmr1-Ak-nuc-ms2nu-al1-k0.4-l0.1-tb5.eps,height=8.7cm}
    %\hspace*{0mm}&\hspace*{-25mm}
      \epsfig{file=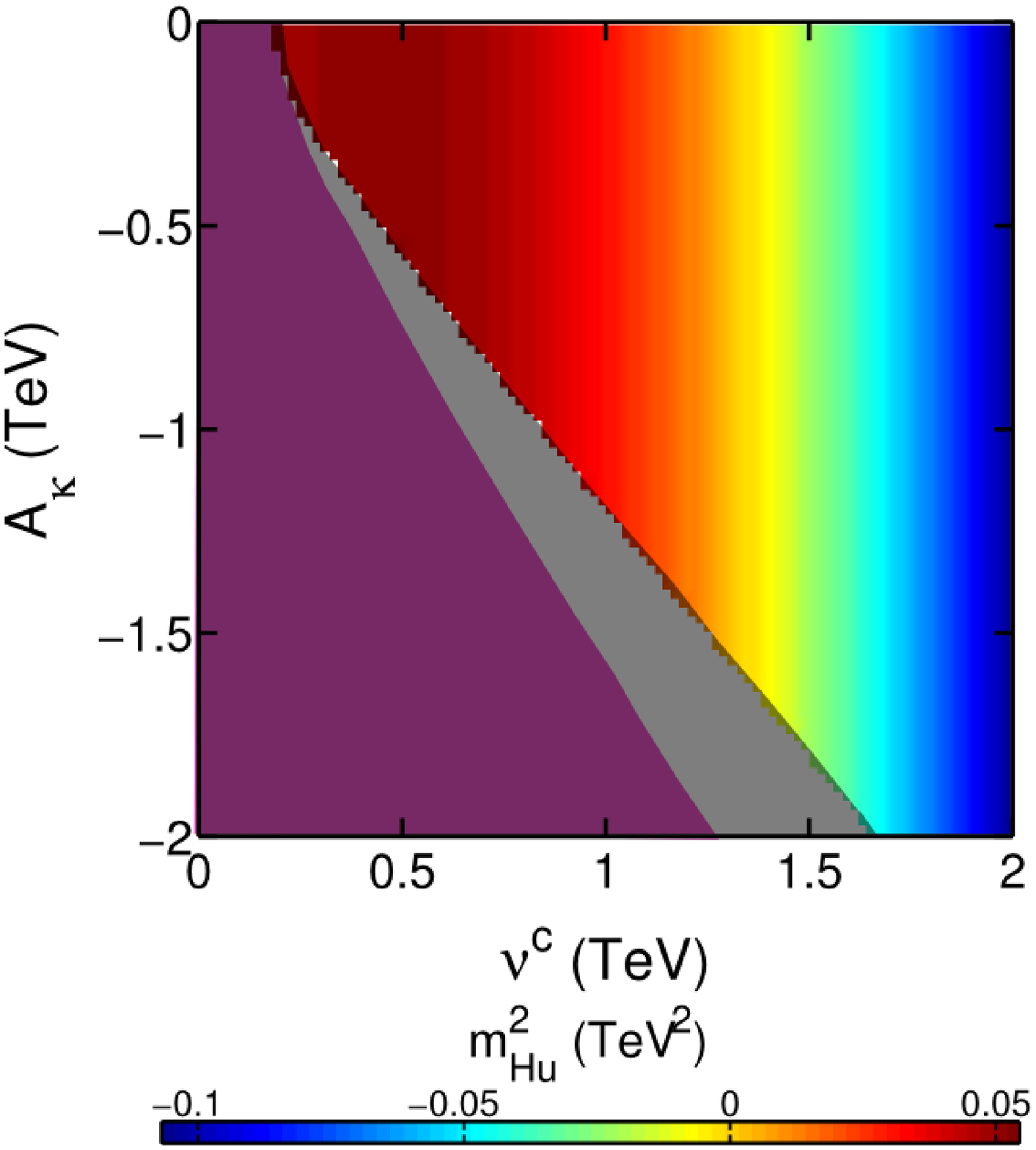,height=8.7cm}
      \hspace*{0mm}&\hspace*{-25mm}
      \epsfig{file=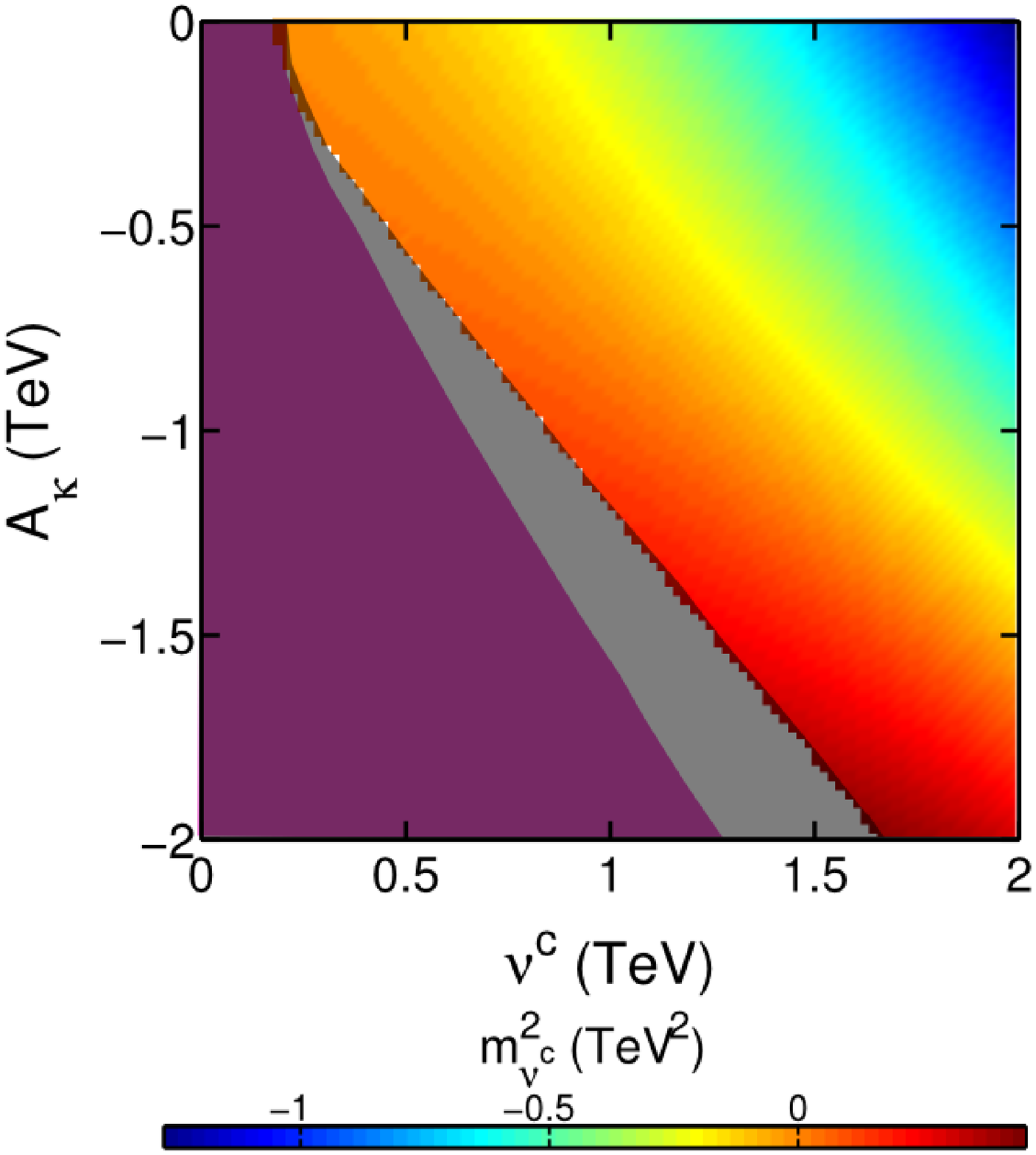,height=8.7cm}
  \vspace*{-0.8cm}       
   \\ & \\
      %(a)\hspace*{8mm} & \hspace*{8mm} (b) 
            (a)\hspace*{8mm} & \hspace*{-3cm} (b)
    \end{tabular}
    \captions{    
($A_\kappa$,\ $\nu^c$) parameter space for $\tan\beta=5$, $\lambda=0.1$, 
$\kappa = 0.4$, and $A_{\lambda}=-A_{\nu}=1$ TeV. 
In both cases the gray and violet areas represent points which are excluded by the
existence of false minima and tachyons, respectively.
In (a) the colours 
indicate different values of the soft mass $m^2_{H_u}$. 
In (b) the colours indicate different
values of the soft masses $m^2_{\tilde\nu^c}$.}
%
% ($A_\kappa$,$\nu^c_i$) parameter space for $\tan(\beta)=5$, $\lambda=0.1$, $\kappa = 0.4$, and $A_{\kappa}=-1$ TeV. (a) We plot  $m^2_{H_u}$. (b) We plot $m^2_{\tilde\nu^c}$. }
    \label{figak}
 \end{center}
\end{figure}

Both effects can be seen in Fig.~\ref{figak}a, where 
the ($A_\kappa$, $\nu^c$) parameter space  (recall our assumption
$\nu^c_i=\nu^c$) is plotted 
for an example  
with $\lambda=0.1$, $\kappa = 0.4$, 
$\tan\beta=5$, and $A_{\lambda}=-A_{\nu}=1$ TeV.
For a given value of $\nu^c$ we see that for $A_{\kappa}$ sufficiently
large and negative one obtains a false minimum (gray area). For larger values
of $\nu^c$ one needs values more negative of $A_{\kappa}$ to obtain the false minimum.
Let us remark that 
%$m^2_{H_u}$ is independent on the value of $A_\kappa$ as can be deduced 
%from eq. (3.5). This is also obvious from Fig.~\ref{figak}a.
although $m^2_{\tilde\nu^c}$ depend on $A_\kappa$, as can be obtained from eq. (3.5),
we can see in Fig.~\ref{figak}b that this variation is not crucial for the discussion above. Notice that the values of $m^2_{\tilde\nu^c}$ for points of the parameter space close to the false minimum area do not vary in a relevant way.

In Fig.~\ref{figak} we can also see that part of the parameter space is excluded due to the
occurrence of tachyons in the CP-even neutral scalar sector.
Thus the realistic direction with all VEVs non-vanishing is not even
a local mimimum.
This happens in general when 
the off-diagonal values $|M_{h_{d}(\widetilde{\nu}_i^c)^R }^{2}|$ or 
$|M_{h_{u}(\widetilde{\nu}_i^c)^R }^{2}|$ 
of the CP-even neutral scalar matrix (see Appendix
\ref{appendixA.1.1})
become significantly larger 
than $|M_{(\widetilde{{\nu}}^{c}_{i})^R (\widetilde{{\nu}}^{c}_{j})^R}^{2}|$
in some regions of the parameter space, thus leading to the
appearance of a negative eigenvalue. 
%This will typically happen for
%moderate to large values of $\lambda$ and small $\kappa$, for which
%$m_{h^0_1}$ is small. 
The violet area
in Fig.~\ref{figak} corresponds to this
situation.
% [ESTO SE DECIA EN EL NMSSM, LO PONEMOS TAMBIEN AQUI?: On the other hand, the eigenvalues of the CP-odd Higgs mass matrix
% are never negative.
% The CP-odd Higgs masses also decrease for values of the parameter space,
%large $\lambda$ and small $\kappa$, 
% but their
% minimum value is bounded by the appearance of tachyons in the CP-even
% sector.
% AQUI NO ES CIERTO, HAY TAQUIONES EN EL SECTOR PSEUDOESCALAR EN LA ZONA DE LAMBDA PEQUENIA, PERO SIEMPRE ACOMPANIANADO A LOS TAQUIONES ESCALARES]
%
%
%  If one increase $\tan\beta$ the mixing becomes more importante and is more easy to get tachyons. We will explained
% this below when the $\lambda$ vs $\kappa$ plane is analysed. 
In particular, notice that the relevant terms in the off-diagonal
pieces are linear in $\nu^c$, whereas in
$M_{(\widetilde{{\nu}}^{c}_{i})^R (\widetilde{{\nu}}^{c}_{j})^R}^{2}$
they are quadratical.
Thus, for a given value of $A_{\kappa}$, the smaller the value of $\nu^c$, the smaller the latter terms become
giving rise to the possibility of tachyons.
Notice also that there is a term proportional to $a_{\kappa}$ in 
$M_{(\widetilde{{\nu}}^{c}_{i})^R (\widetilde{{\nu}}^{c}_{j})^R}^{2}$,
implying that, for a given value of $\nu^c$, the more negative 
the value of $A_{\kappa}$, the smaller
$M_{(\widetilde{{\nu}}^{c}_{i})^R (\widetilde{{\nu}}^{c}_{j})^R}^{2}$ become.
This is also reflected in Fig.~\ref{figak}.

% [NO ENTIENDO DEL TODO QUE SE QUIERE DECIR AQUI: 
% On the left of Fig.~\ref{figal1} we can see that moving to points where the different between the trivial minimun and our candidate increase appears the tachyons region (in violet).

% On the other hand, the eigenvalues of the CP-odd Higgs mass matrix
% are never negative.
% The CP-odd Higgs masses also decrease for
% large $\lambda$ and small $\kappa$, but their
% minimum value is bounded by the appearance of tachyons in the CP-even
% sector.

\begin{figure}[t!]
%  \begin{center}
\hspace*{-8mm}
    \begin{tabular}{cc}
\epsfig{file=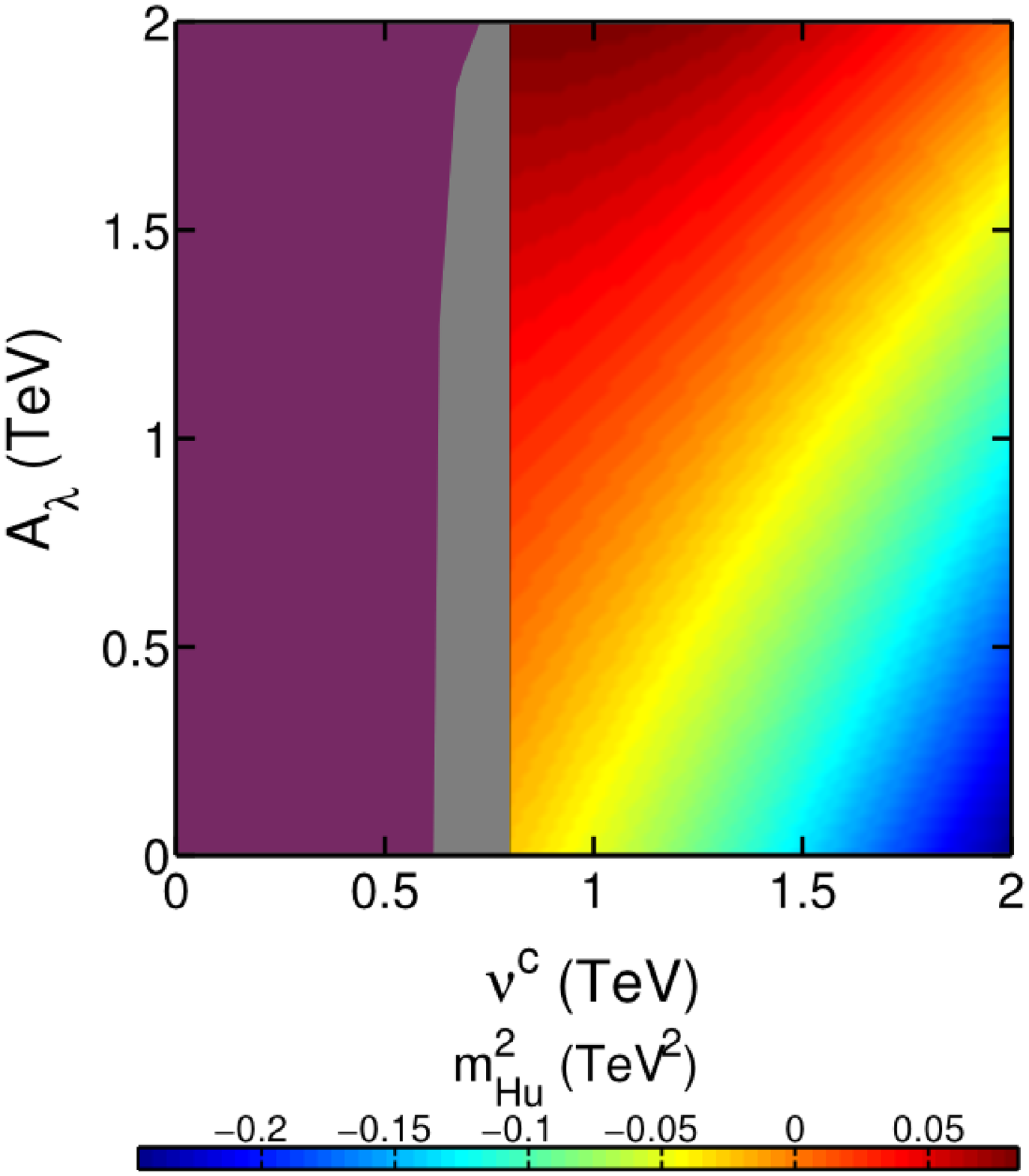,height=8.7cm}
      \hspace*{0mm}&\hspace*{-25mm}
  \epsfig{file=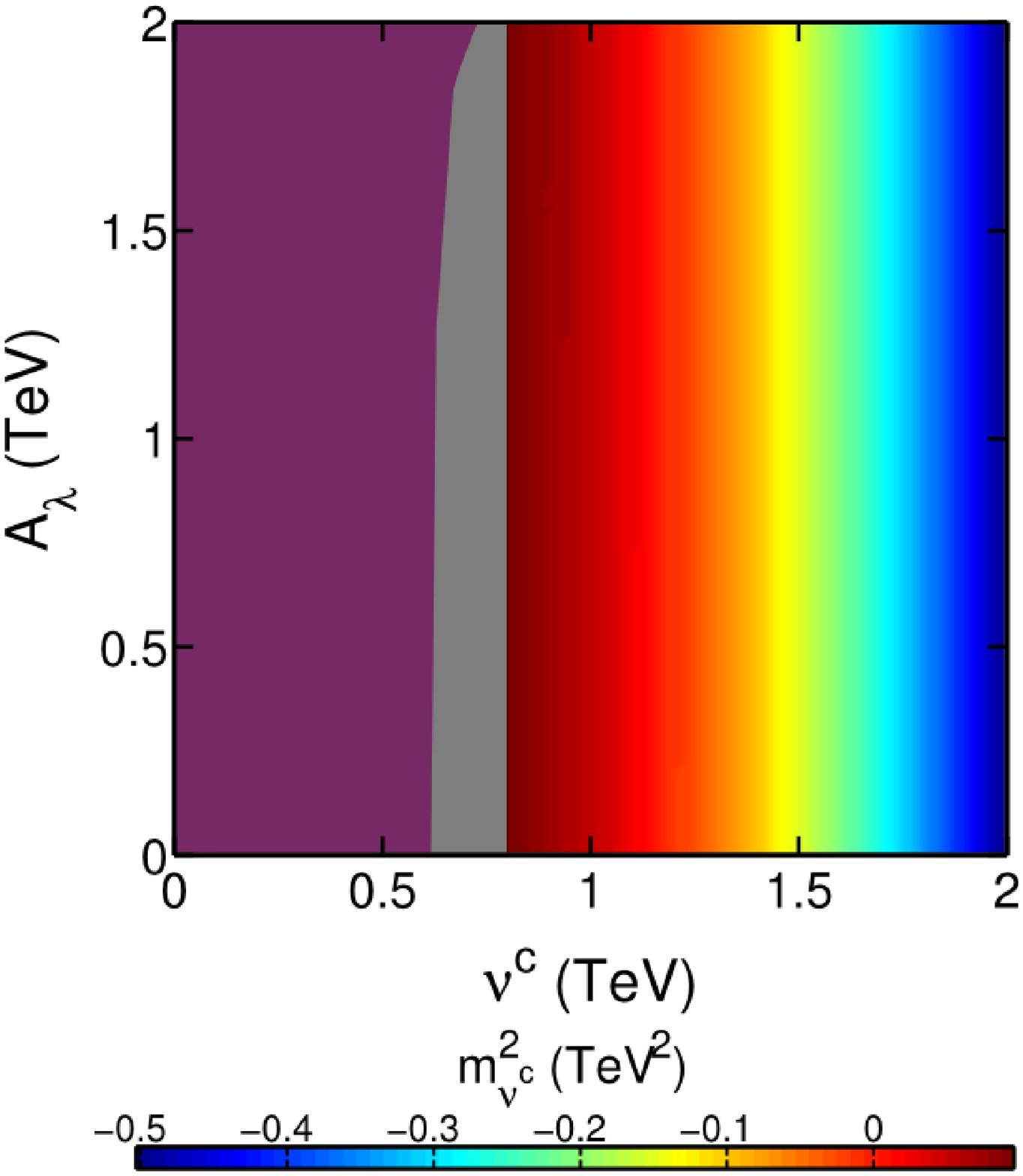,height=8.7cm}
  \vspace*{-0.8cm}       
   \\ & \\
      %(a)\hspace*{8mm} & \hspace*{8mm} (b) 
            (a)\hspace*{8mm} & \hspace*{-3cm} (b)
%     \\ & \\
%      (a)\hspace*{8mm} & \hspace*{8mm} (b)
    \end{tabular}
\captions{
($A_\lambda$,\ $\nu^c$) parameter space for $\tan\beta=5$, $\lambda=0.1$, 
$\kappa = 0.4$, and $A_{\kappa}=A_{\nu}=-1$ TeV. 
In both cases the gray and violet areas represent points which are excluded by the
existence of false minima and tachyons, respectively. In (a) the colours 
indicate different values of the soft mass $m^2_{H_u}$. 
In (b) the colours indicate different
values of the soft masses $m^2_{\tilde\nu^c}$.}
    \label{figal1}
%  \end{center}
\end{figure}

\begin{figure}[t!]
%  \begin{center}
\hspace*{-8mm}
    \begin{tabular}{cc}
       \epsfig{file=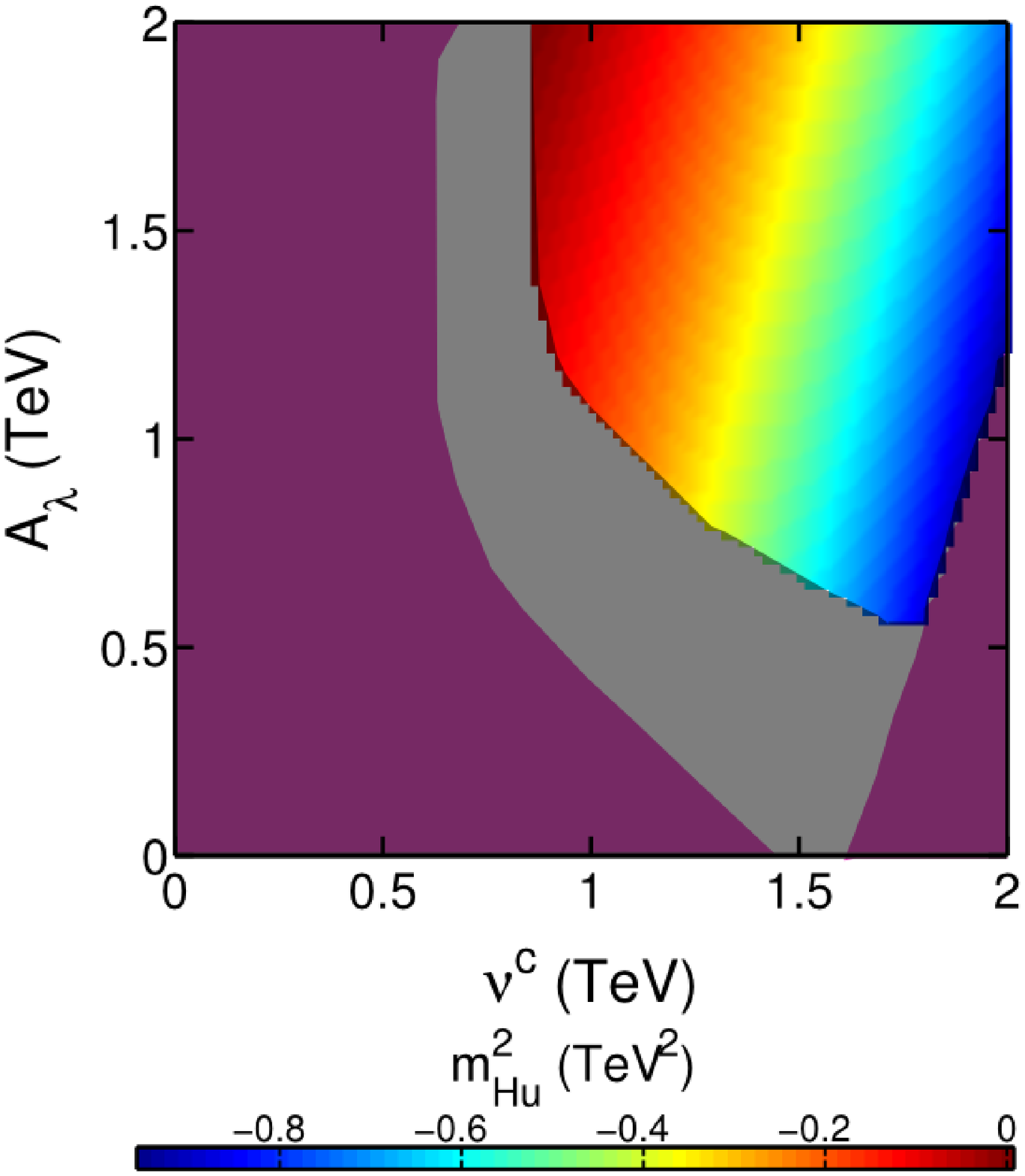,height=8.7cm}
      \hspace*{0mm}&\hspace*{-25mm}
\epsfig{file=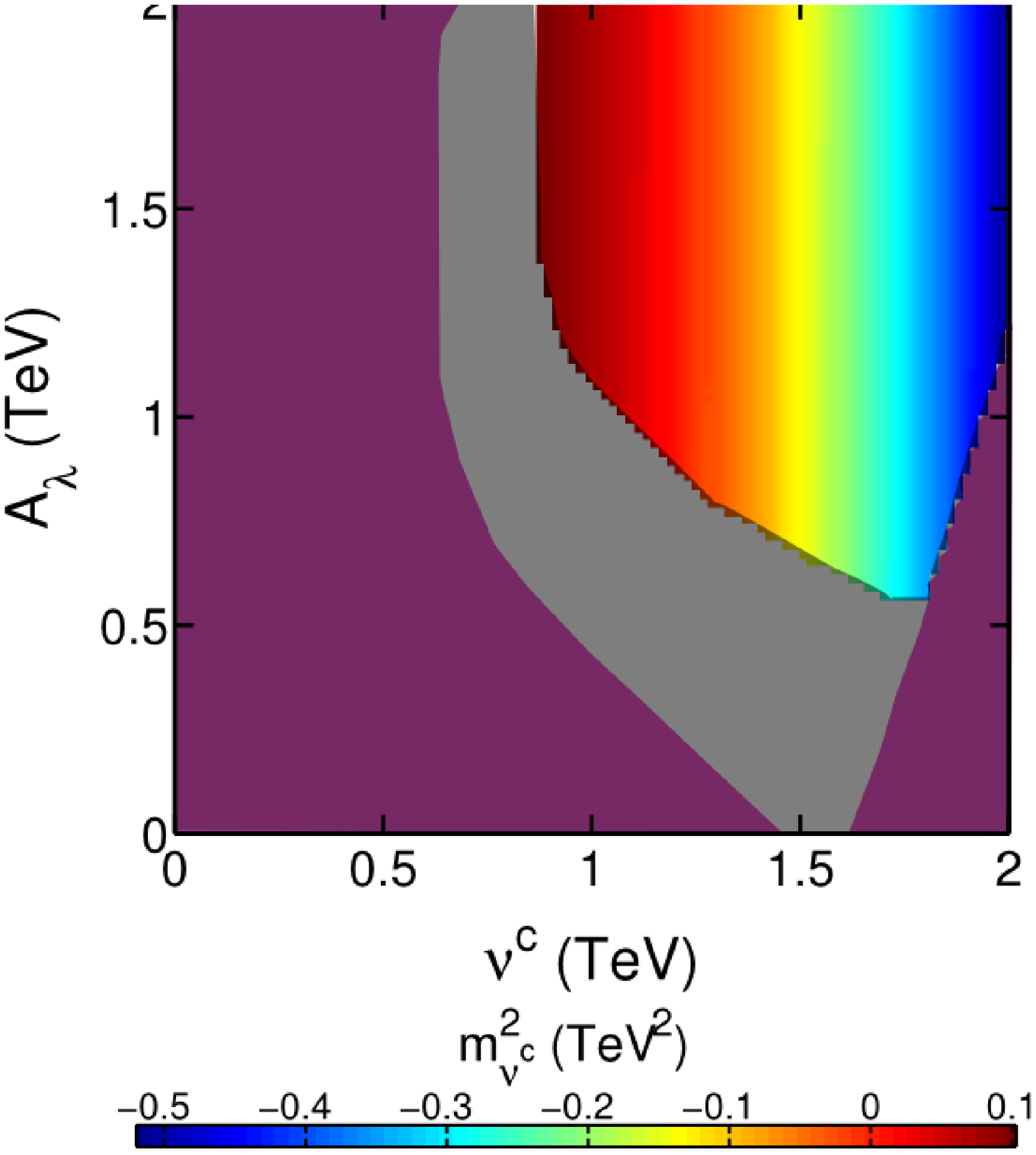,height=8.7cm}  
  \vspace*{-0.8cm}       
   \\ & \\
      %(a)\hspace*{8mm} & \hspace*{8mm} (b) 
            (a)\hspace*{8mm} & \hspace*{-3cm} (b)
%     \\ & \\
%       (a)\hspace*{8mm} & \hspace*{8mm} (b)
    \end{tabular}
\captions{
The same as in Fig. 3 but for $\lambda=0.2$.
}
% In addition we take trilinear universal soft parameters: 
    \label{figal2}
%  \end{center}
\end{figure}

% =========
% To clarify the above comments, let us shown in Fig.~\ref{figal1}a 
% the ($A_\lambda$, $\nu^c$) parameter space (recall our assumption
% $\nu^c_i=\nu^c$)
% for an example with $\lambda=0.1$, $\kappa = 0.4$, 
% $\tan\beta=5$, $A_{\kappa}=-1$ TeV, and $A_{\nu}=-1$ TeV.

% The colours 
% indicate different values of the soft masses $m^2_{\tilde\nu^c}$
% (SUPONGO QUE LAS TRES MASAS ADQUIEREN EL MISMO VALOR DE LAS CONDICIONES DE MINIMO). 
% As discussed, for a fixed $\nu^c$ we can see 
% that $m^2_{\tilde\nu^c}$ is basically independent on $A_\lambda$.
% Also we see that 
% $m^2_{\tilde\nu^c}$ becomes large for $\nu^c$ small, making easy the appearance of a global minimum in the $v_u$ direction of the potential.
% This is the origin of the gray area representing points excluded by the
% existence of false minima. 

Let us now discuss the possibility of minima deeper than the realistic one in the direction
with only $v_u\neq 0$.
When the values of $\nu^c$ are large,
we can see from eq. (3.5) 
that $m_{{\tilde\nu^c}}^2$ must be negative in order to cancel the cubic term
in  $\nu^c$. However, when the values of $\nu^c$ are small,
$m_{{\tilde\nu^c}}^2$ must be positive in order to cancel the quadratic term
in $\nu^c$ proportional to $a_{\kappa}$, which is now the relevant one. This may give rise for small $\nu^c$ to a value of the potential 
(\ref{finalpotential}) 
in the direction with only $v_u \neq 0$, more negative than
the one produced in the realistic direction with all VEVs non vanishing.
This situation is shown in Fig.~\ref{figal1}, where
the ($A_\lambda$, $\nu^c$) parameter space is plotted
for an example with $\lambda=0.1$, $\kappa = 0.4$, 
$\tan\beta=5$, $A_{\kappa}=A_{\nu}=-1$ TeV.
We can see in Fig.~\ref{figal1}b that 
the smaller $\nu^c$, the larger $m^2_{\tilde\nu^c}$ become, making it easy
the appearance of a false minimum.
Let us remark that the points in the gray area above $A_\lambda\approx 1$ TeV
are actually forbidden by minima deeper than the realistic one with all VEVs vanishing.
Notice to this respect in the figure that those points correspond to
positive values of $m^2_{H_u}$ and $m_{{\tilde\nu^c}}^2$.
This is also true for Fig.~\ref{figal2} discussed below, but for points
above $A_\lambda\approx 2$ TeV.

It is worth noticing here 
that $m^2_{\tilde\nu^c}$ is essentially independent on the value of $A_\lambda$, as can be easily deduced 
from eq. (3.5).
On the other hand, we can see from eq. (3.4) that $m^2_{H_u}$ does depend
on $A_\lambda$ through the term proportional to $a_\lambda$. In particular, if we 
decrease $A_\lambda$, $m^2_{H_u}$ also decreases,
as shown in Fig.~\ref{figal1}a.
Although this might in principle contribute to produce a 
minimum deeper than the realistic one in the direction with only $v_u\neq 0$, we see in the figure that for the parameter space studied the effect is negligible.  
Nevertheless, increasing the value of $\lambda$, $a_\lambda$ also increases, and this effect can be more important.
This is shown in 
Fig.~\ref{figal2}a, where $\lambda=0.2$ is considered.
We can see that the parameter space is now more constrained.
We also show in Fig.~\ref{figal2}b the values of $m^2_{\tilde\nu^c}$ 
in the allowed region.

Actually, there is a new tachyonic region for large values of $\nu^c$.
This happens because the off-diagonal value  
$|M_{h_{d}h_{u}}^{2}|$ in Appendix (\ref{appendixA.1.1})
has a quadratic dependence on $\nu^c$, thus leading to the appearance of a negative
eigenvalue.
Notice in this respect that a similar dependence in the diagonal pieces
$|M_{h_{d}h_{d}}^{2}|$ and $|M_{h_{u}h_{u}}^{2}|$
is canceled once we substitute the value of the soft masses using eqs. (3.3) and (3.4).

%DECIR ALGO DE LOS TAQUIONES O PASAR YA?

% =========

%\end{document}

% \end{document}
% \begin{figure}[t!]
% \begin{center}
% \hspace*{-8mm}
%     \begin{tabular}{cc}
%     \epsfig{file=plots/elmr1-k-l-ms2hu-al1-nuc2-ak-1-tb5.eps,height=8.7cm}
%     \hspace*{0mm}& \hspace*{-25mm}
%     \epsfig{file=plots/elmr1-k-l-ms2nu-al1-nuc2-ak-1-tb5.eps,height=8.7cm}
%     \\ & \\
%      (a)\hspace*{8mm} & \hspace*{8mm} (b)
%     \end{tabular}
% \captions{The same as in Fig. 4 but for the case ${\nu}^c=2$~TeV.}
%     \label{figkl2}
% \end{center}
% \end{figure}

% ($\lambda$, $\kappa$) parameter space for 
% $\tan\beta=5$,
% $A_\lambda=1$~TeV, $A_\kappa=-1$ TeV (SUPONGO??), and $\nu^c=2$~TeV. 
% In both cases the gray and magenta (over Gray or Orange when 
% overlap other regions ????????) areas 
%represent points which are excluded by the
% existence of false minima and tachyons, respectively.
% The orange area (HAY UNA ORANGE Y OTRA ROSA ???????)
% represent points which are excluded due to the occurrence 
%of a Landau pole.
% In (a) the colours 
% indicate different values of the soft mass $m^2_{H_u}$. 
% In (b) the colours indicate different
% values of the soft masses $m^2_{\tilde\nu^c}$.}
%%%

For each point in the parameter space, one also requires perturbativity, i.e.
the absence of Landau singularities for the couplings. Let us discuss now in detail the case of $\lambda$, since this is the relevant coupling when
discussing the upper bound on the lightest Higgs mass in the next Subsection.

Once perturbativity is imposed, the value of 
$\lambda$ 
is actually bounded.
%, producing 
%In addition, assuming 
%perturbativity of the $\lambda_i$ couplings also imposes 
%a bound
%on the value in (\ref{boundHiggs}). 
To obtain a rough estimation
we can use eq. (\ref{lambda}) in the Appendix neglecting $Y_{\nu_{ij}}$,
and taking $\kappa_{iii}=\kappa$ and  $\kappa_{ijk}=0$ if $i= j =k$ is not satisfied.
Then we can write that equation as
\begin{equation}
\frac{d}{dt} \bm \lambda^2= \frac{2}{16 \pi^2}( C - 4 \bm \lambda^2)\ \bm \lambda^2\ ,
\label{gg}
\end{equation}
where we have defined 
${\bm \lambda}^2\equiv\lambda_i \lambda_i$, $i=1,...,n$, with
$n$ the number of singlets,
and
$C$ is a quantity independent on $\lambda_i$. 
It is worth noticing here that the RGE for the relevant parameter ${\bm \lambda}^2$
is clearly independent on $n$. 
Thus we could in principle expect a bound for ${\bm \lambda}^2$ similar to the one of the NMSSM for $\lambda$. Recall that in the NMSSM there is only one singlet, and
$\lambda^2\lsim (0.7)^2$.
To complete the discussion we can solve a simplified version of eq. (\ref{gg})
neglecting the piece proportional to $C$, with the result
% \bea
% \frac{d}{dt} {\bm \lambda}^2= -\frac{{\bm \lambda}^4}{2\pi^2}\ ,
% \eea
% with the result
\bea
{\bm \lambda}^2 (Q)=\frac{{\bm \lambda}^2 (Q_0)}
{1+\frac{{\bm \lambda}^2 (Q_0)}{2\pi^2}\ln(\frac{Q_0}{Q})}\ , 
\label{rgel2}
\eea
where $Q$ is the renormalization scale, 
and $Q_0$ the scale of the high-energy theory.
% The most conservative upper bound can be found when
% ${\bm \lambda}^2 (Q_0)\rightarrow \infty$, implying 
% \bea
% {\bm \lambda}^2 (Q) = \frac{2\pi^2}{\ln(\frac{Q_0}{Q})}\ , 
% \label{rgep23}
% \eea
At the high-energy scale the Landau pole condition
for each coupling can be imposed as
$\lambda_i^2(Q_0) < 4\pi$, implying 
${\bm \lambda}^2(Q_0)<4 \pi n$, and therefore one obtains the following upper bound:
\bea
{\bm \lambda}^2 (Q)<\frac{4n\pi}
{1+\frac{2n}{\pi}\ln(\frac{Q_0}{Q})}\ .
\label{rgep28}
\eea
%Obviously, for
%$n \to \infty$ the upper bounds (\ref{rgep23}) and (\ref{rgep28}) coincide.
For $Q_0$ sufficiently large the second term in the denominator 
%of (\ref{rgep28}) 
is much larger than one, and the equation can be 
approximated as 
\bea
{\bm \lambda}^2 (Q) < \frac{2\pi^2}{\ln(\frac{Q_0}{Q})}\ . 
\label{rgep2345}
\eea
For example, 
if the high-energy theory is a typical GUT with $Q_0\sim 10^{16}$ GeV, then 
from eq. (\ref{rgep2345}) with $Q\sim 100$ GeV one obtains the low-energy bound
${\bm \lambda}^2 < (0.78)^2$.
% Using (\ref{rgep28})
% for $n=1$ (NMSSM like) one obtains 
% ${\lambda}^2 < (0.76)^2$.
% For the $\mu\nu$SSM where $n=3$ we get ${\bm \lambda}^2<(0.77)^2$.
% Obviously, $n \to \infty$ one recovers ${\bm \lambda}^2 < (0.78)^2$.
Taking into account that $C$ in eq. (\ref{gg}) gets a
negative(positive) contribution from the top(gauge) coupling, 
one should expect a final
bound slightly stronger.  
%Let us remark nevertheless that since gauge couplings enter with a 
%positive sign (SUPONGO QUE QUIERES DECIR NEGATIVO) in $c$ in eq. (\ref{gg}),
%these bounds can be relaxed.
The numerical analysis indicates that this is the case, with 
%In the NMSSM one finds typically  $\lambda^2<(0.7)^2$, and we can expect
%as a good approximation for any number of singlets, 
${\bm \lambda}^2\lsim (0.7)^2$ as expected.
%Of course, in the analysis of the spectrum in the next subsection we will carry %out the exact computation.
Thus in our case where $i=1,2,3$, we obtain the bound for each coupling $\lambda\equiv \lambda_i\lsim 0.7 / \sqrt{3} \approx 0.4$.

Although in the numerical analysis below we will 
impose the Landau pole constraint assuming that the perturbative description
of the model is valid up to the GUT scale, it is worth noticing here that
intermediate scales like $10^{11}$ GeV
seem also to be interesting to explain several experimental
observations. In addition, it has been found that
the string scale may be anywhere between the weak and the Planck scale \cite{strings}.
Also NMSSM-like models restricted to be perturbative up to about 10-100 TeV have been 
studied \cite{barbieri}.
Considering these possible uncertainties in the unification scale, 
and using e.g. $Q_0\sim 10^{11}$ GeV,
from eq. (\ref{rgep2345}) we would obtain 
${\bm \lambda}^2 < (0.95)^2$.
Taking into account as above the other contributions to the RGE, one can find the final bound ${\bm \lambda}^2 \lsim  (0.88)^2$, and therefore 
$\lambda_i\lsim  0.88 / \sqrt{3} \approx 0.5$. It is worth noticing then that,
for intermediate scales 
the allowed parameter space is larger than in the
case of a typical GUT. Obviously, smaller scales
would imply even larger allowed regions.
For example, with $Q_0\sim 10$ TeV, one obtains a final bound
${\bm \lambda}^2 \lsim  (1.91)^2$, 
implying $\lambda_i\lsim  1.1$.
Another modification will be related to the lightest
Higgs mass. As will be discussed in the next Subsection, its upper bound 
is also larger for smaller unification scales.

\begin{figure}[h!]
\begin{center}
\vspace*{-1,1cm}
\hspace*{-8mm}
    \begin{tabular}{cc}
    \epsfig{file=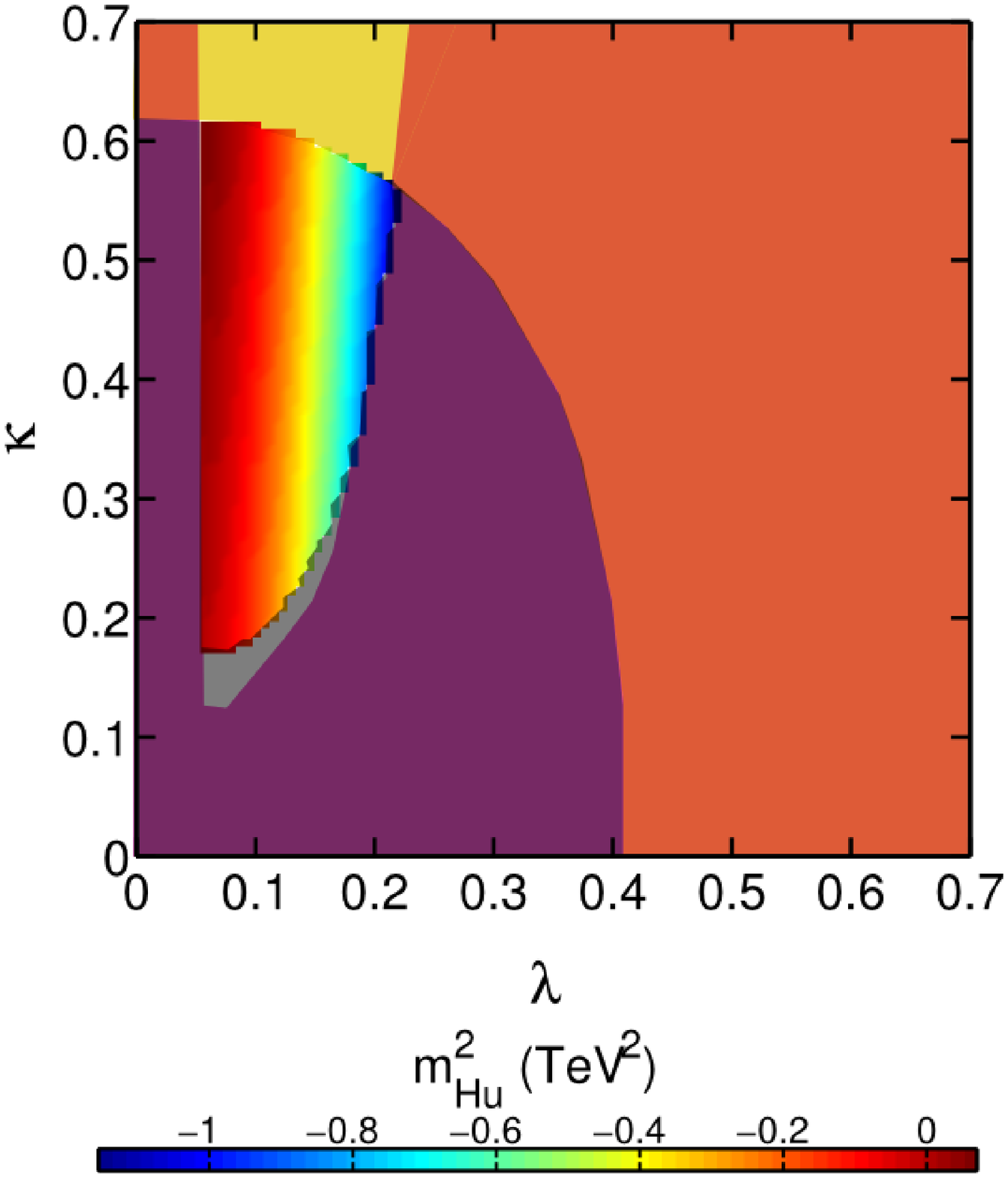,height=8.7cm}
    \hspace*{0mm}& \hspace*{-20mm}
    \epsfig{file=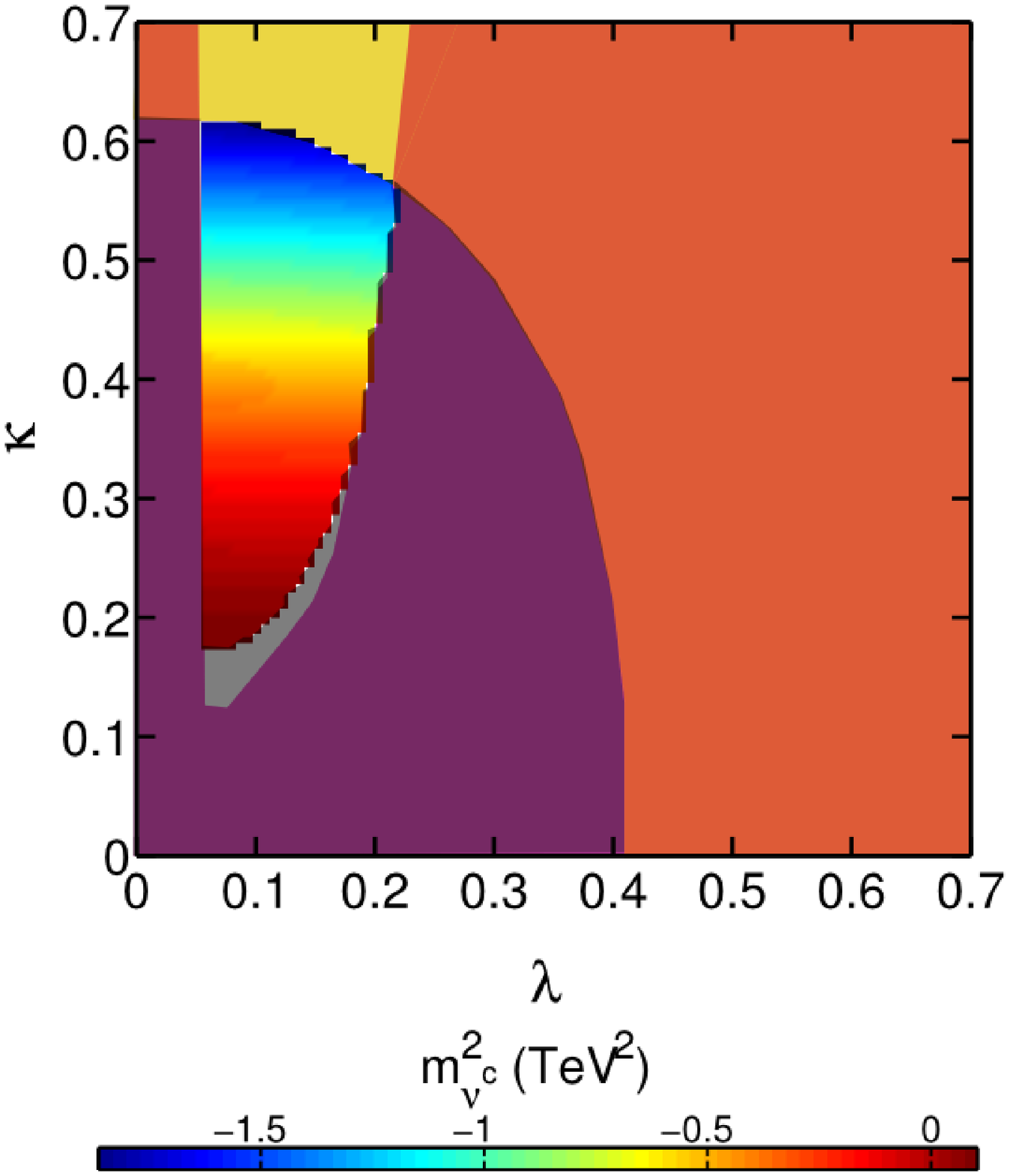,height=8.7cm}
  \vspace*{-0.8cm}       
   \\ & \\
      %(a)\hspace*{8mm} & \hspace*{8mm} (b) 
            (a)\hspace*{8mm} & \hspace*{-3cm} (b)
%     \\ & \\
%      (a)\hspace*{8mm} & \hspace*{8mm} (b)
    \end{tabular}
\captions{
($\lambda$, $\kappa$) parameter space for 
$\tan\beta=5$,
$A_\lambda=1$~TeV, $A_\kappa=A_{\nu}=-1$ TeV, and $\nu^c=2$~TeV. 
In both cases the gray and violet areas represent points which are excluded by the
existence of false minima and tachyons, respectively.
The yellow area
represents points which are excluded due to the occurrence of a Landau pole.
The orange area is excluded by both, Landau pole and tachyons. 
In (a) the colours 
indicate different values of the soft mass $m^2_{H_u}$. 
In (b) the colours indicate different
values of the soft masses $m^2_{\tilde\nu^c}$.}
    \label{figkl1}
\end{center}
\vspace*{1.1cm}
\end{figure}

\begin{figure}[!h!]
\begin{center}
\vspace*{-1,5cm}
\hspace*{-8mm}
    \begin{tabular}{cc}
     \epsfig{file=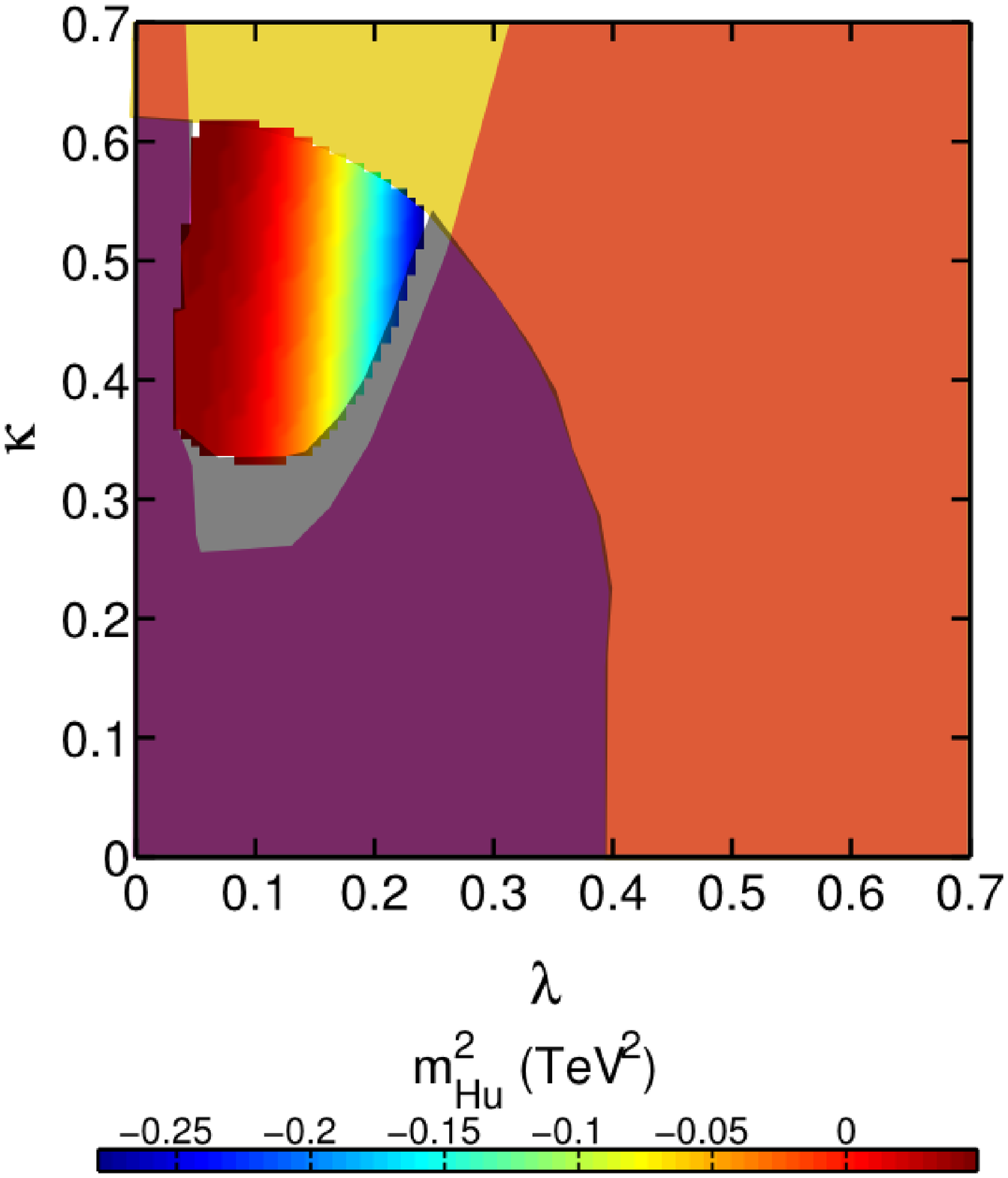,height=8.7cm}
       \hspace*{0mm}&\hspace*{-25mm}
       \epsfig{file=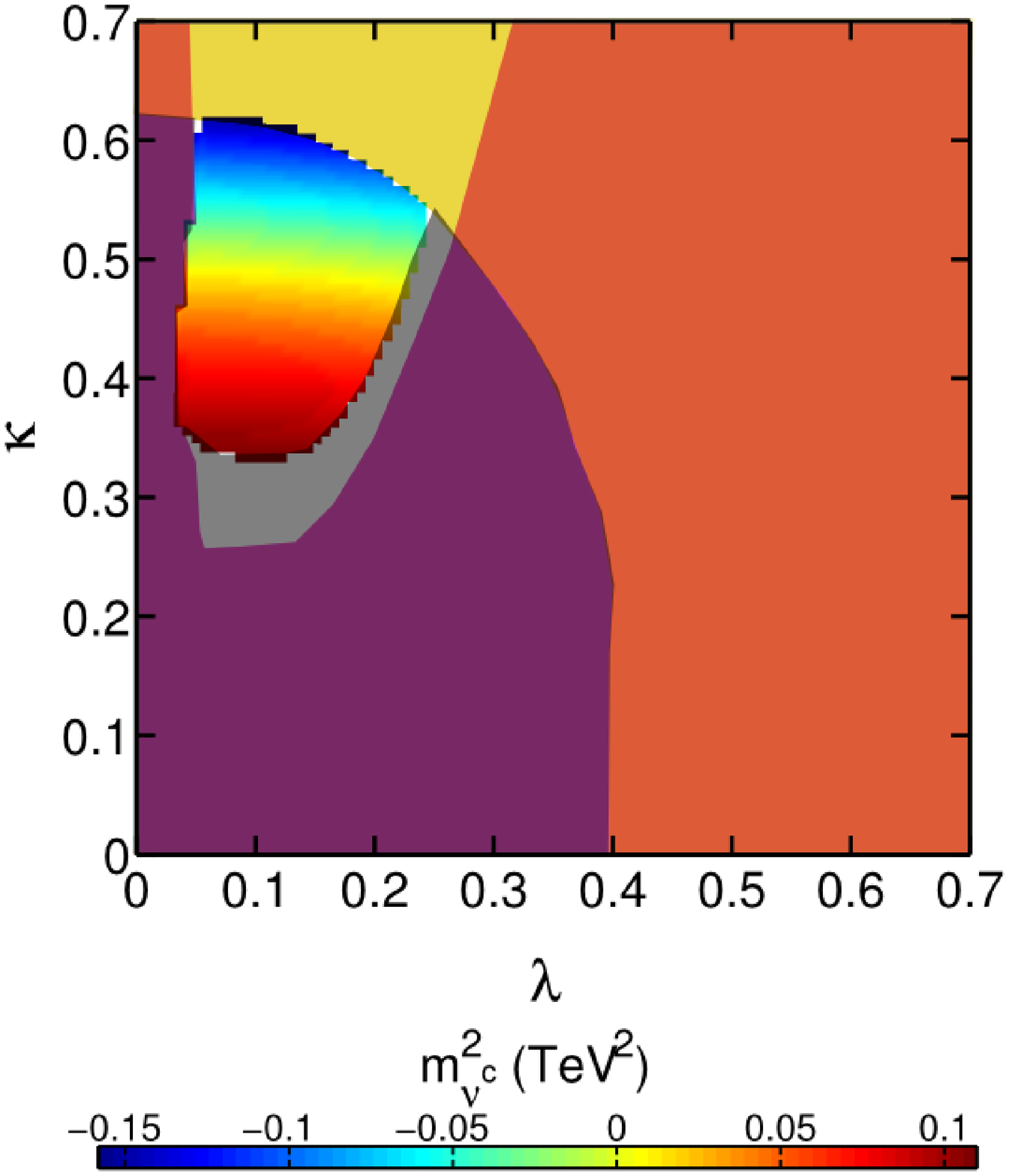,height=8.7cm}
     \vspace*{-0.8cm}      
\\ & \\
      (a)\hspace*{8mm} & \hspace*{-3cm} (b)
    \end{tabular}
    \captions{The same as in Fig. 5 but for the case ${\nu}^c=1$~TeV.}
    \label{figkl2}
\end{center}
\end{figure}

\begin{figure}[t!]
  \begin{center}
\hspace*{-8mm}
    \begin{tabular}{cc}
\epsfig{file=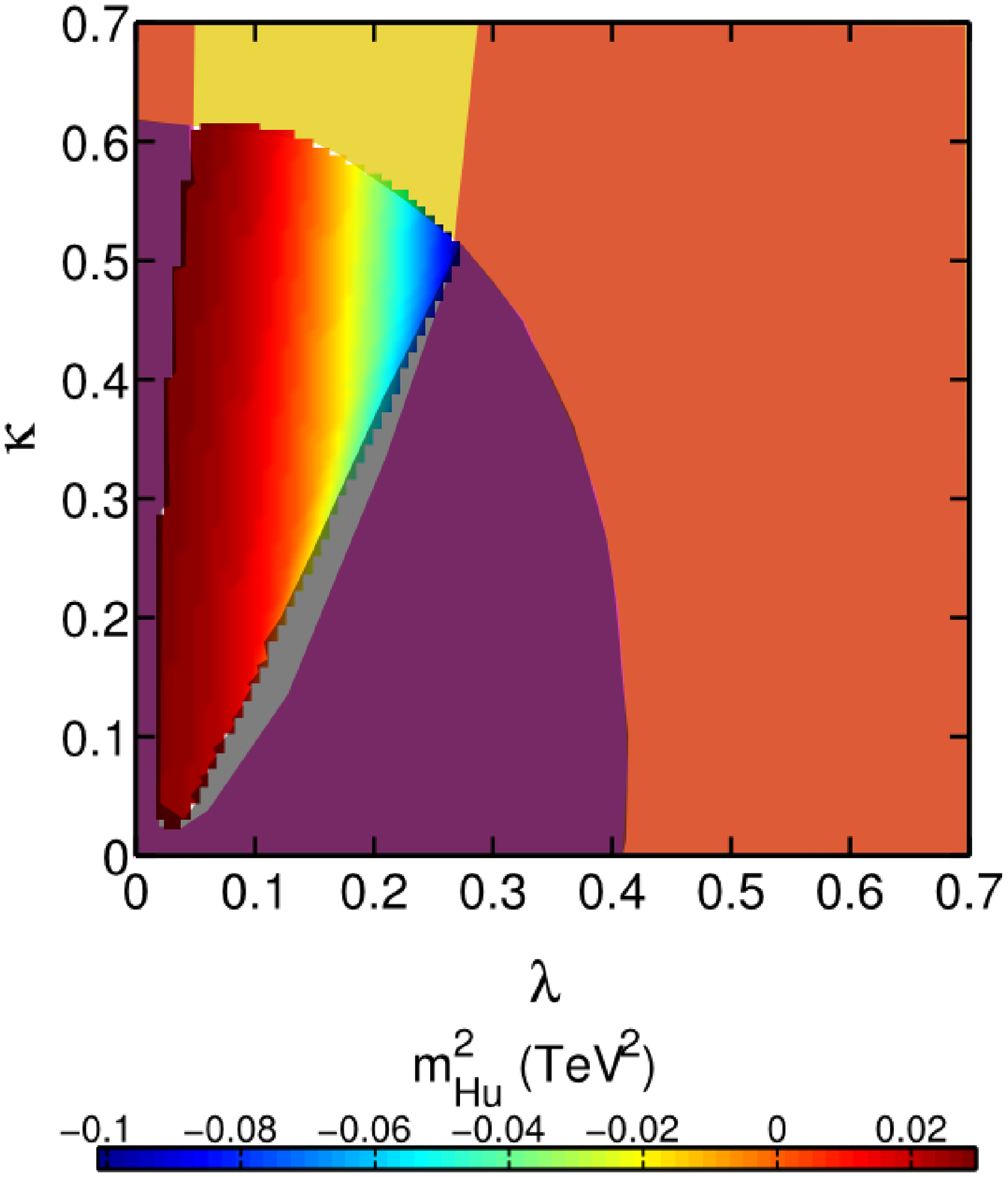,height=8.7cm} 
      \hspace*{0mm}&\hspace*{-25mm}
       \epsfig{file=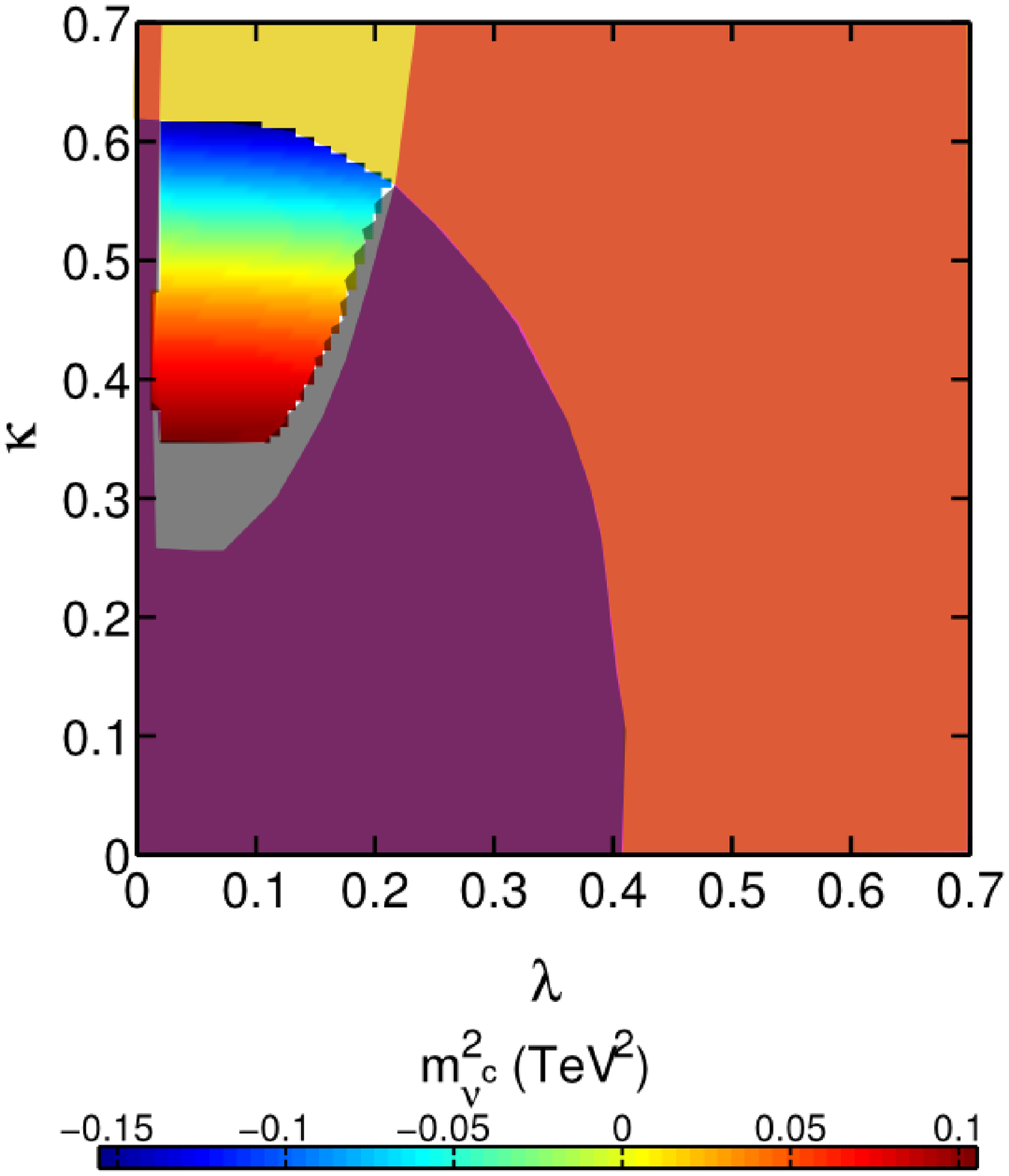,height=8.7cm}
 \vspace*{-0.8cm}     
      \\ & \\
(a) \hspace*{8mm} & \hspace*{-3cm} (b)
    \end{tabular}
    \captions{
    The same as in Fig. 5 but for the cases 
(a) $\tan\beta=5$,
$A_{\lambda}=200$~GeV,  $A_{\kappa}=-50$ GeV, $A_{\nu}=-1000$ GeV, and
$\nu^c=500$~GeV.  The colours 
indicate different values of the soft mass $m^2_{H_u}$.
(b) $\tan\beta=20$,
$A_{\lambda}=1000$~GeV,  $A_{\kappa}=A_{\nu}=-1000$ GeV, and
$\nu^c=1000$~GeV.  
The colours indicate different values of the soft masses $m^2_{\tilde\nu^c}$.}
%  $\lambda$ (0.01,07) vs $\kappa$ (0.01,0.7) plane. On Orange 
%we plot the region that is excluded by the Landau pole 
%condition up to GUT scale. On Gray we plot the false minimun, 
%on Magenta (over Black and Orange) the ones that are tachyons 
%are plotted . (a) $A_\lambda=200$~GeV,  $\nu^c=500$~GeV 
%$A_\kappa=-50$ $\tan(\beta)=5$.  Colours for the soft mass 
%$M^2_{H_u^0}$ and (b)
% $A_\lambda=1000$~GeV,  $\nu^c=1000$~GeV $A_\kappa=-1000~GeV$ 
%$\tan(\beta)=20$.  Colours for the soft mass $M^2_{\nu^c}$. }
    \label{figkl3}
 \end{center}
\end{figure}

In Figs.~\ref{figkl1}-\ref{figkl3} we study
the ($\lambda$, $\kappa$) parameter space.
As expected from the above discussion, 
$\lambda\lsim 0.4$.
Concerning the value of $\kappa$, we also see that perturbativity up to the GUT
scale imposes 
the bound $\kappa\lsim 0.6$, similarly to the NMSSM.
In Fig.~\ref{figkl1} we show an example with
$\tan\beta=5$, $A_{\lambda}=-A_{\kappa}=-A_{\nu}=1$ TeV, 
%$A_{\kappa}=-1$ TeV, 
and $\nu^c=2$~TeV.
For $\lambda\gsim 0.05$ a false minimum region appears. 
As we can deduce from Fig.~\ref{figkl1}a, the reason is that
$m^2_{H_u}$ becomes large and negative, producing as a consequence 
a minimum deeper than the realistic one in the
direction with only $v_u\neq 0$.

It is clear from  Fig.~\ref{figkl1} that the presence of tachyons increases for large values of $\lambda$ (see e.g. the orange area).
The reason is that the off-diagonal value  
$|M_{h_{d}h_{u}}^{2}|$ in Appendix (\ref{appendixA.1.1})
has a dependence on $a_{\lambda}$, thus leading to the appearance of a negative
eigenvalue.
We can also see to the left of the figure, for very small values of 
$\lambda$, a narrow band with tachyons.
The relevant off-diagonal piece is now
$|M_{h_{u}(\widetilde{\nu}_i^c)^R }^{2}|$.
Notice that there are terms with opposite signs producing a cancellation of the mixing for particular values of $\lambda$. However for 
very small values the cancellation disappears and a 
large mixing producing negative eigenvalues arises.


%\vspace{-5cm}

%%%

In Fig.~\ref{figkl2} we show the modifications produced by a   
decrease in the value of $\nu^c$.
In particular, we consider the same values of the parameters as in Fig.~\ref{figkl1} 
but with $\nu^c=1$ TeV instead of 2 TeV. 
% For $\lambda\gsim 0.15$ a false minimum region appears. As discussed above,
% we can deduce from Fig.~\ref{figkl2}a that the reason is that
% $m^2_{H_u}$ becomes large and negative.
% On the other hand, for smaller values of $\lambda$ we can see in 
% Fig.~\ref{figkl2}b that $m^2_{\tilde\nu^c}$ become large and positive contributing
% to make again the direction with only $v_u\neq 0$ the global minimum.
The allowed region is now reduced.
%smaller than in Fig.~\ref{figkl1}.
Notice that $m^2_{\tilde\nu^c}$ becomes positive for
larger values of $\kappa$, producing the presence of minima deeper than the
realistic one 
in the direction with only $v_u\neq 0$.
Let us also remark here 
that the points in the gray area about $\lambda\approx 0.05$ and
$\kappa\approx 0.35$
are actually forbidden by minima deeper than the realistic one with all VEVs vanishing.
% Notice to this respect in the figure that those points correspond to
% positive values of $m^2_{H_u}$ and $m_{{\tilde\nu^c}}^2$.
% This is also true for Fig.~\ref{figal2} discussed below.

Decreasing further $\nu^c$ the allowed region decreases, and in particular for
$\nu^c\approx 500$ GeV, and the same values of the parameters as above, we find that the whole region disappears.
Nevertheless this situation can be improved if we modify the values of
$A_{\kappa}$ and $A_{\lambda}$.
In particular, decreasing $A_{\lambda}$, and increasing (decreasing in modulus)
$A_{\kappa}$, the terms in the potential proportional to them contribute to generate a realistic minimum. 
This can be seen in  Fig.~\ref{figkl3}a, where we take $\nu^c=500$ GeV,  
$A_{\lambda}= 200$ GeV, and $A_{\kappa}=-50$ GeV.
The allowed region is even larger than in 
Fig.~\ref{figkl2} where $\nu^c=1000$ GeV.

Let us finally discuss the variation in $\tan\beta$.
Larger values of
$\tan\beta$ lead to an increase of the mixing in the CP-even neutral scalar matrix,
and as a consequence the
tachyonic region is larger.
We show this effect in  Fig. \ref{figkl3}b for $\tan\beta=20$.
Although the allowed region is smaller
than in Fig.~\ref{figkl2}, the effect is not very important. 
This is also true for larger values of $\tan\beta$.
The reason being that the large value of $\nu^c=1$ TeV produces a heavy
right-handed sneutrino, and therefore a large entry 
$|M_{(\widetilde{\nu}_i^c)^R (\widetilde{\nu}_j^c)^R }^{2}|$.
Since the other relevant entries, $|M_{h_{u} h_u}|$ and
$|M_{h_{u}(\widetilde{\nu}_i^c)^R }^{2}|$, are generically much smaller, 
it turns out to be difficult to generate a negative eigenvalue.
As for $\tan\beta=5$, decreasing further $\nu^c$ for the same value of the parameters,
the allowed region decreases. 
Both effects, the generation of false minima and tachyons, are contributing significantly to forbid points of the parameter space. 
In particular, the latter effect also contributed to forbid 
the whole region for $\tan\beta=5$ and $\nu^c \approx 500$ GeV. This is obvious, since the potential is bounded from below, and, as a consequence, the existence of tachyons implies the existence of a deeper minimum.   
%In particular, as for $\tan\beta=5$, 
%for $\tan\beta=20$ and $\nu^c\approx 500$ GeV 
%we find that the whole region disappears. 
The whole region is also fobidden for $\tan\beta$ larger than 5 when $\nu^c \approx 500$ GeV.

\subsection{Analysis of the spectrum}
\label{analisisspectrum}

%\subsubsection{The spectrum}
%\label{restofspectrum}

Let us now discuss general characteristics of the particle spectrum 
of the $\mu\nu$SSM.
The breaking of $R$-parity generates a peculiar structure for the mass matrices.
The presence of right and left-handed sneutrino VEVs leads to mixing of the
neutral Higgses with the sneutrinos producing the 
$8\times 8$ neutral scalar mass matrices for the 
CP-even 
and CP-odd states written in eqs. (\ref{matrix1}) 
and (\ref{matrix2}), respectively.
Note that after rotating away the CP-odd would be Goldstone boson,
we are left with seven states.
It is also worth noticing here that
the $5\times 5$ Higgs--right handed sneutrino submatrix is basically decoupled from the 
$3\times 3$ left handed sneutrino submatrix, since the mixing occurs only through terms proportional
to $\nu_i$ or $Y_{\nu_{ij}}$, which
are therefore negligible.

Given the interest of the lightest Higgs boson mass in the analysis of SUSY models,
it is worth discussing here its upper bound in the $\mu\nu$SSM.
%\subsubsection{The lightest Higgs mass and the Landau pole constraint}
%\label{Landaupole}
Let us recall that
for an extension of the MSSM with singlets $S_i$, $i=1,...,n$, generating the $\mu$ term through the couplings $\epsilon_{ab} \lambda_{i} \, \hat S_i\,\hat H_d^a \hat H_u^b$, 
one can obtain
a tree-level upper bound on the lightest neutral Higgs 
mass \cite{espinosa,drees} using
the $2\times 2$ submatrix defined by $m_{H_u}$ and $m_{H_d}$
(see Appendix \ref{appendixA.1.1}),
\bea
m^2_{h} \leq M^2_Z \left ( \cos^2 2\beta + \frac{2 {\bm \lambda}^2 
\cos^2 \theta_W}{g^2_2} \sin^2 2\beta \right ) \approx  M^2_Z \left ( \cos^2 2\beta + 3.62 \, {\bm \lambda}^2  \sin^2 2\beta \right ) \label{cota},
\label{boundHiggs}
\eea
where 
${\bm \lambda}^2=\lambda_i \lambda_i$ was defined in the previous Subsection.
Neglecting the small neutrino Yukawa couplings $Y_{\nu_{ij}}$ and
with the substitutions $S_i\rightarrow  \tilde \nu_i^c, i=1,2,3$, the superpotential
of the $\mu\nu$SSM (\ref{superpotential}) is
equivalent to the above extension, and therefore 
we can use the same bound (\ref{boundHiggs}) in our computation.

Clearly, one can optimise this bound choosing $\tan\beta$ as small as possible, as well as 
$\bm \lambda$ as large as possible.
Concerning the latter, let us recall our discussion in the previous
Subsection: the value of 
$\bm \lambda$ 
is actually bounded once perturbativity 
is imposed, and, in particular,
we found
${\bm \lambda}^2\lsim (0.7)^2$ for a typical GUT.
Now, using this bound
%${\bm \lambda}^2<(0.7)^2$ 
one can write (\ref{boundHiggs}) as%Perturbativity respect to the  $\lambda_i$ parameters impose a  bound on (\ref{cota}).
%Since the behaviour of  $\lambda^2$ is similar independently of the number of singlets we expect a similar bound as that of the NMSSM. In such case imposing perturbativity up to the GUT scale $10^{16}$~GeV as we already mention at EW scale we need $\lambda^2<(0.7)^2$.
%${\bm \lambda}^2<(0.7)^2$, one obtains
%can write (\ref{boundHiggs}) as
\bea
m^2_{h} \lsim  M^2_Z \left ( \cos^2 2\beta + 1.77 \, \sin^2 2\beta \right ) \label{cota22}\ ,
\eea
which indicates that for small values of $\tan\beta$ (i.e. large values 
of $\sin 2\beta$) one might obtain in principle large tree-level values for the 
lightest Higgs mass, unlike the MSSM where the second term in 
(\ref{cota22})
is absent.
%%%%%     se puede agrgar esto:      %%%%%%%%%%%%%%%
%%%%%%%   Ejemplos que quite          %%%%%%%%%%
%%%%%%%%%%%%%%%%%%%%%%%%%%%%
For example, for $\tan\beta=2 (4)$ one obtains 
$m_{h} \lsim   1.22(1.08)\times M_Z \approx 111(98)$ GeV.

Of course, in order to get masses close to the upper bound, choosing a certain range of values for other parameters of the model in (\ref{freeparameters25}) is also necessary. 
In particular, we must avoid
as much as possible the mixing of the light eigenstate $h$ of the
$2\times 2$ Higgs submatrix in Appendix (\ref{appendixA.1.1}) with the
right-handed sneutrinos (see eqs. ($\ref{Adr}$) and ($\ref{Aur}$)).
Since this submatrix
is essentially diagonalized by the angle $\frac{\pi}{2} - \beta$, it is easy to check that one has to impose
\begin{equation}
\lambda [6\lambda \nu^c -(A_{\lambda} + 2 \kappa \nu^c)  \sin 2\beta ]
\rightarrow 0  \ .
\label{mixingg}
\end{equation}
% is small compared with 
% $M_Z$ (which is the typical scale of $m_h$), and/or the entries 
% in eq. ($\ref{evenrr}$) large.

On the other hand,
it is well known that the one-loop correction 
to the lightest Higgs mass can be very important. One can check that, similarly to the
NMSSM \cite{higgs}, the upper bound for the lightest doublet-like Higgs mass of the $\mu\nu$SSM is 
of the order of $140$ GeV for $\tan\beta\sim 2$.

As discussed in the previous Subsection, for high-energy theories with
smaller fundamental scales than the GUT one, the upper bound for the coupling
turns out to be larger. In particular, for an intermediate scale 
of the order of $10^{11}$ GeV 
we found
${\bm \lambda}^2 \lsim (0.88)^2$.
Thus, from (\ref{boundHiggs}), one is also able to get a larger tree-level upper bound on the Higgs mass,
%larger tree-level uppper bounds
%on the Higgs mass.
%For the $\mu\nu$SSM 
%where $n=3$
%this the bound would be ${\bm \lambda}^2 < (0.97)^2$, implying
\bea
m^2_{h} \lsim  M^2_Z \left ( \cos^2 2\beta + 2.8 \, \sin^2 2\beta \right ) \label{cota222}\ ,
\eea
generating more flexibility with respect to the experimental data. For example,
for $\tan\beta=2(4)$ one obtains $m_{h} \lsim  1.47(1.18)\times M_Z\approx 134(107)$ GeV.
Using the above mentioned possibility of $10$ TeV for the high-energy, 
scale \cite{barbieri}, producing  
${\bm \lambda}^2 \lsim (1.96)^2$, the result would be
% \bea
% m^2_{h} \lsim  M^2_Z \left ( \cos^2 2\beta + 13.9 \, \sin^2 2\beta \right ) \label{cota2222}\ ,
% \eea
$m^2_{h} \lsim  M^2_Z \left ( \cos^2 2\beta + 13.2 \, \sin^2 2\beta \right )$.
In this case,
for $\tan\beta=2(4)$ one obtains $m_{h} \lsim  2.96(1.92)\times M_Z\approx 270(175)$ GeV.

Concerning the rest of the spectrum, 
the charged Higgses are mixed with the charged sleptons generating the $8\times 8$ charged scalar mass matrix written
in eq. (\ref{matrixchargedscalars}). Nevertheless,
similarly to the neutral scalar mass matrices where some sectors are decoupled,
the $2\times 2$ charged Higgs submatrix is decoupled from the 
$6\times 6$ charged slepton submatrix.
%, since the terms proportional
%to $\nu_i$ and $Y_{\nu_{ij}}$
%are negligible.

The
neutralinos are mixed
with the right- and left-handed neutrinos producing the 10$\times$10 neutral
fermion mass matrix
written in eq. (\ref{matrixneutralinos}). 
As discussed 
in Section \ref{sec:strategy}, three eigenvalues are very small
corresponding to the neutrino masses. The other seven eigenvalues 
arise from the mixing of neutralinos and right-handed neutrinos.

As discussed also in Section \ref{sec:strategy},
although the charginos
mix with the charged leptons giving rise to the 5$\times$5 charged fermion mass matrix
written in eq. (\ref{matrixcharginos}),
the $2\times 2$ chargino submatrix is basically decoupled from the 
$3\times 3$ charged lepton submatrix.
The former is like the one of the MSSM provided that one uses
$\mu=\lambda_i \nu^c_i$.

Let us finally mention that the squark mass matrices are written in
eq. (\ref{matrixsquarks}).
When compared to the MSSM case, their structure is essentially unaffected, 
provided that one uses
$\mu=\lambda_i \nu^c_i$, and neglects the terms proportional to $Y_{\nu}$.

\begin{figure}[t!]
\begin{center}
\hspace*{-8mm}
    \begin{tabular}{cc}
      \epsfig{file=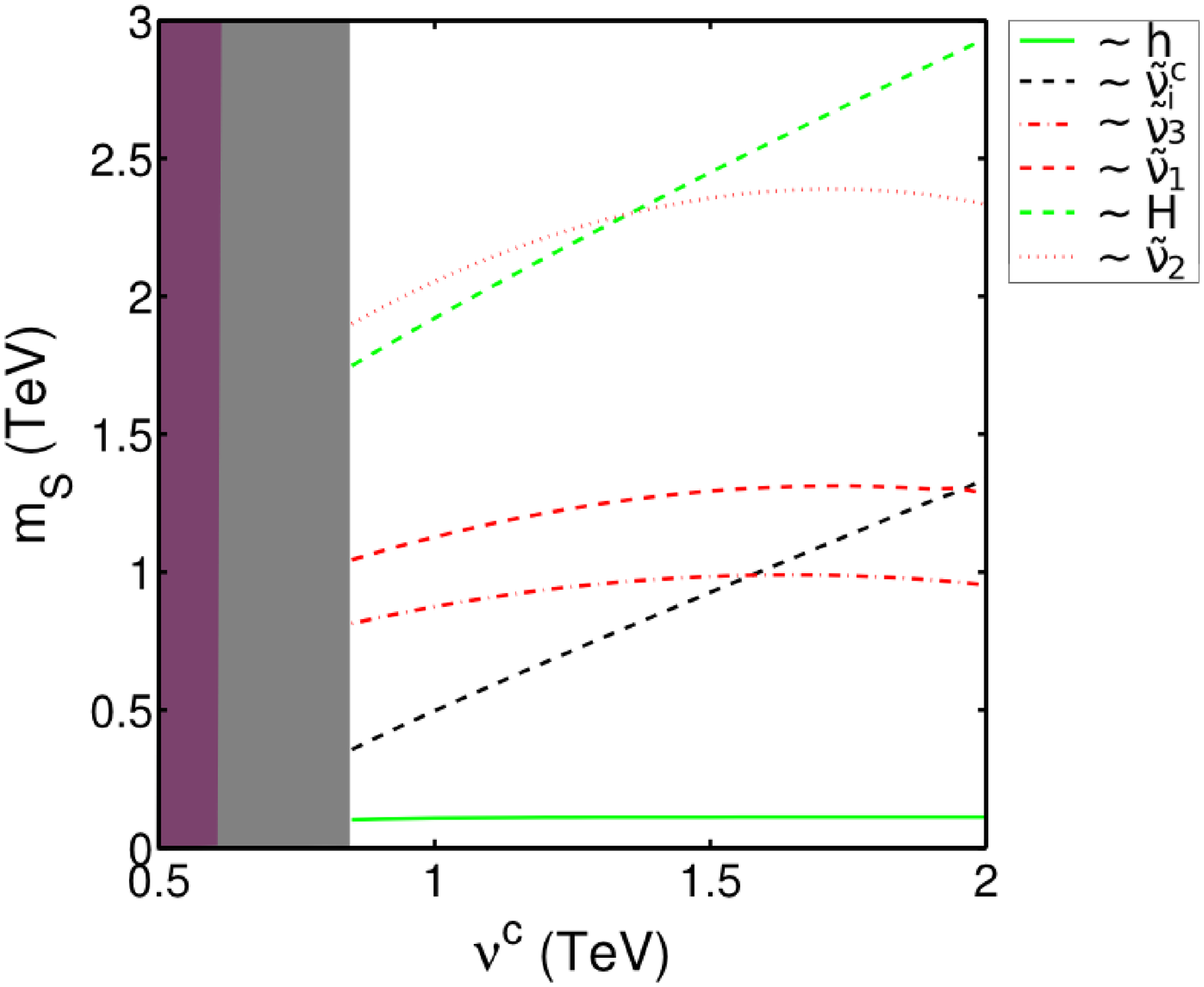,height=8.7cm}
       %\hspace*{0mm}&\hspace*{-25mm}
        % \epsfig{file=plots/elmr1-k-l-ms2nu-al1-nuc2-ak-1-tb5.eps,height=8.7cm}
     % \\ & \\
 %     (a)\hspace*{8mm} & \hspace*{8mm} (b)
    \end{tabular}
    \captions{Masses of the CP-even neutral scalars as a function of the right-handed
sneutrino VEVs, for the parameter space of Fig. 7b with $\lambda=0.1$ and
$\kappa=0.4$. The gray and violet areas are excluded by the existence of false minima and tachyons, respectively.}
    \label{figscalarss}
\end{center}
\end{figure}

For a more detail discussion of the characteristics of the spectrum we need more information about the parameter space. As an example, let us consider the viable region studied in Fig. \ref{figkl3}b with
$\lambda=0.1$ and $\kappa=0.4$.
%, and different values of the right-handed sneutrino VEVs$\nu^c$.
We show first in Fig. \ref{figscalarss} the masses of the CP-even neutral scalars as a function of the right-handed sneutrino VEVs.
For this parameter space we can see from Appendix \ref{appendixA.1.1}
that the mixing between the Higgses and the right-handed sneutrinos is of the order of $a_{\lambda_i}v_u=A_\lambda \lambda v_u$, and therefore small compared 
with the relevant diagonal terms
$\lambda_i\lambda_j v^c_i v^c_j=9 \lambda^2 {v^c}^2$. Thus we have essentially doublet-like Higgses and the LEP bound 
%of 114 GeV 
for the lightest Higgs mass applies.
The masses of the heavy and light Higgses, $H$ and $h$, are shown in the figure with green dashed and solid lines, respectively. Concerning the former, its mass varies between 1748 and 2935 GeV.
Concerning the latter, since $\tan\beta=20$ the upper bound is like in the MSSM,
as discussed above.
%in Section \ref{Landaupole}.
For the values of the parameters used in this example, we obtain
$m_h \approx 115.5$ GeV.
If instead of $A_t=1$ TeV, we would have consider the 'maximal mixing' 
scenario \cite{espinoso}, which in our case is obtained for $A_t\approx 2.4$ TeV, we would have obtained
$m_h \approx 126$ GeV.
As discussed also in eq. (\ref{mixingg}),
larger values can be obtained avoiding as much as possible the small mixing of the light Higgs $h$ with the right-handed sneutrinos. For example, for $\lambda=0.05$ one obtains 
$m_h \approx 117.5$ GeV. Imposing in addition the maximal mixing scenario,
$m_h \approx 128$ GeV.

\begin{figure}[t!]
\begin{center}
\hspace*{-8mm}
    \begin{tabular}{cc}
      \epsfig{file=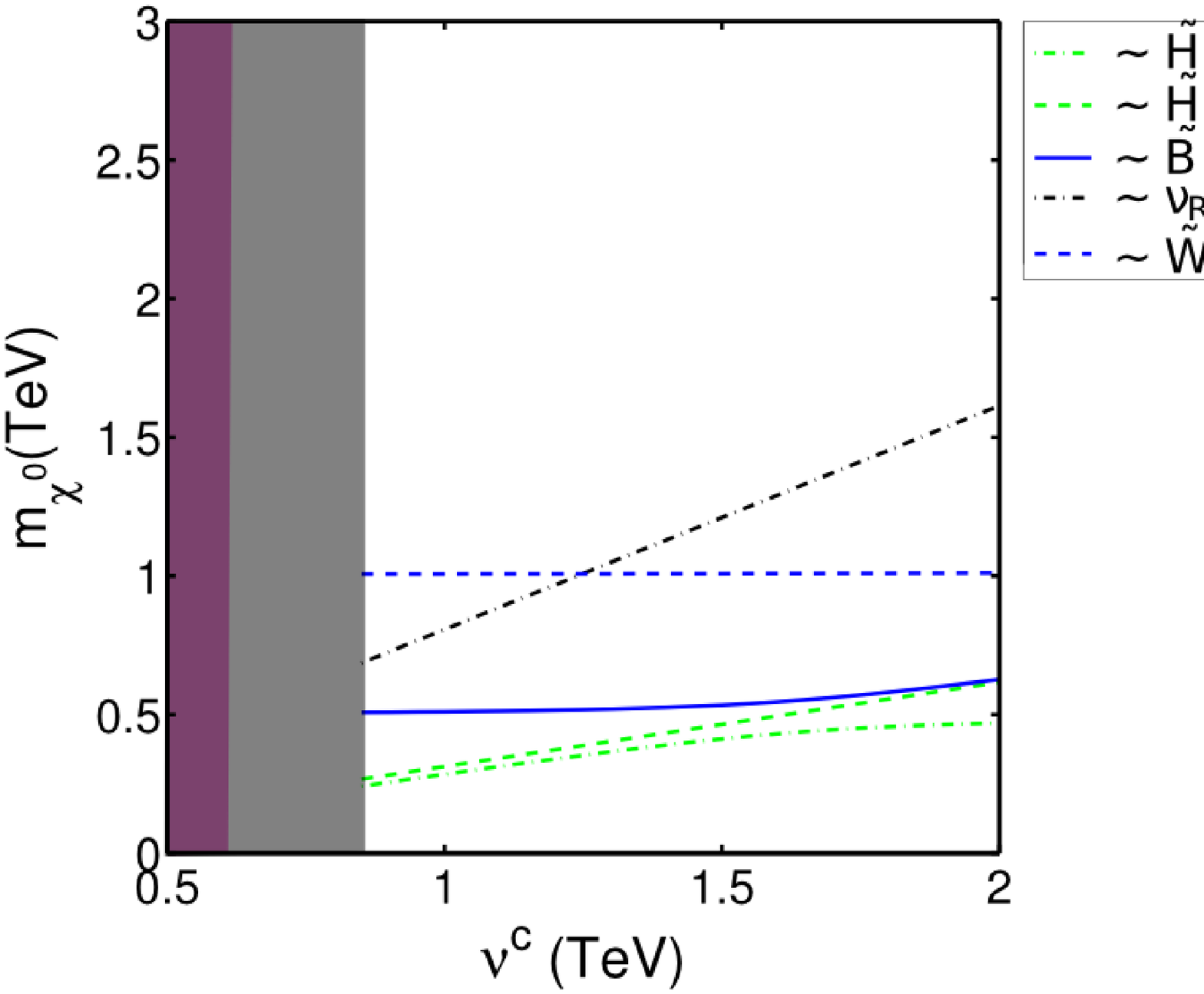,height=8.7cm}
       %\hspace*{0mm}&\hspace*{-25mm}
        % \epsfig{file=plots/elmr1-k-l-ms2nu-al1-nuc2-ak-1-tb5.eps,height=8.7cm}
     % \\ & \\
 %     (a)\hspace*{8mm} & \hspace*{8mm} (b)
    \end{tabular}
    \captions{The same as in Fig. 8 but for the masses of the neutral fermions.
%Masses of the neutral fermions as a function of the right-handed
%sneutrino VEVs, 
%for the parameter space of Fig. 6b with $\lambda=0.1$ and $\kappa=0.4$.
}
    \label{figneutralinos}
\end{center}
\end{figure}

The three right-handed sneutrinos are essentially degenerated (up to small contributions due to neutrino Yukawas), and we show their masses  with a black dashed line which varies approximately between 357 and 1346 GeV.
%and increase for larger values of the VEVs as expected.
Let us remark that in general
to obtain singlet-like Higgses, thus scaping detection and being in agreement
with accelerator data, is also possible for small values of $\kappa$.
This can be qualitatively understood from the expresion of the corresponding mass matrix. In particular, the terms 
$M_{(\widetilde{{\nu}}^{c}_{i})^R (\widetilde{{\nu}}^{c}_{i})^R}^{2}$
are of the order of $\kappa^2 {\nu^c}^2$, and
become very small when $\kappa$ decreases.

Concerning the left-handed sneutrinos $\tilde\nu_i$ in Fig. \ref{figscalarss}, we see in the Appendix
that their masses are basically determined by the corresponding soft masses,
$m_{\widetilde{L}_{i}}$. Notice that the other terms in
$M_{(\widetilde{{\nu}}_{i})^R (\widetilde{{\nu}}_{j})^R}^{2}$
are proportional
to $\nu_i$ or $Y_{\nu_{ij}}$, and therefore 
negligible. On the other hand, the values of
$m_{\widetilde{L}_{i}}^2$ are fixed by
the minimisation conditions (\ref{eq:tadpoles}), and as a consequence they are
essentially proportional to 
$(Y_{{\nu}_i}/\nu_i)\nu^c$ for the viable region of the parameter space studied here.
For example, for $\nu^c=1$ TeV in the figure, the values of the Yukawa couplings are given by $Y_{\nu_1}=1.64\times 10^{-7}$, $Y_{\nu_2}= 5.43\times 10^{-7}$ and 
$Y_{\nu_3}= 9.85\times 10^{-7}$.  
Using the VEVs $\nu_i$ discussed above eq. (\ref{freeparameters25}), one obtains from the previous formula
$m_{\tilde\nu_2}\sim 1.8 m_{\tilde\nu_1}$, and $m_{\tilde\nu_3}\sim 0.77 
m_{\tilde\nu_1}$.
This can be checked with the figure.

Let us finally remark that for the region of the parameter space discussed here,
to work with other values of $\tan\beta$ would not modify the spectrum obtained, with the exception of the masses of $h$ and $H$.
This is also true for the rest of the spectrum discussed below.
For example, for $\tan\beta=5$ we obtain essentially the same spectrum but
with $m_H$ varying approximately between $1310$ and $2332$ GeV, and $m_h\approx 112$ (124 GeV for maximal mixing).

It is straightforward to see from Appendix \ref{A.1.2} that  
the masses of the CP-odd neutral scalars are very similar to those of the 
CP-even neutral scalars discussed above.
In particular, the masses of the pseudoscalar and left-handed sneutrinos are similar to the masses 
of the heavy Higgs $H$, and left-handed sneutrinos in Fig. \ref{figscalarss}.
The only differences appear for the right-handed sneutrino masses.
Note e.g. that the terms
$2a_{\kappa_{ijk}}{\nu^c_{k}}$ and $2\kappa_{ijk}\kappa_{lmk}\nu^{c}_l\nu^{c}_m$ 
have different signs in eqs. (\ref{evenrr}) and (\ref{oddrr}), implying that now the  masses
vary approximately between 1 and 1.5 TeV.

Concerning the charged scalars, we can see in Appendix \ref{A.1.3} that the mass of the charged Higgs is
very similar to the ones of the pseudoscalar and heavy Higgs.
As mentioned above, the right- and left-handed charged sleptons are decoupled from the charged Higgs. In the Appendix we see that their masses are essentially determined by the corresponding soft masses,
$m_{\widetilde{e}^c_{i}}, m_{\widetilde{L}_{i}}$.
Although the former are free at the electroweak scale in our computation, 
the latter are fixed by
the minimization conditions (\ref{eq:tadpoles}), and therefore we obtain
the same masses as for the left-handed sneutrinos.

In Fig. \ref{figneutralinos} we show the seven eigenvalues corresponding 
to the mixing of neutralinos and right-handed neutrinos. As mentioned in the 
previous Subsection,
we have taken values for the soft gaugino masses
that mimic at low scale the results from a hypothetical
unified value at the GUT scale. In particular, we have assumed
$M_2=1$ TeV and consequently $M_1\approx 500$ GeV.
As we can see in the figure, and can be deduced from the matrix (\ref{neumatrix}),
for the values of the parameters analysed we obtain almost pure Wino, Bino,
Higgsino and right-handed neutrino states.
%Indeed lowering the values of $M_2$ or $\lambda_i$ the mixing would have been
%more important. 
The blue dashed (solid) line corresponds to the Wino (Bino) mass, which is determined approximately by the soft mass $M_2$ ($M_1$).
The Higgsino masses are determined approximately
by the effective $\mu$ term, $\lambda_i \nu^c_i=3\lambda \nu^c$.
We show with a green dashed (dot-dashed) line the heaviest (lightest) Higssino
$\tilde H_2$ ($\tilde H_1$). 
Their masses vary between 267 (242) and 617 (464) GeV.
Finally, the three right-handed neutrinos ${\nu_R}_i$ are degenerated
with a mass that can be approximated as $2\kappa \nu^c$.
This is shown with a black dot-dashed line in the figure varying 
between 686 and 1620 GeV.
Although in the present case the lightest neutralino is a Higgsino, due to our choice of input values with $M_1>3\lambda \nu^c$, this can easily be modified 
by choosing other values of the parameters.
The lightest neutralino can also be essentially a 
right-handed neutrino for small $\kappa$. 
Let us finally remark 
that varying the values of the parameters also the mixing of states can be augmented.
This can be obtained by making the diagonal entries similar to each other and/or increasing the off diagonal entries.  

On the other hand,
from the $2\times 2$ chargino submatrix in eq. (\ref{submatrix})
we can easily deduced that the mass of the charged Wino is approximately
given by $M_2$, and the mass of the charged Higgsino by
the effective $\mu$ term, $\mu=\lambda_i \nu_i^c$.

% Actually, concerning gaugino masses we will only use $M_2$ as input,
% since we will assume for the others the relations of 
% a typical GUT with universal gaugino masses,
% $M_1 = \frac{g_1^2}{g_2^2} M_2,
% M_3 =  \frac{g_3^2}{g_2^2} M_2$.
% \begin{equation}
% M_1 = \frac{g_1^2}{g_2^2} M_2\ , \qquad
% M_3 =  \frac{g_3^2}{g_2^2} M_2\ .
% \end{equation}

Finally,
the eigenvalues of the squark mass matrices depend on the soft masses.
As for the right-handed sleptons, in our computation these are free parameters at the electroweak scale.

\section{Conclusions and outlook}
\label{conclusions}

We have performed the first detailed
analysis of the $\mu\nu$SSM. As explained in the Introduction,
this model was proposed \cite{NuMSSM} as a SUSY standard model for solving 
the crucial $\mu$ problem of SUSY constructions,
generating at the same time 
the small neutrino masses through a dynamical see-saw
at the electroweak scale.
This is due to the inclusion of three 
generations of right-handed neutrino superfields
and the corresponding new gauge invariant couplings,
$\epsilon_{ab} \lambda_{i} \, \hat \nu^c_i\,\hat H_d^a \hat H_u^b$
and
$\kappa_{ijk} \hat \nu^c_i\hat \nu^c_j\hat \nu^c_k$.
The latter couplings break R-parity explicitly and therefore the phenomenology
of the $\mu\nu$SSM is very peculiar and different from other models,
not only from those conserving R-parity, but also from those were
R-parity is also broken.

In this work we have extended the analysis of ref. \cite{NuMSSM}, where the
characteristics of the $\mu$$\nu$SSM were only introduced, and
several approximations were considered in the
phenomenological discussion. In particular, only one generation of
sneutrinos were assumed to acquire VEVs.
Here we have worked with the full three generations.
We have written for the first time the corresponding scalar potential and minimized it
in order to study the electroweak symmetry breaking.
One-loop corrections have been taken into account in the computation.
In total eight fields acquire VEVs. They are, in addition to the
usual Higgses, the right- and left-handed sneutrinos.
Notice that minima with some or all of the VEVs
vanishing are in principle possible, and therefore one has to check that 
the minimum with non-vanishing VEVs 
breaking the electroweak symmetry, and generating
the $\mu$ term and neutrino masses spontaneously, is the global one.

Obviously, due to the many VEVs and the new couplings, the parameter space of $\mu$$\nu$SSM is very involved.
After discussing in detail the strategy to follow in the low-energy analysis, we 
have studied viable regions of the parameter space which are left after imposing
several constraints. 
In addition to discard regions with the false minima mentioned above, we
have discarded also regions with tachyons, as well as those where the Landau pole constraint
on the couplings at the GUT scale is not fulfiled.
Of course, reproducing neutrino data is also used as a constraint in the parameter space. Results are shown in Figs. 2-7.

Finally, we have discussed the particle spectrum.
% associated to the viable regions.
The breaking of $R$-parity generates complicated mass matrices and
mass eigenstates.
The presence of right and left-handed sneutrino VEVs leads to mixing of the
neutralinos
with the neutrinos producing a 10$\times$10 matrix. 
Indeed three
eigenvalues of this matrix are very small, reproducing the experimental results on neutrino masses.
On the other hand, the charginos
mix with the charged leptons giving rise to a 5$\times$5 matrix.
Nevertheless, there will always be three light eigenvalues corresponding to the 
electron, muon and tau.
Concerning the scalar mass matrices,
the neutral Higgses are mixed with the sneutrinos,
and the charged Higgses with the charged sleptons,
and we are left with fifteen (eight CP-even and seven CP-odd) neutral states 
and seven charged states.
Notice however that the three left handed sneutrinos are basically decoupled 
from the Higgs-right handed sneutrinos,
and also the six charged sleptons are decoupled from the charged Higgses.

Given the interest of the lightest Higgs boson mass in the analysis of SUSY models,
we have discussed in detail the mass of the lightest CP-even neutral scalar
in our model.
The upper bound turns out to be similar to the one of the NMSSM, about 140 GeV
after imposing the Landau pole constraint up to the GUT scale.
For the precise masses of the Higges and of the rest of the spectrum, it is not possible to give a result valid for the whole parameter space, given the complicated structure of the model. Nevertheless, we have pointed out several interesting characteristics, and analysed particular regions and possible variations. An example of a possible spectrum is shown
in Figs. 8-9.

Once we have checked explicitly that the parameter space of our model
contains viable solutions and the associated spectrum 
is interesting, and given the hope that the LHC will be able to test SUSY,
it is then important to study in detail
the collider phenomenology of the $\mu\nu$SSM.
In particular, the impact of the new couplings on the usual SUSY searches, and 
indeed
novel signals that might facilitate the confirmation of the 
$\mu\nu$SSM as the adequate SUSY Standard Model.
This necessary task will be the subject of a forthcoming publication.

\vspace{1cm}

\noindent {\bf Acknowledgments} 

\noindent D.E. L\'opez-Fogliani thanks the Science and Technology Facilities Council (STFC) for financial support.
C. Mu\~noz and R. Ruiz de Austri were supported
in part by the
MICINN under Proyectos Nacionales FPA2006-05423 and
FPA2006-01105, and by
the European Union under the RTN program
MRTN-CT-2004-503369. 
We thank 
the project HEPHACOS P-ESP-00346 
of the Comunidad de Madrid. 
The
use of the ciclope cluster of the IFT-UAM/CSIC is also acknowledged.
D.E. L\'opez-Fogliani also wants to thank S. Fauquier for her support.

\vspace{0.5cm}

%\newpage

%%%%%%%%%%%%%%%%%%%%%%%%%%%%%%%%%%%%%%%%%%%%%%%%%%%%%%%%%%%%%%%%
\appendix

%%%%%%%%%%%%%%%%%%%%%%%%%%%%%%%%%%%%%%%%%%%%%%%%%%%%%%%%%%%%%%%%
\section{Mass matrices \label{appx:mass}}
%%%%%%%%%%%%%%%%%%%%%%%%%%%%%%%%%%%%%%%%%%%%%%%%%%%%%%%%%%%%%%%%

%================

%We redefine the files as 
%$H^0_u=h_u+\,iP_u+v_u$, $\,H_d^0=h_d+\,iP_u+v_d$,  
%$\, \widetilde{\nu}^c_i=(\widetilde{\nu}^c_i)^R+\,
%                           i(\widetilde{\nu}_i^c)^I+\nu_i^c$,  
%$\, \widetilde{\nu}_i=(\widetilde{\nu}_i)^R  +\, 
%                           i (\widetilde{\nu}_i)^I+ \nu_i$.
%==============0

In this Appendix we will study the general mass matrices generated in the
$\mu\nu$SSM. 
For this study we will use the indices $i,j,k,l,m=1,2,3$, and $\alpha,\beta,\gamma,\delta=1,...,8$.

\subsection{Scalar mass matrices}

Here we study the scalar mass matrices.
Let us recall that concerning the Higgses,
the neutral ones are mixed with the sneutrinos, and the
charged ones with the charged sleptons.
%For this study we will use the indices $i,j,k,l,m=1,2,3$, and $\alpha,\beta,\gamma,\delta=1,...,8$.

\subsubsection{CP-even neutral scalars}
\label{appendixA.1.1}

The quadratic potential includes
\begin{align}
V_{\text{quadratic}}= \mathbf{S'}_{\alpha} M^2_{s_{\alpha\beta}}  \mathbf{S'}_{\beta} +... \ ,
\label{matrix1}
\end{align}
where  $\mathbf{S}'_\alpha=(h_d, h_u,(\widetilde{\nu}^c_i)^R, (\widetilde\nu_i)^R)$ is in the unrotated basis, and
% \begin{align}
% V_{\text{quadratic}}=(h_d,h_u,,(\widetilde{\nu}^c_1)^R,(\nu_1)^R,(\widetilde{\nu}^c_2)^R,(\nu_2)^R,(\widetilde{\nu}^c_3)^R,(\nu_3)^R)M^2_s \left( \begin{array}{c} h_d \\h_u\\ (\widetilde{\nu}^c_1)^R \\ (\nu_1)^R\\ (\widetilde{\nu}^c_2)^R\\ (\nu_2)^R\\ (\widetilde{\nu}^c_3)^R\\ (\nu_3)^R  \end{array}\right)+...
% \end{align}
below we give the expressions for the independent coefficients of $M^2_{s_{\alpha\beta}}$\\
% The independent $m_{hh}$ terms:
\begin{align}
M_{h_{d}h_{d}}^{2}=m_{H_d}^{2}+\frac{G^2}{4}\{3v_{d}^{2}-v_{u}^{2}+\nu_{i}\nu_{i}\}+\lambda_{i}\lambda_{j}\nu_i^c \nu_j^c 
+\lambda_{i}\lambda_{i}v_{u}^{2}\ ,
\end{align}
\begin{align}
M_{h_{u}h_{u}}^{2}=m_{H_u}^{2}+\frac{G^2}{4}(-v^2_{d}+3v_{u}^{2}-\nu_{i}\nu_{i})+ \lambda_{i}\lambda_{j} \nu_i^c \nu_j^c +\lambda_{i}\lambda_{i}v_{d}^2 \nonumber \\ 
-2Y_{\nu_{ij}} \lambda_{j}v_{d}\nu_{i}+ Y_{\nu_{ik}} Y_{\nu_{ij}}\nu_j^c \nu_k^c+Y_{\nu_{ik}}Y_{\nu_{jk}} \nu_{i}\nu_{j}\ ,
\end{align}
\begin{align}
M_{h_{d}h_{u}}^{2}=-a_{\lambda_{i}}\nu_i^c-\frac{G^2}{2}v_{d}v_{u}+2v_{d}v_{u}\lambda_{i}\lambda_{i}-(\lambda_{k}\kappa_{ijk} \nu_i^c \nu_j^c + 2Y_{\nu_{ij}}\lambda_{{j}}v_{u}\nu_{i})\ , 
\end{align}
% The independent $m_{h\nu^c}^{2}$ terms:
\begin{align}
M_{h_{d}(\widetilde{\nu}_i^c)^R }^{2}=-a_{\lambda_{i}}v_{u}+2\lambda_{i}\lambda_{j}v_{d}\nu_j^c -2\lambda_{k} \kappa_{ijk}v_{u}\nu_j^c-Y_{\nu_{ji}}\lambda_{k}\nu_{j}\nu_k^c -Y_{\nu_{jk}}\lambda_{i}\nu_{j}\nu_k^c\ ,
\label{Adr}
\end{align}
\begin{align}
M_{h_{u}(\widetilde{\nu}_i^c)^R }^{2}=-a_{\lambda_{i}}v_{d}+a_{\nu_{ji}}\nu_{j}+2\lambda_{i}\lambda_{j}v_{u}\nu^c_{j}-2\lambda_{k}\kappa_{ilk}v_{d}\nu^c_{l}+2Y_{\nu_{jk}}\kappa_{ilk}\nu_{j}\nu^c_{l}
+2Y_{\nu_{jk}}Y_{\nu_{ji}}v_{u}\nu^c_{k}\ ,
\label{Aur}
\end{align}
\begin{align}
M_{h_{d}(\widetilde{\nu}_i)^R}^{2}=\frac{1}{2} G^2v_{d}\nu_{i}-(Y_{\nu_{ij}}\lambda_{j}v_{u}^{2}+Y_{\nu_{ij}}\lambda_{k}\nu^c_{k}\nu^c_{j})\ ,
\end{align}
\begin{align}
M_{h_{u}(\widetilde{\nu}_i)^R}^{2}=a_{\nu_{ij}}\nu^c_{j}-\frac{G^2}{2}v_{u}\nu_{i}-2Y_{\nu_{ij}}\lambda_{j}v_{d}v_{u}+Y_{\nu_{ik}}\kappa_{ljk}\nu^c_{l}\nu^c_{j}+2Y_{\nu_{ij}}Y_{\nu_{kj}}v_{u}\nu_{k}\ ,
\end{align}
% The independent $m_{\widetilde{\nu}^{c}\widetilde{\nu}}^{2}$terms:
\begin{align}
M_{(\widetilde{\nu}_{i})^R (\widetilde{\nu}_{j})^R}^{2}=m_{\tilde{L}_{ij}}^{2}+\frac{G^2}{2}\nu_{i}\nu_{j}+\frac{1}{4}
G^2(\nu_{k}\nu_{k}+v_{d}^{2}-v_{u}^{2})\delta_{ij}+Y_{\nu_{ik}}Y_{\nu_{jk}}v^2_{u}+Y_{\nu_{ik}}Y_{\nu_{jl}}\nu^c_{k}{\nu}^{c}_l\ ,
\end{align}
\begin{align}
M_{(\widetilde{\nu}_{i})^R (\widetilde{\nu}^{c}_{j})^R}^{2}=a_{\nu_{ij}}v_{u}-Y_{\nu_{ij}}\lambda_{k} v_{d}{\nu}^{c}_k-Y_{\nu_{ik}}\lambda_{j}v_{d}{\nu}^{c}_k
+2Y_{\nu_{ik}}\kappa_{jlk}v_{u}{\nu}^{c}_l\nonumber\\
+Y_{\nu_{ij}}Y_{\nu_{kl}}\nu_{k}{\nu}^{c}_l
+Y_{\nu_{il}}Y_{\nu_{kj}}\nu_{k}{\nu}^{c}_l\ , 
\end{align}
\begin{align}
M_{(\widetilde{{\nu}}^{c}_{i})^R (\widetilde{{\nu}}^{c}_{j})^R}^{2}=m_{\widetilde{{\nu}}^{c}_{ij}}^{2}+2a_{\kappa_{ijk}}{\nu^c_{k}}-2\lambda_{k}\kappa_{ijk}v_{d}v_{u}+ 2\kappa_{ijk}\kappa_{lmk}\nu^{c}_l\nu^{c}_m+4\kappa_{ilk}\kappa_{jmk}\nu^{c}_l\nu^{c}_m\nonumber \\
+\lambda_{i}\lambda_{j}(v_{d}^{2}+v_{u}^{2})+2Y_{\nu_{lk}}\kappa_{ijk}v_{u}\nu_{l}-(Y_{\nu_{kj}}\lambda_{i}+Y_{\nu_{ki}}\lambda_{j})v_{d}\nu_{k}+Y_{\nu_{ki}}Y_{\nu_{kj}}v_{u}^{2}
+Y_{\nu_{ki}}Y_{\nu_{lj}}\nu_{k}\nu_{l}\ .
\label{evenrr}
\end{align}

Then the mass eingenvectors are
\bea
 \mathbf{S}_{\alpha}=R^s_{\alpha\beta} \mathbf{S'}_\beta\ ,
\eea
with the diagonal mass matrix 
\bea
(M^{\text{diag}}_{s_{\alpha \beta}})^2=R^s_{\alpha \gamma} M^2_{s_{\gamma \delta}} R^s_{\beta \delta}\ .
\eea

\subsubsection{CP-odd neutral scalars}
\label{A.1.2}

In the unrotated basis $\mathbf{P'}_{\alpha}=\left( P_d,P_u,(\widetilde{\nu}^c_i)^I,(\widetilde{\nu}_i)^I \right)$ we have

% \begin{align}
% V_{\text{quadratic}}=\left(P_d,P_u,(\widetilde{\nu}^c_1)^I,(\nu_1)^I,(\widetilde{\nu}^c_2)^I,(\nu_2)^I,(\widetilde{\nu}^c_3)^I,(\nu_3)^I \right)M^2_s \left( \begin{array}{c} h_d \\h_u\\ (\widetilde{\nu}^c_1)^I \\ (\nu_1)^I\\ (\widetilde{\nu}^c_2)^I\\ (\nu_2)^I\\ (\widetilde{\nu}^c_3)^I\\ (\nu_3)^I  \end{array}\right)+...
% \end{align}
\begin{align}
V_{\text{quadratic}}=\mathbf{P'}_{\alpha} M^2_{P_{\alpha \beta}} \mathbf{P'}_{\beta}+...
\label{matrix2}
\end{align}
Below we give the expressions for the independent cofficients of $M^2_{P_{\alpha \beta}}$\\
% 
% Below we give the expresion for the independent cofficient\\
% The independent $m_{PP}$terms (For simplify the notation we don't
% writte the supraindex I in $\widetilde{\nu}_{i}^{I}$and $(\nu^c_{i})^{I}$):
\begin{align}
M_{P_{d}P_{d}}^{2}=m_{H_{d}}^{2}+\frac{G^2}{4}(v_{d}^{2}-v_{u}^{2}+\nu_{i}\nu_{i})+\lambda_{i}\lambda_{j}\nu^c_i \nu^c_{j}
+\lambda_{i}\lambda_{i}v_{u}^{2}\ ,
\end{align}
\begin{align}
M_{P_{u}P_{u}}^{2}=m_{H_{u}}^{2}+\frac{G^2}{4}(v_{u}^{2}-v_{d}^{2}-\nu_{i}\nu_{i})+\lambda_{i}\lambda_{j}\nu^c_i\nu^c_{j}+\lambda_{i}\lambda_{i}v_{d}^2\nonumber \\
-2Y_{\nu_{ij}}\lambda_{j}v_{d}\nu_{i}+Y_{\nu_{ik}}Y_{\nu_{ij}}\nu^c_k\nu^c_{j}+Y_{\nu_{ik}}Y_{\nu_{jk}}\nu_{i}\nu_{j}\ ,
\end{align}
\begin{align}
M_{P_{d}P_{u}}^{2}=a_{\lambda_{i}}\nu^c_{i}+\lambda_{k}\kappa_{ijk}\nu^c_{i}\nu^c_{j}\ ,
\end{align}
%\\
% The independent $m_{h\widetilde{\nu}^c}^{2}$ terms:
\begin{align}
M_{P_{d}(\widetilde{\nu}^c_{i})^{I}}^{2}=a_{\lambda_{i}}v_{u}-2\lambda_{k}\kappa_{ijk}v_{u}\nu^c_{j}-Y_{\nu_{ji}}\lambda_{k} \nu^c_{k}\nu_{j}+Y_{\nu_{jk}}\lambda_{i}\nu^c_{k}\nu_{j}\ ,
\end{align}
\begin{align}
M_{P_{d}(\widetilde{\nu}_{i})^{I}}^{2}=-Y_{\nu_{ij}}\lambda_{j}v_{u}^{2}-Y_{\nu_{ij}}\lambda_{k}\nu^c_{k}\nu^c_{j}\ ,
\end{align}
\begin{align}
M_{P_{u}(\widetilde{\nu}^c_{i})^{I}}^{2}=&a_{\lambda_{i}}v_{d}-a_{\nu_{ji}} \nu_{j}-2\lambda_{k}\kappa_{ilk}v_{d}\nu^c_{l}+2Y_{\nu_{jk}}\kappa_{ilk} \nu_{j}\nu^c_{l}\ ,
\end{align}
\begin{align}
M_{P_{u}(\widetilde{\nu}_{i})^{I}}^{2}=-a_{\nu_{ij}}\nu^c_{j}-Y_{ik}\kappa_{ljk}\nu^c_{l}\nu^c_{j}\ ,
\end{align}
% The independent $m_{\widetilde{\nu}^c\widetilde{\nu}}^{2}$terms :
\begin{align}
M_{(\widetilde{\nu}_{i})^{I}(\widetilde{\nu}_{j})^{I}}^{2}=m_{\widetilde{L}_{ij}}^{2}+\frac{1}{4}G^2(\nu_{k}\nu_{k}+v_{d}^{2}-v_{u}^{2})\delta_{ij}+Y_{\nu_{ik}}Y_{\nu_{jk}}v^2_{u}+Y_{\nu_{ik}}Y_{\nu_{jl}}\nu^c_{k}\nu^c_{l}\ ,
\end{align}
\begin{align}
M_{(\widetilde{\nu}_{i})^{I}(\widetilde{\nu}^c_{j})^{I}}^{2}=&-a_{\nu_{ij}}v_{u}-Y_{\nu_{ik}}\lambda_{j}v_{d}\nu^c_{k}-Y_{\nu_{ij}}Y_{\nu_{lk}}\nu_{l}\nu^c_{k}+Y_{\nu_{ik}}Y_{\nu_{lj}}\nu_{l}\nu^c_{k}+Y_{\nu_{ij}}\lambda_{k}v_{d}\nu^c_{k}+2Y_{\nu_{il}}\kappa_{jlk}v_{u}\nu^c_{k}\ ,
\end{align}
\begin{align}
&&M_{(\widetilde{\nu}^c_{i})^{I}(\widetilde{\nu}^c_{j})^{I}}^{2}=m_{\widetilde{\nu}^c_{ij}}^{2}-2a_{\kappa_{ijk}}\nu^c_{k}+2\lambda_{k}\kappa_{ijk}v_{d}v_{u}
-2\kappa_{ijk}\kappa_{lmk}\nu^c_{l}\nu^c_{m}
+4\kappa_{imk}\kappa_{ljk}\nu^c_{l}\nu^c_{m}\nonumber\\ &+&\lambda_{i}\lambda_{j}(v_{d}^{2}+v_{u}^{2})-(Y_{\nu_{ki}}\lambda_{j}+Y_{\nu_{kj}}\lambda_{i})v_{d}\nu_{k}-2Y_{\nu_{lk}}\kappa_{ijk}v_{u}\nu_{l}+Y_{\nu_{ki}}Y_{\nu_{kj}}v_{u}^{2}	
+Y_{\nu_{li}}Y_{\nu_{kj}}\nu_{k}\nu_{l}\ .
\label{oddrr}
\end{align}

Then the mass eingenvectors are 
\bea
\mathbf{P}_\alpha=R^P_{\alpha \beta} \mathbf{P'}_\beta\ ,
\eea
with the diagonal mass matrix 
\bea
(M^{\text{diag}}_{P_{\alpha \beta}})^2=R^P_{\alpha \gamma} M^2_{P_{\gamma \delta} }R^P_{\beta \delta}\ .
\eea

\subsubsection{Charged scalars}
\label{A.1.3}

We give here the mass matrix coefficients for the charged scalars which follows from the quadratic term in the potential
\begin{align}
V_{\text{quadratic}}=\mathbf{S'}^-_{\alpha } M^2_{s^\pm_{\alpha \beta}} \mathbf{S'_{\beta}}^+\ .
\label{matrixchargedscalars}
\end{align}
The unrotated charged scalars are $\mathbf{S'}^+_{\alpha}=(H^+_d,H^+_u,\tilde{e}^+_L,\widetilde{\mu}^+_L,\widetilde{\tau}^+_L,\widetilde{e}^+_R,\mu^+_R,\tau^+_R) $,
and
\begin{align}
M_{H_{d}H_{d}}^{2}=m_{H_{d}}^{2}+\frac{1}{2}g_2^{2}({v_{u}}^2-\nu_{i}\nu_{i})+\frac{G^{2}}{4}(\nu_{i}\nu_{i}+v_{d}^{2}-v_{u}^{2})+\lambda_{i}\lambda_{j}\nu^c_{i}\nu^c_{j}+Y_{e_{ik}}Y_{e_{jk}}\nu_{i}\nu_{j}\end{align}
\begin{align}
M_{H_{u}H_{u}}^{2}=m_{H_{u}}^{2}+\frac{1}{2}g_2^{2}(v_{d}^{2}+\nu_{i}\nu_{i})-\frac{G^2}{4}(v_{i}v_{i}+v_{d}^{2}-v_{u}^{2})+\lambda_{i}\lambda_{j}\nu^c_{i}\nu^c_{j}+Y_{\nu_{ij}}Y_{\nu_{ik}}\nu^c_{j}\nu^c_{k}\end{align}
\begin{align}
M_{H_{d}H_{u}}^{2}=a_{\lambda_i}\nu^c_{i}+\frac{1}{2}g_2^{2}v_{d}v_{u}-\lambda_i \lambda_i  v_d v_u+ \lambda_{k} \kappa_{ijk}\nu^c_i \nu^c_j+Y_{\nu_{ij}}\lambda_jv_u \nu_i  \end{align}
%+Y_\nu^{ij}v_u \nu_i \lambda^*_j+\kappa_{ijk}\lambda_{k}^* \nu^c_i \nu^c_j
% \begin{align}
% M_{\widetilde{e} \widetilde{e}}^{2}=\left(\begin{array}{cc}
% M_{\widetilde{e}_L\widetilde{e}_L}^{2} & M_{\widetilde{e}_L \widetilde{e}_R}^{2}\\
% M_{\widetilde{e}_R \widetilde{e}_L}^{2} & M_{\widetilde{e}_R \widetilde{e}_R}^{2}\end{array}\right)\end{align}
\begin{align}
M_{\widetilde{e}_{L_i} \widetilde{e}_{L_j}}^{2}=&m_{\widetilde{L}_{ji}}^{2}+\frac{g_2^{2}}{2}(- \nu_{k} \nu_k-v_{d}^{2}+v_{u}^{2})\delta_{ij}+\frac{1}{2}g_2^{2}\nu_{i}\nu_{j} +\frac{1}{4}G^2(\nu_{k}\nu_k+v_{d}^{2}-v_{u}^{2})\delta_{ij}\nonumber\\ &+Y_{\nu_{il}}Y_{\nu_{jk}}\nu^c_{l}\nu^c_{k}+Y_{e_{il}}Y_{e_{jl}}v_{d}^{2}\end{align}
\begin{align}
M_{\widetilde{e}_{L_i} \widetilde{e}_{R_j}}^{2}=a_{e_{ij}}v_{d}-Y_{e_{ij}}\lambda_{k}v_{u}\nu^c_{k}\end{align}
\begin{align}
M_{\widetilde{e}_{R_j} \widetilde{e}_{L_i}}^{2}={M_{\widetilde{e}_{L_i} \widetilde{e}_{R_j}}^2}\end{align}
\begin{align}
M_{\widetilde{e}_{R_i}\widetilde{e}_{R_j}}^2=m_{\widetilde{e}^c_{ij}}^{2}+\frac{g_1^{2}}{2}(-\nu_{k} \nu_k-v_{d}^{2}+v_{u}^{2})\delta_{ij}+Y_{e_{ki}}Y_{e_{kj}}v_{d}^{2}+Y_{e_{li}}Y_{e_{kj}}\nu_{k}\nu_{l}
\end{align}
\begin{align}
M_{\widetilde{e}_{L_i} H_d}^2=\frac{g_2^{2}}{2}v_{d}\nu_{i} -Y_{\nu_{ij}}\lambda_{k}\nu^c_{k}\nu^c_{j}-Y_{e_{ij}}Y_{e_{kj}}v_{d}\nu_{k}
\end{align}
\begin{align}
M_{\widetilde{e}_{L_i} H_u}^2=-a_{\nu_{ij}}\nu^c_{j}+\frac{g_2^{2}}{2}v_{u}\nu_{i}-Y_{\nu_{ij}} \kappa_{ljk} \nu^c_{l} \nu^c_{k}+Y_{\nu_{ij}} \lambda_j v_d v_u-Y_{\nu_{ik}} Y_{\nu_{kj}}v_u \nu_j
\end{align}
\begin{align}
M_{\widetilde{e}_{R_i} H_d}^2=- a_{e_{ji}}\nu_{j}-Y_{e_{ki}}Y_{\nu_{kj}}v_{u}\nu^c_{j}
\end{align}
\begin{align}
M_{\widetilde{e}_{R_i} H_u}^2=-Y_{e_{ki}}(\lambda_{j}\nu_{k}\nu^c_{j}+Y_{\nu_{kj}}v_{d}\nu^c_{j})\ ,
\end{align}
where  $a_{e_{ij}}\equiv (A_eY_e)_{ij}$. Then the mass eigenvectors are
\bea
\mathbf{S}^\pm_{\alpha}=R^{s^\pm}_{\alpha \beta} \mathbf{S'}^{\pm}_{\beta}\ ,
\eea
with the diagonal mass matrix 
\bea
(M^{\text{diag}}_{s^\pm})^2_{\alpha \beta}=R^{s^\pm}_{\alpha \gamma} M^2_{s^{\pm}_{\gamma \delta}} R^{s^{\pm}}_{\beta \delta}\ .
\eea

It is worth noticing here that if we allow the presence of the lepton number violating terms in the superpotential, 
$\lambda_{ijk}\hat{L}_{i} \hat{L}_{j}\hat{e}_{k}^{c}$,
discussed in the Introduction,
they would contribute to the above charged scalar masses.

%%%%%%%%%%%%%%%%%%%%%% Squarks %%%%%%%%%%%%%%%%%%%%

\subsubsection{Squarks}
\label{squarkss}

% ===============

% CHEQUEAR ESTO:
% On the other hand,
% when compared to the MSSM case, the structure of squark mass terms
% is essentially unaffected, provided that one uses $\mu = \lambda\nu^c
% $,
% and neglects the contribution of the fourth term in (\ref{superpotential}).

% =========

In the unrotated basis,
$\widetilde{u'}_i=(\widetilde{u}_{L_i},\widetilde{u}^*_{R_i})$ and 
$\widetilde{d'}_i=(\widetilde{d}_{L_i},\widetilde{d}^*_{R_i} )$, we get
\begin{align}
V_\text{quadratic}=
     \frac{1}{2}\widetilde{u'}^\dag M_{\widetilde{u}}^2\ \widetilde{u'}
     +\frac{1}{2}\widetilde{d'}^{\dag} M_{\widetilde{d}}^2\ \widetilde{d'}\ ,
     \label{matrixsquarks}
\end{align}
where
\begin{align}
M_{\widetilde{q}_{ij}}^2= 
  \left( \begin{array}{cc} M^2_{\widetilde{q}_{L_iL_j}}&M^2_{\widetilde{q}_{L_iR_j}}\\
         M^2_{\widetilde{q}_{R_iL_j}} & M^2_{\widetilde{q}_{R_iR_j}}
         \end{array} \right)\ ,  
\end{align}
with $\widetilde{q}=(\widetilde{u'},\widetilde{d'})$. 
The blocks are different for up and down quarks, and we have
\begin{eqnarray}
M^2_{\widetilde{u}_{L_iL_j}}&=&
    m^2_{\widetilde{Q}_{ij}} + \frac{1}{6}(\frac{3g^2_2}{2} 
    - \frac{g_1^2}{2})(v_d^2-v_u^2 + \nu_k\nu_k) + 
    Y_{u_{ik}} Y_{u_{jk}} v_u^2\ ,  \nonumber\\ 
M^2_{\widetilde{u}_{R_iR_j}}&=&
    m^2_{\widetilde{u}_{ij}}+ \frac{g^2_1}{3}(v_d^2-v_u^2+\nu_k\nu_k)
    + Y_{u_{ki}} Y_{u_{kj}}v_u^2\ ,  \nonumber\\
M^2_{\widetilde{u}_{L_iR_j}}&=& 
    a_{u_{ij}}v_u -Y_{u_{ij}}\lambda_k v_d\nu_k^c  
    + Y_{\nu_{lk}}Y_{u_{ij}}\nu_l\nu^c_k\ ,\nonumber\\
M^2_{\widetilde{u}_{L_iR_j}}&=&
    m^2_{\widetilde{u}_{R_jL_i}}\ ,
\end{eqnarray} 
and
\begin{eqnarray}
M^2_{\widetilde{d}_{L_iL_j}}&=&
    m^2_{\widetilde{Q}_{ij}}-\frac{1}{6} (\frac{3g^2_2}{2}
    + \frac{g_1^2}{2})(v_d^2-v_u^2+\nu_k\nu_k) 
    + Y_{d_{ik}} Y_{d_{jk}}  v_d^2 \nonumber\\ 
M^2_{\widetilde{d}_{R_iR_j}}&=& 
    m^2_{\widetilde{d}_{ij}}- \frac{g^2_1}{6}(v_d^2-v_u^2+\nu_k\nu_k)
    + Y_{d_{ik}} Y_{d_{jk}}v_d^2  \nonumber\\
M^2_{\widetilde{d}_{L_iR_j}}&=&
    a_{d_{ij}} v_d -Y_{d_{ij}}\lambda_k v_u\nu^c_k  \nonumber\\
M^2_{\widetilde{d}_{L_iR_j}}&=&
    m^2_{\widetilde{d}_{R_jL_i}}\ ,
\end{eqnarray} 
where $a_{u_{ij}}\equiv (A_uY_u)_{ij}$ and $a_{d_{ij}}\equiv (A_dY_d)_{ij}$. For the mass state $\widetilde{\mathbf{q}}_i$ we have
\begin{align}
\widetilde{\mathbf{q}}_i = R^{\widetilde{q}}_{ij} \widetilde{q}_j\ , \end{align}
% The rotation matrix are obtained from
% \begin{align}
% R^{\widetilde{q}}_{ki} (M^{\text{diag}}_{\widetilde{q}})^2_{kl} R^{\widetilde{q}}_{lj} 
%    = M^2_{\widetilde{q}_{ij}} \end{align}
with the diagonal mass matrix
\begin{align}
 (M^{\text{diag}}_{\widetilde{q}})^2_{ij} 
   = R^{\widetilde{q}}_{il}  M^2_{\widetilde{q}_{lk}} R^{\widetilde{q}}_{jk}\ .\end{align}

It is worth noticing here that if we allow
the presence of the baryon number violating terms in the superpotential
discussed in the Introduction, 
$\lambda'_{ijk}\hat{L}_{i}\hat{Q}_{j}\hat{d}^c_{k}$,
they would contribute to the above squark masses.
Actually, even if they are set to zero,
one-loop corrections will generate them, as discussed in Appendix
\ref{appx:rges}. However, these contributions are negligible.

% We must comment that a term 
%$\frac{\lambda_{ijk}}{2}\hat{L}_{i} \hat{L}_{j} \hat{e}_{k}^{c} %+\frac{\lambda'_{ijk}}{2}\hat{L}_{i}\hat{Q}_{j}\hat{d}^c_{k}$ in 
%the superpotential will appear at the loop level and by the RGE affected 
%the charged masses, but we neglected this contributions since are 
%really very small.]

%%%%%%%%%%%%%% Charginos %%%%%%%%%%%%%%%%%%%%%%%%%%%%%%%%
 
\subsection{
Charged fermion mass matrix
%Chargino mass matrix
}
\label{cfms}

Charginos mix with the charged leptons and therefore in a basis where ${\Psi^+}^T =(-i \tilde \lambda^{+}, \tilde H_u^+ ,  e_R^{+} , \mu_R^{+}, \tau_R^{+})$
and 
${\Psi^-}^T = (-i \tilde \lambda^-, \tilde H_d^- ,  e_L^-,  \mu_L^-,  \tau_L^-)$,
one obtains the matrix 
\begin{align}
-\frac{1}{2} ({\psi^+}^T,{\psi^-}^T) \left( \begin{array}{cc} 0&M^T_C\\M_C & 0  
\end{array} \right) \left( \begin{array}{cc} {\psi^+}^T\\{\psi^-}^T\\  
\end{array} \right)\ ,  
\label{matrixcharginos}
\end{align}
where
\begin{align}
M_C=
\left(\begin{array}{ccccc}
M_{2} & g_{2}v_{u} & 0 & 0 & 0\\
g_{2}v_{d} & \lambda_{i}\nu^c_{i} & -Y_{e_{i1}}\nu_{i} & -Y_{e_{i2}}\nu_i & -Y_{e_{i3}}\nu_{i}\\
g_{2}\nu_{1} & -Y_{\nu_{1i}}\nu^c_{i} & Y_{e_{11}}v_{d} &  Y_{e_{12}}v_{d} &  Y_{e_{13}}v_{d}\\
g_{2}\nu_{2} & -Y_{\nu_{2i}}\nu^c_{i} &  Y_{e_{21}}v_{d}& Y_{e_{22}}v_{d} &  Y_{e_{23}}v_{d}\\
g_{2}\nu_{3} & -Y_{\nu_{3i}}\nu^c_{i} &  Y_{e_{31}}v_{d} &  Y_{e_{32}}v_{d} & Y_{e_{33}}v_{d}
\end{array}\right)\ .
\label{submatrix}
\end{align}

\subsection{
Neutral fermion mass matrix
%Neutralino mass matrix
}
\label{nfms}

% $\chi^{0}=(\tilde{B^{0}},\tilde{W^{0}},\tilde{H_{d}},
%\tilde{H_{u}},\tilde{\nu_{R_{1}}},
%\tilde{\nu_{R_{2}}},\tilde{\nu_{R_{3}}},
%\tilde{\nu_{1}},\tilde{\nu_{2}},\tilde{\nu_{3}}$)

Neutralinos mix with the neutrinos and therefore in a basis where
${\chi^{0}}^T=(\tilde{B^{0}},\tilde{W^{0}},\tilde{H_{d}},\tilde{H_{u}},\nu_{R_i},\nu_{L_i})$,
one obtains the following neutral fermion mass terms in the Lagrangian
%$\mathcal{L}_{\mathrm{neutral}}^{\mathrm{mass}}
%^{\tilde \chi^0} 
%=
\begin{equation}
-\frac{1}{2} (\chi^0)^T \mathcal{M}_{\mathrm{n}}
%_{\tilde \chi^0} 
\chi^0 + \mathrm{c.c.}\ ,
\label{matrixneutralinos}
\end{equation}
where
\begin{align}
%\textrm{\cal{M}}_n
{\cal M}_n=\left(\begin{array}{cc}
M & m\\
m^{T} & 0_{3\times3}\end{array}\right),\end{align}
with
{\small  \begin{align}
\hspace*{-2.5cm} \hspace{.2mm} M=\hspace{-.2mm}
\left(
\begin{array}{ccccccc}
M_{1} & 0 & -A v_{d} & A v_{u} & 0 & 0 & 0\\
0 & M_{2} &B v_{d} & -B v_{u} & 0 & 0 & 0\\
-A v_{d} & B v_{d} & 0 & -\lambda_{i}\nu^c_{i} & -\lambda_{1}v_{u} & -\lambda_{2}v_{u} & -\lambda_{3}v_{u}\\
A v_{u} & -B v_{u} & \: \: -\lambda_{i}\nu^c_{i} & 0 & -\lambda_{1}v_{d}+Y_{\nu_{i1}}\nu_{i} & -\lambda_{2}v_{d}+Y_{\nu_{i2}}\nu_{i} & -\lambda_{3}v_{d}+Y_{\nu_{i3}}\nu_{i}\\
0 & 0 &  -\lambda_{1}v_{u} & \: \:-\lambda_{1}v_{d}+Y_{\nu_{i1}}\nu_{i} & 2\kappa_{11j}\nu^c_{j} & 2\kappa_{12j}\nu^c_{j} & 2\kappa_{13j}\nu^c_{j}\\
0 & 0 & -\lambda_{2}v_{u} &  \: \: -\lambda_{2}v_{d}+Y_{\nu_{i2}}\nu_{i} & 2\kappa_{21j}\nu^c_{j} & 2\kappa_{22j}\nu^c_{j} & 2\kappa_{23j}\nu^c_{j}\\
0 & 0 & -\lambda_{3}v_{u} & \: \:-\lambda_{3}v_{d}+Y_{\nu_{i3}}\nu_{i} & 2\kappa_{31j}\nu^c_{j} & 2\kappa_{32j}\nu^c_{j} & 2\kappa_{33j}\nu^c_{j}\end{array}
\right)\ ,
\label{neumatrix}
\end{align}
}
where $A=\frac{G}{\sqrt{2}} \sin\theta_W$, $B=\frac{G}{\sqrt{2}} \cos\theta_W$,
and
\begin{align}
m^{T}=\left(\begin{array}{ccccccc}
-\frac{g_{1}}{\sqrt{2}}\nu_{1} \: & \: \frac{g_{2}}{\sqrt{2}}\nu_{1} & \: 0 & \: Y_{\nu_{1i}}\nu^c_{i} & \: Y_{\nu_{11}}v_{u} & \: Y_{\nu_{12}}v_{u} & \: Y_{\nu_{13}}v_{u}\\
\: -\frac{g_{1}}{\sqrt{2}}\nu_{2} & \: \frac{g_{2}}{\sqrt{2}}\nu_{2} & \: 0 & \: Y_{\nu_{2i}}\nu^c_{i} & \: Y_{\nu_{21}}v_{u} & \: Y_{\nu_{22}}v_{u} & \: Y_{\nu_{23}}v_{u}\\
\: -\frac{g_{1}}{\sqrt{2}}\nu_{3}\: & \: \frac{g_{2}}{\sqrt{2}}\nu_{3} & \: 0 & \: Y_{\nu_{3i}}\nu^c_{i} & \: Y_{\nu_{31}}v_{u} & \: Y_{\nu_{32}}v_{u} & \: Y_{\nu_{33}}v_{u}\end{array}\right)\ .\end{align}

%%%%%%%%%%%%%%%%%%%%%%%%%%%%%%%%%%%%%%%%%%%%%%%%%%%%%%%%%%%%%%%%
\section{Couplings \label{appx:couplings}}
%%%%%%%%%%%%%%%%%%%%%%%%%%%%%%%%%%%%%%%%%%%%%%%%%%%%%%%%%%%%%%%%

In this Appendix we show the relevant couplings involved in the computation of the one-loop radiative corrections to the scalar potential tadpoles and the CP-even scalars masses.

\subsection{Scalar--up squarks--up squarks}

With the definition
\beqa
{\cal L}= g_{\alpha ij}^{S'^0\widetilde{u}'\widetilde{u}'^*}\,
S'^0_\alpha\, \widetilde{u}'_i\, \widetilde{u}'^*_j + \cdots,
\eeqa
%
%where $S'^0=(H_d^0, H_u^0, \tilde{\nu}^L_i, \tilde{\nu}^R_i)$ and $i=1..3$
%
we get
\beqa
g_{\alpha ij}^{S'^0\widetilde{u}'\widetilde{u}'^*}=\left(
\begin{matrix}
g_{\alpha L_i L_j}^{S'^0\widetilde{u}'\widetilde{u}'^*}&
g_{\alpha L_i R_j}^{S'^0\widetilde{u}'\widetilde{u}'^*}\cr
\vb{18}
g_{\alpha R_i L_j}^{S'^0\widetilde{u}'\widetilde{u}'^*}&
g_{\alpha R_i R_j}^{S'^0\widetilde{u}'\widetilde{u}'^*}\cr
\end{matrix}
\right)\ ,
\eeqa
where
\beqa
g_{\alpha L_i L_j}^{S'^0\widetilde{u}'\widetilde{u}'^*}
&=& u_\beta\, \hat \delta_{\alpha\beta} 
\left(-\smallfrac{1}{2}\, g^2 +\smallfrac{1}{6}\, g'^2
\right)
-2 \, \delta_{i2} \, v_u \, Y_{u_{jl}} Y_{u_{kl}}\ , \cr
\vb{18}
g_{\alpha L_i R_j}^{S'^0\widetilde{u}'\widetilde{u}'^*}
&=& - \,\delta_{\alpha 2}\, (A_uY_u)_{ij} + \delta_{\alpha 1} \, \nu^c_l \lambda_l \, Y_{u_{ij}}
- \delta_{\alpha-2,l}  \,  Y_{\nu_{lm}} \nu^c_m \, Y_{u_{ij}}  
+ \delta_{\alpha-5,l} \, (v_d \lambda_l -  Y_{\nu_{ml}} \nu_m) \, Y_{u_{ij}}\ , \cr
\vb{18}
g_{\alpha R_i L_j}^{S'^0\widetilde{u}'\widetilde{u}'^*}
&=& 
g_{\alpha L_jR_i}^{S'^0\widetilde{u}'\widetilde{u}'^*}\ ,\cr
\vb{18}
g_{\alpha R_i R_j}^{S'^0\widetilde{u}'\widetilde{u}'^*}
&=& -\smallfrac{2}{3} u_\beta\, \hat \delta_{\alpha\beta} \, g'^2  
-2 \, \delta_{\alpha 2} \, v_u Y_{u_{li}} Y_{u_{lj}}\ , 
\eeqa
and we have defined 
\beq
u_\beta \equiv (v_d,v_u,\nu_1,\nu_2,\nu_3,\nu^c_1,\nu^c_2,\nu^c_3)\ ; \hskip 10mm
\hat \delta_{ij}\equiv \hbox{diag}(+,-,+,+,+,0,0,0)
\eeq
while $\delta_{ij}$ is equal to one for $i=j$, and zero for $i\neq j$.

%and we have defined 
\noindent
%
%\beq
%u_m \equiv (v_d,v_u,v_1,v_2,v_3,v^c_1,v^c_2,v^c_3) \hskip 10mm ; \hskip 10mm
%\hat \delta_{ij}\equiv \hbox{diag}(+,-,+,+,+,0,0,0),
%\eeq
%
%$\mu_{eff} \equiv \lambda_i v^c_i$, $\epsilon_i \equiv Y_{\nu_{ij}} v^c_j$ 
%and  $\eta_i \equiv Y_{\nu_{ji}} v_j$. 
%
%While $\delta_{ij}$ without the {\it hat} is the usual Kronecker delta.

%
%%%%%%%%%%%%%% Scalar-Squark-Squark %%%%%%%%%%%%%%%%

\subsection{Scalar--down squarks--down squarks}

With the definition
\beqa
{\cal L}= g_{\alpha ij}^{S'^0\widetilde{d}'\widetilde{d}'^*}\,
S'^0_\alpha\, \widetilde{d}'_i\, \widetilde{d}'^*_j + \cdots \ ,
\eeqa
we get
\beqa
g_{\alpha ij}^{S'^0\widetilde{d}'\widetilde{d}'^*}=\left(
\begin{matrix}
g_{\alpha L_i L_j}^{S'^0\widetilde{d}'\widetilde{d}'^*}&
g_{\alpha L_i R_j}^{S'^0\widetilde{d}'\widetilde{d}'^*}\cr
\vb{18}
g_{\alpha R_i L_j}^{S'^0\widetilde{d}'\widetilde{d}'^*}&
g_{\alpha R_i R_j}^{S'^0\widetilde{d}'\widetilde{d}'^*}\cr
\end{matrix}
\right)\ ,
\eeqa
where
\beqa
g_{\alpha L_i L_j}^{S'^0\widetilde{d}'\widetilde{d}'^*}
&=& u_\beta \, \hat \delta_{\alpha\beta}\,
\left( \smallfrac{1}{2}\, g^2 +\smallfrac{1}{6}\, g'^2
\right)
-2 \, \delta_{\alpha 1} \, v_d  \, Y_{d_{il}} Y_{d_{jl}}\ , \cr
\vb{18}
g_{\alpha L_i R_j}^{S'^0\widetilde{d}'\widetilde{d}'^*}
&=& - \delta_{\alpha 1}\, (A_{d}Y_{d})_{ij} +  \delta_{\alpha 2} \, \nu^c_l \lambda_l \, Y_{d_{ij}} 
+ \delta_{\alpha -5,l} \, \lambda_l \, v_u \, Y_{d_{ij}}\ ,  \cr
\vb{18}
g_{\alpha R_i L_j}^{S'^0\widetilde{d}'\widetilde{d}'^*}
&=& 
g_{\alpha L_jR_i}^{S'^0\widetilde{d}'\widetilde{d}'^*}\ ,\cr
\vb{18}
g_{\alpha R_i R_j}^{S'^0\widetilde{d}'\widetilde{d}'^*}
&=& \smallfrac{1}{3}\, u_\beta \,\hat \delta_{\alpha\beta} \, g'^2 
-2 \, \delta_{\alpha 1} \, v_d Y_{d_{li}} Y_{d_{lj}}\ .
\eeqa
We find the couplings in the squark $\widetilde{q}_{1,2}$ basis via
$g_{\alpha ij}^{S'^0\widetilde{q}\widetilde{q}^*} = 
 R^{\widetilde{q}}_{il} (g_{\alpha lm}^{S'^0\widetilde{q}'\widetilde{q}'^*}) 
R^{\widetilde{q}}_{jm}$.
%
%%%%%%%%%%%%%% Scalar-Quark-Quark %%%%%%%%%%%%%%%%

\subsection{Scalar--quark--quark}

With the definition
\beqa
{\cal L}=  g_{\alpha ij}^{S'^0\overline{u}u}\,
S'^0_\alpha\, \overline{u}_i\, u_j +
g_{\alpha ij}^{S'^0\overline{d}d}\,
S'^0_\alpha\, \overline{d}_i\, d_j + \cdots \ ,
\eeqa
we get
\beqa
g_{\alpha ij}^{S'^0\overline{u}u}= - \delta_{\alpha 2} \, Y_{u_{ij}}\ , 
\eeqa
and
\beqa
g_{\alpha ij}^{S'^0\overline{d}d}= - \delta_{\alpha 1} \, Y_{d_{ij}}\ .
\eeqa

%
%%%%%%%%%%%%%% Scalar-Squark-Squark-Scalar %%%%%%%%%%%%%%%%

\subsection{Scalar--scalar--up scalars--up scalars}

With the definition
\beqa
{\cal L}=  g_{\alpha\beta ij}^{S'^0S'^0\widetilde{u}'\widetilde{u}'^*}\,
S'^0_\alpha\, S'^0_\beta \, \widetilde{u}'_i\, \widetilde{u}'^*_j + \cdots \ ,
\eeqa
we get
\beqa
g_{\alpha\beta ij}^{S'^0S'^0\widetilde{u}'\widetilde{u}'^*}=\left(
\begin{matrix}
g_{\alpha\beta L_iL_j}^{S'^0S'^0\widetilde{u}'\widetilde{u}'^*}&
g_{\alpha\beta L_iR_j}^{S'^0S'^0\widetilde{u}'\widetilde{u}'^*}\cr
\vb{18}
g_{\alpha\beta R_iL_j}^{S'^0S'^0\widetilde{u}'\widetilde{u}'^*}&
g_{\alpha\beta R_iR_j}^{S'^0S'^0\widetilde{u}'\widetilde{u}'^*}\cr
\end{matrix}
\right)\ ,
\eeqa
where
\beqa
g_{\alpha\beta L_iL_j}^{S'^0S'^0\widetilde{u}'\widetilde{u}'^*}
&=& \hat \delta_{\alpha\beta} 
\left(-\smallfrac{1}{4}\, g^2 +\smallfrac{1}{12}\, g'^2
\right)
- \delta_{\alpha 2} \, \delta_{\beta 2} \, Y_{u_{il}} Y_{u_{jl}}\ , \cr
\vb{18}
g_{\alpha\beta L_iR_j}^{S'^0S'^0\widetilde{u}'\widetilde{u}'^*}
&=& \frac{1}{2} \left ( \delta_{\alpha 1} \, \delta_{\beta-5,l} \lambda_l \, Y_{u_{ij}}
- \delta_{\alpha-2,l}  \, \delta_{\beta-5,m} \,  Y_{\nu_{lm}} \, Y_{u_{ij}} 
\right )\ , \cr
\vb{18}
g_{\alpha\beta R_iL_j}^{S'^0S'^0\widetilde{u}'\widetilde{u}'^*}
&=& 
g_{\alpha\beta L_jR_i}^{S'^0\widetilde{u}'\widetilde{u}'^*}\ , \cr
\vb{18}
g_{\alpha\beta R_iR_j}^{S'^0S'^0\widetilde{u}'\widetilde{u}'^*}
&=& -\smallfrac{1}{3} \hat \delta_{\alpha\beta} \, g'^2  
- \delta_{\alpha 2} \, \delta_{\beta 2} \, Y_{u_{il}} Y_{u_{jl}}\ . 
\eeqa
%

%%%%%%%%%%%%%% Scalar-Squark-Squark-Scalar %%%%%%%%%%%%%%%%

\subsection{Scalar--scalar--down scalars--down scalars}

With the definition
\beqa
{\cal L}=  g_{\alpha\beta ij}^{S'^0S'^0\widetilde{d}'\widetilde{d}'^*}\,
S'^0_\alpha \, S'^0_\beta \, \widetilde{d}'_i\, \widetilde{d}'^*_j + \cdots\ ,
\eeqa
we get
\beqa
g_{\alpha\beta ij}^{S'^0S'^0\widetilde{d}'\widetilde{d}'^*}=\left(
\begin{matrix}
g_{\alpha\beta L_iL_j}^{S'^0S'^0\widetilde{d}'\widetilde{d}'^*}&
g_{\alpha\beta L_iR_j}^{S'^0S'^0\widetilde{d}'\widetilde{d}'^*}\cr
\vb{18}
g_{\alpha\beta R_iL_j}^{S'^0S'^0\widetilde{d}'\widetilde{d}'^*}&
g_{\alpha\beta R_iR_j}^{S'^0S'^0\widetilde{d}'\widetilde{d}'^*}\cr
\end{matrix}
\right)\ ,
\eeqa
where
\beqa
g_{\alpha\beta L_iL_j}^{S'^0S'^0\widetilde{d}'\widetilde{d}'^*}
&=& \hat \delta_{\alpha\beta} 
\left(\smallfrac{1}{4}\, g^2 +\smallfrac{1}{12}\, g'^2
\right)
- \delta_{\alpha 1} \, \delta_{\beta 1} \, Y_{d_{il}} Y_{d_{jl}}\ , \cr
\vb{18}
g_{\alpha\beta L_iR_j}^{S'^0S'^0\widetilde{d}'\widetilde{d}'^*}
&=& \frac{1}{2} \delta_{\alpha 2} \, \delta_{\beta-5,l} \lambda_l \, Y_{d_{ij}}\ , \cr
\vb{18}
g_{\alpha\beta R_iL_j}^{S'^0S'^0\widetilde{d}'\widetilde{d}'^*}
&=& 
g_{\alpha\beta L_jR_i}^{S'^0\widetilde{d}'\widetilde{d}'^*}\ , \cr
\vb{18}
g_{\alpha\beta R_iR_j}^{S'^0S'^0\widetilde{d}'\widetilde{d}'^*}
&=& \smallfrac{1}{6} \hat \delta_{\alpha\beta} \, g'^2  
- \delta_{\alpha 1} \, \delta_{\beta 2} \, Y_{d_{il}} Y_{d_{jl}}\ .
\eeqa
%

%%%%%%%%%%%%%%%%%%%%%%%%%%%%%%%%%%%%%%%%%%%%%%%%%%%%%%%%%%%%%%%%
%\section{Potential \label{appx:potential}}
%%%%%%%%%%%%%%%%%%%%%%%%%%%%%%%%%%%%%%%%%%%%%%%%%%%%%%%%%%%%%%%%

%$V^1 =  \sum_{f} \sum^2_{i=1} N^f_c G_0(m^2_{\widetilde{f}_i}) 
%   - 2 \, \sum_{f} N^f_c G_0(m^2_f)$

%where $G_0$ is the 0-point Passarino-Veltman function~\cite{Passarino:1978jh}.

%%%%%%%%%%%%%%%%%%%%%%%%%%%%%%%%%%%%%%%%%%%%%%%%%%%%%%%%%%%%%%%%
\section{Tadpoles \label{appx:tadpoles}}
%%%%%%%%%%%%%%%%%%%%%%%%%%%%%%%%%%%%%%%%%%%%%%%%%%%%%%%%%%%%%%%%

In this Appendix we present the leading one-loop $\drbar$ tadpoles
(i.e. the ones involving s(quarks) in the loop) which enter into the 
minimization of the neutral scalar potential (see Fig \ref{fig:Tadpoles}),

%*** Justify why squarks + quarks: may be from old papers by Ellis ***
%*** a picture of the Feynman diagrams ???? ****

\begin{figure}[t]
  \begin{center} %\hspace*{-10mm}
    \begin{tabular}{cc}
	\epsfig{file=finalfigures/Tadpole1.epsi,width=20mm,angle=0,clip=} &
	\epsfig{file=finalfigures/Tadpole2.epsi,width=20mm,angle=0,clip=}
    \end{tabular}
    \captions{Tadpole Feynman diagrams} 
    \label{fig:Tadpoles}
  \end{center}
\end{figure}

\begin{equation}
t^1_{S_\alpha} = \frac{1}{16 \pi^2} \sum_i T_{S_\alpha}^{X_i}\ ,
\end{equation}
where $X = (u, d, \widetilde{u}, \widetilde{d})$, and 
\begin{align}
 T^f_{S'^0_\alpha} & = 
          \sum_{k = 1}^3 3\,g^{S'^0 \bar{f} f}_{\alpha k k} 4\,m_{f_k} A_0(m^2_{f_k})\ , \\
%  T^d_{S'^0_\alpha} & = 
%            \sum_{k = 1}^3 3 g^{S'^0 \bar{d} d}_{\alpha k k} 4 m_k A_0(m^2_k) \\
 T^{\tilde f}_{S'^0_\alpha} & = 
          - \sum_{k = 1}^6 3\,g^{S'^0\widetilde{f}\widetilde{f}^*}_{\alpha k k} 
                                                               A_0(m^2_{f_k})\,,% \\
%  T^{\tilde d}_{S'^0_\alpha} & = 
%           - \sum_{k = 1}^6 3 g^{S'^0\widetilde{d}'\widetilde{d}'^*}_{\alpha k k} 
%                                                                A_0(m^2_k), 
 \end{align}
where $f=u,d$ and $A_0$ is the 1-point Passarino-Veltman 
function \cite{Passarino:1978jh}. 

%%%%%%%%%%%%%%%%%%%%%%%%%%%%%%%%%%%%%%%%%%%%%%%%%%%%%%%%%%%%%%%%
\section{One loop self-energies \label{appx:selfenergies}}
%%%%%%%%%%%%%%%%%%%%%%%%%%%%%%%%%%%%%%%%%%%%%%%%%%%%%%%%%%%%%%%%

Here we list the leading one-loop $\drbar$ self-energies of the CP-even scalar 
mass matrix represented in Fig. (\ref{fig:self_energies}),

%*** may be a footnote ***

\beqa
16 \, \pi^2 \, \Pi_{S'^0_\alpha S'^0_\beta}(p^2) &=& 
 \sum_{f=u,d}\, \sum_{k=1}^3 N^f_c \left(g^{S'^0 \bar{f} f}_{\alpha  k k} \right)^2 \delta_{\alpha\beta} 
    \left [ (p^2 - 4 \, m_{f_k}) \, B_0(m_{f_k}, m_{f_k}) - 2 \, A_0(m_{f_k})\right ] \cr
\vb{18} 
 & + & \sum_{f=u,d}\, \sum^6_{k,l=1} N^f_c 
  \left(g^{S'^0 S'^0 \widetilde{f}\widetilde{f}^*}_{\alpha\beta k l} \right)^2 
      A_0(m_k) \cr 
\vb{18}
 & + & \sum_{f=u,d}\, \sum^6_{k,l=1} N^f_c g^{S'^0 \widetilde{f}\widetilde{f}^*}_{\alpha  k l} 
      g^{S'^0 \widetilde{f}\widetilde{f}^*}_{\beta k l}  
      B_0(m_{f_k}, m_{f_l})\ ,
\eeqa
where $N^f_c$ is the number of colours, which is 3 for a (s)quark
and $B_0$ is the 2-point Passarino-Veltman function \cite{Passarino:1978jh}.

\begin{figure}[t]
  \begin{center} %\hspace*{-10mm}
    \begin{tabular}{ccc}
	\epsfig{file=finalfigures/SE1.epsi,width=40mm,angle=0,clip=} &
	\epsfig{file=finalfigures/SE2.epsi,width=40mm,angle=0,clip=} &
	\epsfig{file=finalfigures/SE3.epsi,width=40mm,angle=0,clip=}
    \end{tabular}
    \captions{Self-energy diagrams} 
    \label{fig:self_energies}
  \end{center}
\end{figure}

%%%%%%%%%%%%%%%%%%%%%%%%%%%%%%%%%%%%%%%%%%%%%%%%%%%%%%%%%%%%%%%%
\section{Renormalisation group equations of Yukawa couplings \label{appx:rges}}
%%%%%%%%%%%%%%%%%%%%%%%%%%%%%%%%%%%%%%%%%%%%%%%%%%%%%%%%%%%%%%%%

In this Appendix we give the RGEs of Yukawa couplings including $\lambda_i$ and
$\kappa_{ijk}$.
Defining
\begin{equation}
\gamma^{\nu^c_j}_{\nu^c_i}=-2 (\kappa_{ilk} \kappa_{jlk}+  \lambda_i \lambda_j +  Y_{\nu_{ki}} Y_{\nu_{kj}}) \ ,
\end{equation}
\begin{equation}
\gamma^{H_u}_{H_u}=\frac{3}{2}g_2^2+\frac{3}{10}g_1^2-3 Y_{u_{ij}} Y_{u_{ij}}-  \lambda_i \lambda_i -  Y_{\nu_{ij}} Y_{\nu_{ij}} \ ,
\end{equation}
\begin{equation}
\gamma^{H_d}_{H_d}=\frac{3}{2}g_2^2+\frac{3}{10}g_1^2 -Y_{e_{ij}}Y_{e_{ij}}-3 Y_{d_{ij}}Y_{d_{ij}}- \lambda_i \lambda_i \ ,
\end{equation}
\begin{equation}
\gamma^{L_j}_{L_i}=\frac{3}{2}g_2^2+\frac{3}{10}g_1^2 -Y_{e_{ik}} Y_{e_{jk}} - Y_{\nu_{il}} Y_{\nu_{jl}}\ ,
\end{equation}
\begin{equation}
\gamma^{H_d}_{L_i}=\gamma^{L_i}_{H_d}=-Y_{\nu_{ij}} \lambda_j \ ,
\end{equation}
\begin{equation}
\gamma^{e^c_j}_{e^c_i}=\frac{6}{5}g_1^2-2 Y_{e_{ik}} Y_{e_{jk}}\ ,
\end{equation}
\begin{equation}
\gamma^{d^c_j}_{d^c_i}=\frac{8}{3}g_s^2+\frac{2}{15}g_1^2- 2 Y_{d_{ik}} Y_{d_{jk}}\ ,
\end{equation}
\begin{equation}
\gamma^{u^c_j}_{u^c_i}=\frac{8}{3}g_s^2+\frac{8}{15}g_1^2-2 Y_{u_{ik}} Y_{u_{jk}}\ ,
\end{equation}
\begin{equation}
\gamma^{Q_j}_{Q_i} =\frac{8}{3}g_s^2+\frac{3}{2}g_2^2+\frac{1}{30}g_1^2-Y_{u_{ik}} Y_{u_{jk}}-Y_{d_{ik}} Y_{d_{jk}}\ ,
\end{equation}
at one-loop level we have the following RGEs:
\begin{equation}
\frac{d}{dt}\kappa_{ijk}=\frac{1}{16 \pi^2}(\kappa_{ljk}\gamma^{\nu^c_l}_{\nu^c_i} +\kappa_{lik}\gamma^{\nu^c_l}_{\nu^c_j} +\kappa_{lji}\gamma^{\nu^c_l}_{\nu^c_k} )\ ,
\end{equation}
\begin{equation}
\frac{d}{dt}\lambda_i=\frac{1}{16 \pi^2}(\lambda_j\gamma^{\nu^c_j}_{\nu^c_i}+\lambda_i\gamma^{H_u}_{H_u} +\lambda_i\gamma^{H_d}_{H_d} )+\frac{1}{16 \pi^2}Y_{\nu_{ji}} \gamma^{L_j}_{H_d}  \label{lambda}\ ,
\end{equation}
\begin{equation}
\frac{d}{dt} Y_{\nu_{ij}}=\frac{1}{16 \pi^2}( Y_{\nu_{ij}} \gamma^{H_u}_{H_u} + Y_{\nu_{ik}} \gamma^{\nu^c_k}_{\nu^c_j} + Y_{\nu_{kj}} \gamma^{L_k}_{L_i})+\frac{1}{16 \pi^2} \lambda_j \gamma^{H_d}_{L_i}\ ,
\end{equation}
\begin{equation}
\frac{d}{dt} Y_{e_{ij}}=\frac{1}{16 \pi^2}( Y_{e_{ij}} \gamma^{H_d}_{H_d}  +Y_{e_{ik}} \gamma^{e^c_k}_{e^c_j}+ Y_{e_{ik}}\gamma^{L_k}_{L_j} )\ ,
\end{equation}
\begin{equation}
\frac{d}{dt} Y_{d_{ij}}=\frac{1}{16 \pi^2} (Y_{d_{ik}} \gamma^{d^c_k}_{d^c_j}+ Y_{d_{kj}} \gamma^{Q_k}_{Q_i}+ Y_{d_{ij}} \gamma^{H_d}_{H_d})\ ,
\end{equation}
\begin{equation}
\frac{d}{dt} Y_{u_{ij}}=\frac{1}{16 \pi^2} (Y_{u_{ik}} \gamma^{u^c_k}_{u^c_j}+ Y_{u_{kj}} \gamma^{Q_k}_{Q_i}+ Y_{u_{ij}} \gamma^{H_u}_{H_u}) \ ,
\end{equation}
where $t=-\ln Q$, with $Q$ the renormalization scale.

It is worth noticing here
that one-loop contributions in the $\mu\nu$SSM will generate one of the usual
lepton number violating terms mentioned in the introduction,
$\lambda'_{ijk} \hat L_i^a \hat Q_j^b \hat d^c_k$, as shown in Fig. \ref{RGEdiag}.
The corresponding RGEs are:
\begin{equation}
\frac{d}{dt}\lambda'_{ijk}=\frac{1}{16 \pi^2}Y_{d_{jk}}\gamma_{L_i}^{H_d}\ .
\end{equation}
%
% Then $\gamma^{E^c_j}_{E^c_i}$, $\gamma^{d^c_j}_{d^c_i}$,$\gamma^{u^c_j}_{u^c_i}$, $\gamma^{Q_j}_{Q_i} $, are the same as in the MSSM.
However, this contribution is proportional to the neutrino Yukawa coupling, 
and therefore can be neglected in the computation.

\begin{figure}[t]
\begin{center}
\epsfig{file=finalfigures/1loop_lambda.epsi,width=70mm}
\captions{One-loop generation of the $\lambda'_{ijk} \hat L_i^a \hat Q_j^b \hat d^c_k$ term in the superpotential. Note that it is proportional to $Y_{\nu}$, 
$Y_d$, and $\lambda$.} 
  \label{RGEdiag}
\end{center}
\end{figure}

Finally, 
for the VEVs we have
\begin{equation}
\frac{1}{16 \pi^2}\frac{d}{dt} v_u= - v_u \gamma^{H_u}_{H_u}\ ,
\end{equation}
\begin{equation}
\frac{1}{16 \pi^2}\frac{d}{dt} v_d= - v_d \gamma^{H_d}_{H_d} - \nu_i \gamma^{L_i}_{H_d}\ ,
\end{equation}
\begin{equation}
\frac{1}{16 \pi^2}\frac{d}{dt} \nu_i= - \nu_j \gamma^{L_j}_{L_i}- v_d \gamma^{H_d}_{L_i}\ ,
\end{equation}
\begin{equation}
\frac{1}{16 \pi^2}\frac{d}{dt} \nu^c_i= - \nu^c_j \gamma^{\nu^c_j}_{\nu^c_i}\ .
\end{equation}
% --Esto que viene ahora habra que ver como se pone realmente pero es crucial para demostrar que la masa del Higgs esta acotada casi de la misma forma que en el  NMSSM--
% 

%\pagebreak

\end{document}